\documentclass[table]{acmart}

\setcopyright{none}

\usepackage{placeins}
\usepackage{afterpage}

\usepackage{amssymb,amsfonts,amsmath}

\usepackage[mathscr]{eucal}

\usepackage{stmaryrd}

\usepackage{amsbsy}

\usepackage{bm}

\usepackage{braket}

\usepackage{xparse}

\usepackage{subfig}

\usepackage{algorithm,algpseudocode}

\usepackage{xfrac}

\usepackage{multirow}

\usepackage{tabularx}

\usepackage{pgfplots}
\usepackage{tikz}
\usetikzlibrary{3d}
\usetikzlibrary{patterns}
\usetikzlibrary{calc}
\usetikzlibrary{arrows}
\usetikzlibrary{matrix}

\usepackage[nameinlink]{cleveref}
\crefname{algorithm}{Alg.\@}{Algs.\@}
\crefname{figure}{Fig.\@}{Figs.\@}
\crefname{table}{Tab.\@}{Tabs.\@}
\crefname{section}{Sec.\@}{Secs.\@}

\crefformat{equation}{\textup{#2(#1)#3}}
\crefrangeformat{equation}{\textup{#3(#1)#4--#5(#2)#6}}
\crefmultiformat{equation}{\textup{#2(#1)#3}}{ and \textup{#2(#1)#3}}
{, \textup{#2(#1)#3}}{, and \textup{#2(#1)#3}}
\crefrangemultiformat{equation}{\textup{#3(#1)#4--#5(#2)#6}}%
{ and \textup{#3(#1)#4--#5(#2)#6}}{, \textup{#3(#1)#4--#5(#2)#6}}{, and \textup{#3(#1)#4--#5(#2)#6}}

\makeatletter
\newcounter{algorithmicH}%
\let\oldalgorithmic\algorithmic
\renewcommand{\algorithmic}{%
  \stepcounter{algorithmicH}%
  \oldalgorithmic}%
\renewcommand{\theHALG@line}{ALG@line.\thealgorithmicH.\arabic{ALG@line}}
\makeatother

\definecolor{color1}{RGB}{114,147,203}
\definecolor{color2}{RGB}{225,151, 76}
\definecolor{color3}{RGB}{132,186, 91}
\definecolor{color4}{RGB}{211, 94, 96}
\definecolor{color5}{RGB}{128,133,133}
\definecolor{color6}{RGB}{144,103,167}
\definecolor{color7}{RGB}{171,104, 87}
\definecolor{color8}{RGB}{204,194, 16}

\newcommand{\Tra}{{\sf T}} 

\DeclareMathOperator \diag {diag}

\DeclareMathOperator \mymod {mod}
\renewcommand{\mod}{\mymod}

\DeclareMathOperator \mydiv {div}
\renewcommand{\div}{\mydiv}

\newcommand{\qtext}[1]{\quad\text{#1}\quad}

\NewDocumentCommand \RangeSet { G{N} } {[\,#1\,]}

\NewDocumentCommand \LinearSymbol {} {\oast}
\NewDocumentCommand \LeftSymbol {} {\varolessthan}
\NewDocumentCommand \RightSymbol {} {\varogreaterthan}
\NewDocumentCommand \UnfoldSymbol {} {\oslash}

\NewDocumentCommand \SingleSize {s O{J} G{n}} {\IfBooleanTF{#1}{\bar}{} {#2}_{#3}^{\phantom{'}}} %
\NewDocumentCommand \LinearSize {s O{J}} {\IfBooleanTF{#1}{\bar}{} #2^{\LinearSymbol}}
\NewDocumentCommand \LeftSize {s O{J} G{n}} {\IfBooleanTF{#1}{\bar}{} {#2}_{#3}^{\LeftSymbol}}
\NewDocumentCommand \RightSize {s O{J} G{n}} {\IfBooleanTF{#1}{\bar}{} {#2}_{#3}^{\RightSymbol}}
\NewDocumentCommand \UnfoldSize {s O{J} G{n}} {\IfBooleanTF{#1}{\bar}{} {#2}_{#3}^{\UnfoldSymbol}}
\NewDocumentCommand \FullSize{s O{J} G{N}} {%
  \IfBooleanTF{#1}%
  {\SingleSize*[#2]{0} \times \SingleSize*[#2]{1} \times \cdots \times \SingleSize*[#2]{#3-1}}%
  {\SingleSize[#2]{0} \times \SingleSize[#2]{1} \times \cdots \times \SingleSize[#2]{#3-1}}%
}

\NewDocumentCommand \SingleIndex {s O{j} G{n}} {\IfBooleanTF{#1}{\bar}{} {#2}_{#3}} %
\NewDocumentCommand \UnfoldIndex {s O{j} G{n}} {\IfBooleanTF{#1}{\bar}{} {#2}_{#3}'} %

\NewDocumentCommand \FullIndex{s O{j} G{N}} {%
  \IfBooleanTF{#1}%
  {(\SingleIndex*[#2]{0}, \SingleIndex*[#2]{1}, \dots, \SingleIndex*[#2]{#3-1})}%
  {(#2_{0}, #2_{1}, \dots, #2_{#3-1})}%
}

\NewDocumentCommand \IndexVec {s O{j}} {\mathbf{\IfBooleanTF{#1}{\bar}{} {#2}}} %

\NewDocumentCommand \FullSubscript{s O{j} G{N}} {%
  \IfBooleanTF{#1}%
  {\SingleIndex*[#2]{0} \SingleIndex*[#2]{1} \dots \SingleIndex*[#2]{#3-1}}%
  {\SingleIndex[#2]{0} \SingleIndex[#2]{1} \dots \SingleIndex[#2]{#3-1}}%
}

\NewDocumentCommand \FacMat {s O{U} G{n}} {{\mathbf{\IfBooleanTF{#1}{\bar}{} #2}}_{#3}}

\NewDocumentCommand \FacGram {O{U} G{n}} {\FacMat[#1]{#2}^{\phantom{T}}\!\FacMat[#1]{#2}^\Tra}

\NewDocumentCommand \FacComp {G{n}} { \Bigl( \M{I} - \FacGram{#1} \Bigr) }

\NewDocumentCommand \UnfoldGram { } {\Mz{Y}{n}^{\phantom{T}} \Mz{Y}{n}^\Tra}

\NewDocumentCommand \FacSize {O{I} G{n}} {#1_{#2} \times R_{#2}}

\NewDocumentCommand \ColumnBlock {O{\Mz{Y}{n}} G{\ell}} {#1 \left[#2\right]} %

\newcommand{\T}[2][]{{\bm{#1{\mathscr{\MakeUppercase{#2}}}}}}

\newcommand{\V}[2][]{{\bm{#1\mathbf{\MakeLowercase{#2}}}}} 

\newcommand{\M}[2][]{{\bm{#1\mathbf{\MakeUppercase{#2}}}}} 

\newcommand{\Mz}[3][]{\M[#1]{#2}_{(#3)}}

\newcommand{\X}{\T{X}}
\NewDocumentCommand \Y { s } {\IfBooleanTF{#1}{\T[\bar]{Y}}{\T{Y}}}

\newcommand{\G}{\T{G}}
\NewDocumentCommand \Xn {G{n}} {\T[\tilde]{X}_{#1}}
\NewDocumentCommand \Gn {G{n}} {\T[\tilde]{G}_{#1}}
\newcommand{\Xhat}{ \T[\hat]{X} }
\NewDocumentCommand \Jnpn { G{n} G{p_{#1}} } {\mathcal{J}_#1(#2)}
\NewDocumentCommand \Jp { G{\FullIndex[p]} } {\mathcal{J}#1}

\DeclareMathOperator \idxtolin {idx2lin}
\NewDocumentCommand \IdxToLin { G{j} G{J} } { \idxtolin \bigl\{ \, \FullIndex[#1] \, , \, \FullIndex[#2] \, \bigr\} }
\DeclareMathOperator \lintoidx {lin2idx}
\NewDocumentCommand \LinToIdx { } { \lintoidx \bigl\{ \, j' \, , \, \FullIndex[J] \, \bigr\} }
\DeclareMathOperator \lsz {lsz}
\NewDocumentCommand \Lsz { G{J_n} G{\bar p_n} } { \lsz \{\, #2, #1, P_n \} }
\NewDocumentCommand \Dn {} {(J_n \div P_n)}
\NewDocumentCommand \Mn {} {(J_n \mod P_n)}
\DeclareMathOperator \gbltoprc {glb2prc}
\NewDocumentCommand \GblToPrc { } { \gbltoprc \{\, j_n, J_n, P_n \} }
\DeclareMathOperator \prcmap {prcmap}
\NewDocumentCommand \PrcMap { G{J_n} } { \prcmap \{\, \bar p_n, {#1}, P_n \} }
\DeclareMathOperator \gbltolcl {gbl2lcl}
\NewDocumentCommand \GblToLcl { } {\gbltolcl \{\, j_n, \bar p_n, J_n, P_n \} }
\DeclareMathOperator \lcltogbl {lcl2gbl}
\NewDocumentCommand \LclToGbl { } {\lcltogbl \{\, \bar j_n, \bar p_n, J_n, P_n \} }

\NewDocumentCommand \bigO {} {\mathcal{O}}

\newcommand{\local}[1]{\overline{#1}} %
\newcommand{\lt}{\left} %
\newcommand{\rt}{\right} %

\newcommand{\mTTM}{Multi-TTM}

\newcommand{\exsize}{3 \times 4 \times 3 \times 2}
\newcommand{\expsize}{2 \times 2 \times 2 \times 1} %

\newcommand{\spsize}{$500 \times 500 \times 500 \times 11 \times 400$}
\newcommand{\spgrida}{$1 \times 1 \times 40 \times 1 \times 100$}
\newcommand{\spgridb}{$10 \times 8 \times 5 \times 1 \times 10$}
\newcommand{\spgridc}{$40 \times 10 \times 1 \times 1 \times 10$}

\newcommand{\jisize}{$1500 \times 2080 \times 1500 \times 18 \times 10$}
\newcommand{\jigrida}{$1 \times 16 \times 35 \times 1 \times 10$}
\newcommand{\jigridb}{$10 \times 8 \times 7 \times 1 \times 10$}
\newcommand{\jigridc}{$35 \times 16 \times 1 \times 1 \times 10$}

\newcommand{\datafile}{}
\newcommand{\pga}{}
\newcommand{\pgb}{}
\newcommand{\pgc}{}
\newcommand{\pgalabel}{}
\newcommand{\pgblabel}{}
\newcommand{\pgclabel}{}
\newcommand{\ma}{}
\newcommand{\mb}{}
\newcommand{\tola}{}
\newcommand{\tolb}{}
\newcommand{\prep}{}

\newif\iflegend
\legendtrue
\newif\ifylabel
\ylabeltrue

\newcommand{\gramplotoptions}{
  ybar stacked,
  reverse legend,
  bar width=10pt,
  \ifylabel
  	ylabel={Time (seconds)}, 
  \fi
  y label style={yshift=-.5cm},
  ymin=0,
  ymax=150,
  symbolic x coords={old-\ma-\pga,new-\ma-\pga,,old-\mb-\pga,new-\mb-\pga,,,old-\ma-\pgb,new-\ma-\pgb,,old-\mb-\pgb,new-\mb-\pgb,,,old-\ma-\pgc,new-\ma-\pgc,,old-\mb-\pgc,new-\mb-\pgc},
  xticklabels={Old,Old,Old,Old,Old,Old,New,New,New,New,New,New},  %
  xtick=data,
  x tick label style={rotate=270},
  legend pos={north east},
  legend style={draw=none, cells={align=left}, nodes={scale=0.8}},
}

\newcommand{\makegramplot}{
  \begin{axis}[\gramplotoptions
    after end axis/.code={ %
      \node(O11) at (axis cs:old-\ma-\pga,1) {};
      \node(O14) at (axis cs:new-\ma-\pga,1) {};
      \node(O1) at ($(O11)!0.5!(O14)$,1) {};
      \node(N11) at (axis cs:old-\mb-\pga,1) {};
      \node(N14) at (axis cs:new-\mb-\pga,1) {};
      \node(N1) at ($(N11)!0.5!(N14)$,1) {};
      \node(O21) at (axis cs:old-\ma-\pgb,1) {};
      \node(O24) at (axis cs:new-\ma-\pgb,1) {};
      \node(O2) at ($(O21)!0.5!(O24)$,1) {};
      \node(N21) at (axis cs:old-\mb-\pgb,1) {};
      \node(N24) at (axis cs:new-\mb-\pgb,1) {};
      \node(N2) at ($(N21)!0.5!(N24)$,1) {};
      \node(O31) at (axis cs:old-\ma-\pgc,1) {};
      \node(O34) at (axis cs:new-\ma-\pgc,1) {};
      \node(O3) at ($(O31)!0.5!(O34)$,1) {};
      \node(N31) at (axis cs:old-\mb-\pgc,1) {};
      \node(N34) at (axis cs:new-\mb-\pgc,1) {};
      \node(N3) at ($(N31)!0.5!(N34)$,1) {};
      \node(XTL) at (xticklabel cs:0)  {};
      \node[anchor=north] at (O1 |- XTL) {\small mode \ma};
      \node[anchor=north] at (N1 |- XTL) {\small mode \mb};
      \node[anchor=north] at (O2 |- XTL) {\small mode \ma};
      \node[anchor=north] at (N2 |- XTL) {\small mode \mb};
      \node[anchor=north] at (O3 |- XTL) {\small mode \ma};
      \node[anchor=north] at (N3 |- XTL) {\small mode \mb};
      \node(PG1) at ($(O1)!0.5!(N1)$,1) {};
      \node(PG2) at ($(O2)!0.5!(N2)$,1) {};
      \node(PG3) at ($(O3)!0.5!(N3)$,1) {};
      \node[yshift=-10,anchor=north] at (PG1 |- XTL) {\small \pgalabel};
      \node[yshift=-10,anchor=north] at (PG2 |- XTL) {\small \pgblabel};
      \node[yshift=-10,anchor=north] at (PG3 |- XTL) {\small \pgclabel};
      \node[anchor=south] at (axis cs:old-\mb-\pga,151) {$\vdots$};
      \iflegend
      \else
          \node[anchor=south] at (axis cs:old-\ma-\pgc,151) {$\vdots$};
      \fi
    }]
    \pgfplotsset{cycle list={color1, fill=color1 \\ color3, fill=color3 \\ color4, fill=color4 \\ color2, fill=color2 \\ color5, fill=color5 \\}}
    
    \addplot table[x=alg-mode-pg, y expr=(\thisrow{comp})] {\datafile};
    \addplot table[x=alg-mode-pg, y expr=(\thisrow{tcomm}] {\datafile};
    \addplot table[x=alg-mode-pg, y expr=(\thisrow{mcomm})] {\datafile};
    \addplot table[x=alg-mode-pg, y expr=(\thisrow{pack})] {\datafile};
    \addplot table[x=alg-mode-pg, y expr=(\thisrow{other})] {\datafile};
    \iflegend
    	\legend{Computation, Tensor Comm, Matrix Comm, Packing, Other};
    \fi
  \end{axis}
}

\newcommand{\sthosvdplotoptions}{
  ybar stacked,
  reverse legend,
  bar width=20pt,
  ylabel={Time (seconds)}, 
  ymin=0,
  symbolic x coords={\prep-\tola-\pga,\prep-\tolb-\pga,,\prep-\tola-\pgb,\prep-\tolb-\pgb,,\prep-\tola-\pgc,\prep-\tolb-\pgc},
  xticklabels={High,High,High,Low,Low,Low},  %
  xtick=data,
  legend pos={north west},
  legend style={draw=none, cells={align=left}},
}

\newcommand{\makesthosvdplot}{
  \begin{axis}[\sthosvdplotoptions
    after end axis/.code={ %
      \node(AA) at (axis cs:\prep-\tola-\pga,1) {};
      \node(BA) at (axis cs:\prep-\tolb-\pga,1) {};
      \node(AB) at (axis cs:\prep-\tola-\pgb,1) {};
      \node(BB) at (axis cs:\prep-\tolb-\pgb,1) {};
      \node(AC) at (axis cs:\prep-\tola-\pgc,1) {};
      \node(BC) at (axis cs:\prep-\tolb-\pgc,1) {};
      \node(PGA) at ($(AA)!0.5!(BA)$,1) {};
      \node(PGB) at ($(AB)!0.5!(BB)$,1) {};
      \node(PGC) at ($(AC)!0.5!(BC)$,1) {};
      \node(XTL) at (xticklabel cs:0)  {};
      \node[anchor=north] at (PGA |- XTL) {\small \pgalabel};
      \node[anchor=north] at (PGB |- XTL) {\small \pgblabel};
      \node[anchor=north] at (PGC |- XTL) {\small \pgclabel};
    }]
	\pgfplotsset{cycle list={color1, fill=color1 \\ color4, fill=color4 \\ color3, fill=color3 \\}}

	\legend{Gram, Evecs, TTM}

	\addplot table[x=prep-tol-pg, y=gram(0)] {\datafile};
	\addplot table[x=prep-tol-pg, y=eig(0)] {\datafile};
	\addplot table[x=prep-tol-pg, y=ttm(0)] {\datafile};
	\addplot table[x=prep-tol-pg, y=gram(1)] {\datafile};
	\addplot table[x=prep-tol-pg, y=eig(1)] {\datafile};
	\addplot table[x=prep-tol-pg, y=ttm(1)] {\datafile};
	\addplot table[x=prep-tol-pg, y=gram(2)] {\datafile};
	\addplot table[x=prep-tol-pg, y=eig(2)] {\datafile};
	\addplot table[x=prep-tol-pg, y=ttm(2)] {\datafile};
	\addplot table[x=prep-tol-pg, y=gram(3)] {\datafile};
	\addplot table[x=prep-tol-pg, y=eig(3)] {\datafile};
	\addplot table[x=prep-tol-pg, y=ttm(3)] {\datafile};
	\addplot table[x=prep-tol-pg, y=gram(4)] {\datafile};
	\addplot table[x=prep-tol-pg, y=eig(4)] {\datafile};
	\addplot table[x=prep-tol-pg, y=ttm(4)] {\datafile};

\end{axis}
}

\newcommand{\makeweakscalingplot}{
\begin{axis}[
	legend pos = north east,
	ylabel={GFLOPS per Core}, 
	xlabel={Number of Nodes},
	xmode=log,
	log basis x={2},
	xtick=data,
	xticklabels={1,16,81,256,625,1296},
	xmin=-5,
	ymin=0,
	ymax=20.8,
	ytick={5,10,15,20.8},
	y tick label style={rotate=90},
        ylabel near ticks,
        reverse legend
]
	\legend{ST-HOSVD}

	\addplot table[x expr=\thisrow{numNodes}, y=GFLOPS/proc] {\datafile};

\end{axis}
}

\newcommand{\makestrongscalingplot}{
\begin{axis}[
	legend pos = north east,
	reverse legend,
	ylabel={Time (seconds)}, 
	xlabel={Number of Nodes},
	xtick=data,
	xticklabels={1,2,4,8,16,32,64,128,256,512},
	xmode=log,
	log basis x ={2},
	ymode=log,
	log basis y={2},
	y tick label style={rotate=90},
        ylabel near ticks,
]
	\legend{ST-HOSVD}

	\addplot table[x=numNodes, y=runtime] {\datafile};

\end{axis}
}

\acmJournal{TOMS}

\title[TuckerMPI: Large-scale Data Compression via the Tucker Tensor Decomposition]{TuckerMPI: A Parallel C++/MPI Software Package for Large-scale Data Compression via the Tucker Tensor Decomposition}

\author{Grey Ballard}
\affiliation{%
  \institution{Wake Forest University}
  \department{}
  \streetaddress{}
  \city{Winston-Salem}
  \state{North Carolina}
  \postcode{27109}}
\email{ballard@wfu.edu}

\author{Alicia Klinvex}
\affiliation{%
  \institution{Sandia National Laboratories}
  \department{}
  \streetaddress{}
  \city{}
  \state{}
  \postcode{}}
\email{aklinvex@gmail.com}

\author{Tamara G. Kolda}
\affiliation{%
  \institution{Sandia National Laboratories}
  \department{}
  \streetaddress{}
  \city{Livermore}
  \state{California}
  \postcode{94550}}
\email{tgkolda@sandia.gov}

\terms{Data compression, Distributed algorithms}

\begin{abstract}
Our goal is compression of massive-scale grid-structured data, such as the multi-terabyte output of a high-fidelity computational simulation.
For such data sets, we have developed a new software package called TuckerMPI, a parallel C++/MPI software package for compressing distributed data.
The approach is based on treating the data as a tensor, i.e., a multidimensional array, and computing its truncated Tucker decomposition,
a higher-order analogue to the truncated singular value decomposition of a matrix.
The result is a low-rank approximation of the original tensor-structured data.
Compression efficiency is achieved by detecting latent
global structure within the data, which we contrast to most compression methods that are focused on local structure.
In this work, we describe TuckerMPI, our implementation of the truncated Tucker decomposition, including details
of the data distribution and in-memory layouts, the parallel and serial implementations of the key kernels, and analysis of the storage, communication, and computational costs.
We test the software on 4.5~terabyte and 6.7~terabyte data sets distributed across 100s of nodes (1000s of MPI processes), achieving compression ratios between 100--200,000$\times$ which equates to 99-99.999\% compression (depending on the desired accuracy)
in substantially less time than it would take to even read the same dataset from a parallel filesystem.
Moreover, we show that our method also allows for reconstruction of partial or down-sampled data on a single node,  without a parallel computer so long as the reconstructed portion is small enough to fit on a single machine, e.g., in the instance of reconstructing/visualizing a single down-sampled time step or computing summary statistics.
\end{abstract}
\begin{document}

\maketitle
%

\section{Introduction}

The convergence of ever-faster computational platforms and algorithmic
advancements in scientific simulations have led to a data deluge ---
it is now possible to produce very fine-grained and detailed simulations,
resulting in terabyte-sized or larger datasets.
Consider the problem we discuss later on in this work: a simulation on a
three-dimensional rectangular grid of size $500 \times 500 \times 500$,
tracking 11 variables for 400 time steps.
Even this modest-sized simulation yields 4~TB of data in double precision.
With current computational platforms, this data cannot easily be stored, moved,
visualized, or analyzed.
Nevertheless, it is well known to simulation scientists that their massive
datasets have extensive latent structure and are therefore highly compressible.
The problem is how to discover the redundancies automatically. 

Most compression methods focus on compressing \emph{local} structure with
very little loss in precision.
Fout, Ma, and Ahrens~\cite{FoMaAh05} do multivariate volume block data reduction
to take advantage of local multiway structure, achieving up to 70\% compression. 
Likewise, Lindstrom's ZFP compresses data in local blocks  \cite{Li14} with 1.3--2.6$\times$ (23--61\%) compression. More recently, Di and Cappello \cite{DiCa16} report up 3--436$\times$ (66--99.8\%) compression, but again focusing on local blocks.

Our method, in contrast, aims at detecting \emph{global} structure in the data,
yielding much higher compression ratios in exchange
for potential loss in local accuracy.
It does not process the data in blocks but rather considers the data in its entirety.
Before we delve into the mathematical details, we note briefly that lossy
compression need not replace the original data; rather, the compressed version
is a thumbnail or preview of the full dataset, which may reside on long-term
storage or be regenerated.

In this paper, we consider the Tucker tensor decomposition \cite{Tu66}, also
known as the higher-order singular value decomposition (HOSVD) \cite{DeDeVa00}.
This is an effective tool for compression for many application
domains \cite{VaTe02,KaYaMe12,HaCaTe12,AfGiTa14,BaPa15,GaSoVe16}.
The idea is to consider the multiway structure of the data. For instance, the
problem described above is a five-way object, and so we exploit this
structure in the compression procedure.
Specifically, each mode is compressed individually by determining a small
number of vectors that span the \emph{fibers} in that mode, fibers being the
vectors that are the higher-order analogue of matrix rows and columns.
We aim to compress an
$N$-way tensor of size $\FullSize[I]$ to size $\FullSize[R]$ by computing $R_n$
vectors for each mode $n$ that approximately span the range of the mode-$n$ fibers.
We call $R_n$ the \emph{rank} of mode $n$.
The ranks are typically selected to retain a specified relative
accuracy $\epsilon$, i.e., if $\X$ is the data tensor and $\Xhat$ is the reconstruction
from the compressed representation, then we choose the ranks such that 
\begin{equation}\label{eq:epsilon}
  \| \X - \Xhat \| \leq \epsilon \| \X \|.
\end{equation}
Choosing $R_n=I_n$ yields perfect
reconstruction, so viable choices for the ranks always exist.
If we assume $I_n=I$ and $R_n=R$ for all $n$, for sake of exposition,
then the storage is reduced from $I^N$ to
$R^N + NIR$, and so the compression ratio is $\approx (I/R)^N$, which is the size of the original data divided by the size of the compressed representation.

In this paper, we develop a parallel implementation of the
sequentially-truncated HOSVD (ST-HOSVD) \cite{VaVaMe12}.  This papers
builds on past work by Austin, Ballard, and Kolda \cite{AuBaKo16},
which showed initial results for parallel versions of ST-HOSVD as well
as the higher-order orthogonal iteration (HOOI).
Our contributions are as follows:
\begin{enumerate}
\item We describe the details of the TuckerMPI software which were not provided in \cite{AuBaKo16}, including global data
  distributions (even in the case when a tensor dimension is not divisible by the number of processors in that dimension), local data layouts, and sequential and parallel algorithms.
\item We present a new and improved kernel for computing the Gram matrix in
  ST-HOSVD, which is faster and less sensitive to the
  parallel data layout
 than the version in \cite{AuBaKo16}.
\item We present more extensive compression results for terabyte-scale
  data by compressing data sets that are an order of magnitude larger than those in \cite{AuBaKo16}. Specifically, we show that we can compress 4.5 and 6.7~TB datasets by 2--5 orders of magnitude in  O(10-100) seconds, less time then reading the data from the parallel file system.
\item An advantage of the Tucker decomposition that was mentioned but not implemented for \cite{AuBaKo16} is that we can
  reconstruct just a portion of the full tensor. Here, we present an
  efficient method for partial reconstruction, taking care not to create any object
  that is larger than the compressed or reconstructed subtensor so that it can be run on a workstation rather than a parallel system.
  We give experimental results to showcase its efficiency.
\end{enumerate}

%

%
%
%
%
%

\section{Notation and Mathematical Background}
\label{sec:notation}

We use boldface Euler script letters to denote tensors ($\X$)
and boldface uppercase letters to denote matrices ($\M{X}$).
We reserve the uppercase letters $I,J,K,L,M,N,P,R$ to denote sizes and
the corresponding lowercase letters $i,j,k,l,m,n,p,r$ to denote the
indices. 
We use zero-indexing throughout so that if $i$ is the
index corresponding to size $I$, then we have $i=0,1,\dots,I-1$.
For any size $I$,
we use the notation $\RangeSet{I}$ to denote the set $\set{0,1,\dots,I-1}$. 

\subsection{Sizes}
\label{sec:size-computations}

We define a few special quantities with respect to
tuples of sizes, which are used in describing both tensor and processor grid sizes.
For an object with dimensions $\FullSize[I]$,
we define the product of all its sizes (the total size) as
\begin{equation}\label{eq:linear-size}
  \LinearSize[I] = \prod_{n \in \RangeSet} I_n.
\end{equation}
We further define some quantities that depend on the mode $n \in \RangeSet$:
\begin{align}
  \label{eq:unfold-size}
  \UnfoldSize[I] &= \prod_{k \neq n} I_{k} = \LinearSize[I] / I_n, && \text{(product of all sizes except mode $n$)},\\
  \label{eq:left-size}
  \LeftSize[I] & = \prod_{k < n} I_{k}  = \UnfoldSize[I] / \RightSize[I], && \text{(product of sizes below mode $n$)}, \\
  \label{eq:right-size}
  \RightSize[I] & = \prod_{k > n} I_{k} = \UnfoldSize[I] / \LeftSize[I],  && \text{(product of sizes above mode $n$)}.
\end{align}
For the edge cases, we say $\LeftSize[I]{0} = 1$ and $\RightSize[I]{N-1} = 1$.

\subsection{Tensor Operations}
\label{sec:tensor-operations}

We discuss key tensor operations for computing the Tucker decomposition.
Here we assume an $N$-way tensor $\X$ of size $\FullSize[I]$.

\subsubsection{Tensor Unfolding}
\label{sec:tensor-unfolding}

The mode-$n$ unfolding of $\X$ rearranges the elements of the $N$-way tensor into a matrix, denoted
$\Mz{X}{n}$, of size $I_n \times \UnfoldSize[I]$. 
We map tensor element $\FullIndex[i]$ to matrix element
$(i_n,\UnfoldIndex[i])$ where
\begin{equation*}
  \label{eq:unfold-index}
  \UnfoldIndex[i] = \sum_{k<n} i_k \cdot \LeftSize[I]{k} + \sum_{k>n} i_k \cdot \LeftSize[I]{k} / I_n .
\end{equation*}
See also \cref{sec:unfolded_layout} for examples of unfolded tensors
and details of the organization in computer memory.

\subsubsection{Tensor Norm}
\label{sec:tensor-norm}

The norm of a tensor is the square root of the sum of squares of all the elements.
This means it is equivalent to the Frobenious norm of any unfolding, i.e.,
$\|\X\| = \|\Mz{X}{n}\|_F$ for any $n \in \RangeSet{N}$.

\subsubsection{Tensor Times Matrix}
\label{sec:tensor-times-matrix}

The \emph{mode-$n$ product} of $\X$ with a 
matrix $\M{U}$ of size $J \times I_n$
is denoted $\X \times_n \M{U}$, and 
the result is of size $I_{0} \times \cdots \times I_{n-1} \times J
\times I_{n+1} \times \cdots \times I_{N-1}$.
This can be expressed in terms of unfolded tensors, i.e.,
\begin{displaymath}
  \Y = \X \times_n \M{U}
  \quad \Leftrightarrow \quad
  \Mz{Y}{n} = \M{U}\Mz{X}{n}.
\end{displaymath}
This is also known as the tensor-times-matrix (TTM) product.
For different modes, the order is irrelevant so that
\begin{displaymath}
  \X \times_m \M{U} \times_n \M{V} =
  \X \times_n \M{V} \times_m \M{U} \text{ for } m \neq n.
\end{displaymath}
In the same mode, order matters so that 
\begin{equation}\label{eq:same_mode}
  \X \times_n \M{U} \times_n \M{V} =
  \X \times_n \M{VU},
\end{equation}
where $\M{V}$ is $K \times J$.

%
%

%
%

%
%
%
%
%
%
%
%

%
%

%
%
%
%
%
%
%
%
%
%
%
%
%
%
%

%
%

%
%
%
%
%
%
%
%
%
%
%
%
%
%
%
%
%
%
%
%
%
%
%
%

%
%
%
%
%

\section{Review of the Sequentially-truncated higher-order SVD}
\label{sec:sequ-trunc-high}

In this section, we describe the Tucker decomposition and
review the ST-HOSVD method that we parallelize to compute it. Before we
do so, we introduce some notation and basic theory.

Let the $N$-way tensor $\X$ of size $\FullSize[I]$ denote the data tensor to be compressed. 
The goal is to approximate $\X$ as
\begin{equation}\label{eq:tucker}
  \X \approx \Xhat \equiv \G \times_0 \FacMat{0} \times_1 \FacMat{1}
  \cdots \times_{N-1} \FacMat{N-1}.
\end{equation}
The tensor $\G$ is called the \emph{core tensor}, and its size is denoted by
$\FullSize[R]$.
Each \emph{factor matrix} $\FacMat$ is necessarily of size $I_n \times R_n$ for $n \in \RangeSet{N}$,
and we assume throughout that the factor matrices have orthonormal columns.
This means $\FacMat^\Tra\FacMat = \M{I}$ (the $R_n \times R_n$ identity matrix).
The storage of $\X$ is $\LinearSize[I]$
as compared to the storage for $\Xhat$ which is 
$\LinearSize[R] + \sum_{n\in \RangeSet{N}} R_n I_n$.
If, for example, $I_n / R_n = 2$ for all $n$,
then the compression ratio is $\LinearSize[I]/\LinearSize[R] \approx 2^N$.

We review a few relevant facts about Tucker per \cite[Section 4.2]{KoBa09}.
If the factor matrices are given, then it can be shown
that the optimal $\G$ is 
\begin{equation}
  \label{eq:G}
  \G = \X \times_0 \FacMat{0}^\Tra \times_1 \FacMat{1}^\Tra
  \cdots \times_{N-1} \FacMat{N-1}^\Tra.
\end{equation}
Substituting \cref{eq:G} back into \cref{eq:tucker}
and using identity \cref{eq:same_mode}, we have
\begin{equation}\label{eq:Xhat}
  \Xhat = \X
  \times_0 \FacGram{0}
  \times_1 \FacGram{1}
  \cdots
  \times_{N-1} \FacGram{N-1}.
\end{equation}
Thus, the factor matrices determine the projections (one per mode) of
the original tensor down to the reduced space.
It is easy to show that $\|\X - \Xhat\|^2 = \|\X\|^2 - \|\G\|^2$,
which means that $\G$ retains most
of the \emph{mass} of $\X$ but is just represented with respect to a different basis.

It is convenient to make a few special definitions.
Assume that the factor matrices are specified. Then define the \emph{partial core} that is
the result of applying $n+1$ factor matrices to $\X$:
\begin{equation}
  \label{eq:Gn}
  \Gn = \X \times_0 \FacMat{0}^\Tra \cdots \times_{n} \FacMat{n}^\Tra
  \qtext{for} n \in \RangeSet{N}.
\end{equation}
This means that $\Gn$ is only reduced in the first $(n+1)$ modes and so is
of size $R_0 \times \cdots \times R_{n} \times I_{n+1} \cdots \times I_{N-1}$.
By definition, $\G = \T[\tilde]{G}_{N-1}$.
Likewise, we can define the \emph{incremental approximation} using the partial core to be 
\begin{equation}
  \label{eq:Xnhat}
  \Xn = \G \times_0 \FacMat{0} \cdots \times_{n} \FacMat{n}
  = \X \times_0 \FacGram{0} \cdots \times_{n} \FacGram{n}
  \qtext{for} n \in \RangeSet{N}.
\end{equation}
By definition, $\Xhat = \Xn{N-1}$.

\subsection{Key Theorem}
\label{sec:key-theorem}

The following theorem categorizes the error in terms of the incremental approximations
and orthogonal projections based on the factor matrices. We refer the reader to
\cite{VaVaMe12} for the proof. This theorem is key to understanding how to pick the factor matrices
in the ST-HOSVD.

\begin{theorem}[Vannieuwenhoven, Vandebril, and Meerbergen {\cite[Theorem 5.1]{VaVaMe12}}]\label{thm:error}
  Let $\X$ be a tensor of size $\FullSize[I]$ which is approximated by
  $\Xhat$ as defined in \cref{eq:Xhat} in which the factor matrices
  $\FacMat$ of size $R_n \times I_n$ have orthonormal columns.
  The approximation error is then given by
  \begin{equation}
    \label{eq:error-Xnhat}
    \Bigl\|\; \X - \Xhat \;\Bigr\|^2 =
    \Bigl\|\; \X \times_0 \FacComp{0} \;\Bigr\|^2 +
    \Bigl\|\; \Xn{0}  \times_1 \FacComp{1} \;\Bigr\|^2 +
    \cdots +
    \Bigl\|\; \Xn{N-2} \times_{N-1} \FacComp{N-1} \;\Bigr\|^2,
  \end{equation}
  where $\Xn$ is as defined in \cref{eq:Xnhat}.
  Additionally, the error is bounded by
  \begin{equation}
    \label{eq:error_bound}
    \Bigl\|\; \X - \Xhat \;\Bigr\|^2 \leq
    \sum_{n \in \RangeSet{N}} \Bigl\|\; \X \times_n \FacComp{n} \;\Bigr\|^2 .
  \end{equation}
\end{theorem}

\subsection{HOSVD Method}
\label{sec:hosvd-method}

The bound in \cref{eq:error_bound} from \Cref{thm:error} is key to understanding
the HOSVD method, originally known as Tucker1 \cite{Tu66,DeDeVa00}.
Suppose the goal is to find a Tucker decomposition of the form in \cref{eq:tucker}
with relative error no greater then $\epsilon$, i.e.,
\begin{equation}
  \label{eq:error-goal}
 \frac{\| \X - \Xhat \| }{\| \X \|} \leq \epsilon.
\end{equation}
Then the idea is as follows. Let the eigenvalue decomposition of the Gram matrix $\M{S}$ of
the mode-$n$ unfolding be given by
\begin{equation}\label{eq:Xeigprob}
   \M{S} \equiv \Mz{X}{n}^{\phantom{'}} \Mz{X}{n}^\Tra = \M{V}\M{\Lambda}\M{V}^\Tra.
\end{equation}
Here, $\M{\Lambda} = \diag(\set{ \lambda_1, \dots, \lambda_{I_n}})$ and
$\lambda_1 \geq \lambda_2 \cdots \geq \lambda_{I_n} \geq 0$ are the eigenvalues in descending order.
The matrix $\M{V}$ contains the corresponding eigenvectors.
We choose $\FacMat$ and $R_n$ so that
\begin{equation}\label{eq:Un}
  \FacMat = \M{V}(\,:\,, 1\!:\!R_n)
  \qtext{where}
  R_n = \min_{R \in \RangeSet{I_n}} R
  \quad \text{subject to }
  \sum_{i=R+1}^{I_n} \lambda_i \leq \epsilon^2 \| \X \|^2 /N.
\end{equation}
This choice of $\FacMat$ ensures that
\begin{displaymath}
  \Bigl\|\; \X \times_n \FacComp{n} \;\Bigr\|^2 \leq \epsilon^2 \| \X \|^2 /N.
\end{displaymath}
Repeating this procedure for each $n \in \RangeSet{N}$ ensures that the desired error bound \cref{eq:error-goal} holds.
This discussion can be framed equivalently in terms of the leading left singular vectors of $\Mz{X}{n}$,
which is how the HOSVD is usually described. However, we explicitly form the Gram matrices
in our parallelized method, so this presentation is more convenient in our exposition.

\subsection{ST-HOSVD}
\label{sec:st-hosvd}

The HOSVD works with the full tensor at each step.
The idea behind the sequentially-truncated version
introduced in \cite{VaVaMe12} is that we
work with the partial cores.
In other words, at step $n$, we compute the next factor matrix based on $\Gn{n-1}$.
This is possible because we can swap $\Gn{n-1}$ for $\Xn{n-1}$ in the summands
in the error expression \cref{eq:error-Xnhat}
due to the following equivalence:
\begin{equation}\label{eq:error-summand}
  \Bigl\|\; \Xn{n-1} \times_n \FacComp{n} \;\Bigr\|^2 =
  \Bigl\|\; \Gn{n-1} \times_n \FacComp{n} \;\Bigr\|^2 .
\end{equation}
Hence, substituting \cref{eq:error-summand} into \cref{eq:error-Xnhat}
then yields the following key corollary.

\begin{corollary}\label{cor:error}
  Let the conditions of \Cref{thm:error} hold. Then
  \begin{equation}
    \label{eq:error}
    \Bigl\|\; \X - \Xhat \;\Bigr\|^2 =
    \Bigl\|\; \X \times_0 \FacComp{0} \;\Bigr\|^2 +
    \Bigl\|\; \Gn{0}  \times_1 \FacComp{1} \;\Bigr\|^2 +
    \cdots +
    \Bigl\|\; \Gn{N-2} \times_{N-1} \FacComp{N-1} \;\Bigr\|^2,
  \end{equation}
  where $\Gn$ is as defined in \cref{eq:Gn}.
\end{corollary}

\Cref{cor:error} means that we can pick the $(n+1)$st factor matrix, $\FacMat$,
using the partial core $\Gn{n-1}$.
Choosing $\FacMat{0}$ is unchanged.
For $n>1$, however, we replace the eigenvalue problem
in \cref{eq:Xeigprob} with
\begin{equation}
  \label{eq:2}
  \M{S} \equiv \UnfoldGram = \M{V}\M{\Lambda}\M{V}^\Tra \qtext{where} \Y = \Gn{n-1}.
\end{equation}
We can still pick $\FacMat$ according to \cref{eq:Un}, but we are just
working with the eigen-decomposition of a different matrix.
The ST-HOSVD algorithm is given in \cref{alg:sthosvd}.

\begin{algorithm}[ht]
  \caption{Sequentially Truncated Higher-Order SVD (ST-HOSVD) \cite{VaVaMe12}}
  \label{alg:sthosvd}
  \begin{algorithmic}[1]
    \Procedure{$( \G, \mathcal{U} )=$ ST-HOSVD}{$\X$, $\epsilon$}
    \Comment{$\X$ is data tensor of size $I_0 \times \cdots \times I_{N-1}$ and $\epsilon$ is desired accuracy}
    \State $\Y \gets \X$ 
    \For{$n =0,1,\dots,N-1$}\label{line:hosvd:loop}
    \State \label{line:hosvd:gram}
    $\M{S} \gets \UnfoldGram$
    \Comment{$\M{S}$ is referred to as the \emph{Gram matrix}}
    \State{$\left(\V{\lambda},\M{V}\right) \gets \text{eig}(\M{S})$}
    \Comment{Eigenvalues in descending order}
    \State $R_n \gets \min \set{ R \in \RangeSet{I_n} | \sum_{i=R+1}^{I_n} \lambda_i \leq
      \epsilon^2\|\X\|^2/N }$
    \Comment{Choose $R_n$ to satisfy the error bound}
    \State \label{line:hosvd:evecs}
    $\FacMat \gets \M{V}(\,:\, , 1\!:\!R_n)$
    \Comment{$\FacMat$ is the leading $R_n$ eigenvectors of $\M{S}$}
    \State \label{line:hosvd:ttm} $\Y \gets \Y \times_{n}
    \FacMat^\Tra$
    \Comment{Sets $\Y = \Gn$, the partial core}
    \EndFor
    \State $\G \gets \Y$
    \Comment{Core of size $\FullSize[R]$}
    \State $\mathcal{U} \gets \set{ \FacMat{0},\dots,\FacMat{N-1} }$
    \Comment{Factor matrices with $\FacMat{n}$ of size $I_n \times R_n$ for all $n \in \RangeSet{N}$}
    \State \Return $( \G, \mathcal{U} )$
    \Comment{Result satisfies $\| \X - \Xhat \|/\|\X\| \leq \epsilon$ where $\Xhat = \G \times_0 \FacMat{0} \cdots \times_{N-1} \FacMat{N-1}$ }
    \EndProcedure
  \end{algorithmic}
\end{algorithm}

\subsection{Quasi-Optimality}
\label{sec:quasi-optimality}
Neither the HOSVD nor the ST-HOSVD yields an optimal rank-$(R_0,R_1,\dots,R_{N-1})$ decomposition, and neither is necessarily more accurate than the other.
The major advantage of the ST-HOSVD is in terms of computational cost:
the cost of the Gram computation is decreased by a factor of $I_n/R_n$ at each step.
Both methods yield quasi-optimal results with error within a factor of the square root of the number of modes of the best approximation, as specified by the following theorem from \cite{Hackbusch14}.

\begin{theorem}[Hackbusch {\cite[Theorems 6.9-10]{Hackbusch14}}]\label{thm:opt}
  Let $\X$ be a tensor of size $\FullSize[I]$, $\Xhat_{\rm{HOSVD}}$ be the rank $\FullSize[R]$ approximation computed by {\rm HOSVD}, and $\Xhat_{\rm{ST-HOSVD}}$ be the rank $\FullSize[R]$ approximation computed by {\rm ST-HOSVD}.
  Then

  \begin{equation*}
    \Bigl\|\; \X - \Xhat_{\rm{HOSVD}} \;\Bigr\| \leq \sqrt{N} \; \Bigl\|\; \X - \Xhat_{\rm{opt}} \;\Bigr\|
  \end{equation*}
  and 
  \begin{equation*}
    \Bigl\|\; \X - \Xhat_{\rm{ST-HOSVD}} \;\Bigr\| \leq \sqrt{N} \; \Bigl\|\; \X - \Xhat_{\rm{opt}} \;\Bigr\|
  \end{equation*}
  where $\Xhat_{\rm{opt}}$ is the optimal rank-$\FullSize[R]$ approximation of $\X$.
\end{theorem}

\subsection{Pre- and Post-Processing}
\label{sec:pre-post-processing}

Algorithms for computing Tucker models like ST-HOSVD (\cref{alg:sthosvd})
minimize the relative approximation error in a norm-wise sense.
As a result, the relative component-wise errors are generally smaller for large
entries in the data tensor than for small entries.
If the data is not preprocessed, the differences in magnitude are artifacts of the type of
variable or unit of measurement rather than reflecting the importance of the data value.

To address these issues, we pre-process
the original data using hyperslice-wise computations, where a single hyperslice corresponds to the values of a particular variable like pressure or temperature, all with the same physical units.
These computations include gathering statistics, such as mean or maximum value,
for each hyperslice of a particular mode and applying a single linear function to
every element in a hyperslice to uniformly shift and/or scale the values.
The vector of shifting and scaling values are stored so that the inverse
operations can be applied in post-processing after the (partial) reconstruction process.

The hyperslice-wise statistics that can be collected include the mean, standard
deviation, maximum, minimum, vector 1-norm, and vector 2-norm.
Common preprocessing techniques are (a) shifting by the mean value (to center
each hyperslice at 0) and scaling by the standard deviation (to impose a variance of
1) and (b) scaling by the maximum absolute value (to impose a range of ${-}1$ to 1).
We refer to (a) as ``standardization'' and (b) as ``max rescaling.''

%
%
%
%
%

\section{Data Layouts}
\label{sec:data-layouts}

In this section, we describe how tensors are stored in
local memory and how they are distributed for parallel computation.
Factor matrices are stored redundantly on every processor.
We assume throughout that we have a generic $N$-way tensor $\Y$ of size $\FullSize$.
WLOG, this $\Y$ tensor is the one that is updated in  \cref{alg:sthosvd}, so each $J_n$ is either $R_n$ or $I_n$.
We assume that the processors are logically arranged into an 
$N$-way Cartesian processor grid, with dimensions $\FullSize[P]$, as in \cite{AuBaKo16,SK16,BHK18-TR}. 
In the analysis, we assume $P_n \in \set{1,\dots,J_n}$ for each $n \in \RangeSet$.
We use the size shorthands discussed in \cref{sec:size-computations},
e.g., $\LinearSize[P]$ denotes the total number of processors.

\subsection{Tensor Data Layout}
\label{sec:tensor-layout}

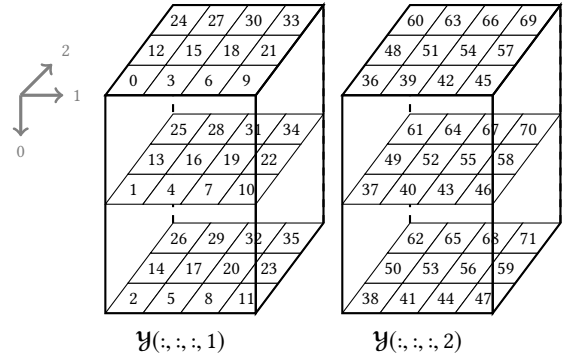
\begin{figure}
  \centering
  %
  %
  %
  %
  %
  %
  %
  %
  %
\begin{tikzpicture}[%
  scale=0.5,
  ]
 \begin{scope}[xshift=0cm]
 \draw[dashed,thick] (1.8,2.4) -- ++(0,5.8);
 \draw[dashed,thick] (5.8,2.4) -- ++(0,5.8);
  \draw[fill=white](0,5.8) -- ++(1,0) -- ++(0.6,0.8) -- ++(-1,0) -- cycle;
  \node at (0.75,6.2) {\footnotesize 0};
  \draw[fill=white](0,2.9) -- ++(1,0) -- ++(0.6,0.8) -- ++(-1,0) -- cycle;
  \node at (0.75,3.3) {\footnotesize 1};
  \draw[fill=white](0,0) -- ++(1,0) -- ++(0.6,0.8) -- ++(-1,0) -- cycle;
  \node at (0.75,0.4) {\footnotesize 2};
  \draw[fill=white](1,5.8) -- ++(1,0) -- ++(0.6,0.8) -- ++(-1,0) -- cycle;
  \node at (1.75,6.2) {\footnotesize 3};
  \draw[fill=white](1,2.9) -- ++(1,0) -- ++(0.6,0.8) -- ++(-1,0) -- cycle;
  \node at (1.75,3.3) {\footnotesize 4};
  \draw[fill=white](1,0) -- ++(1,0) -- ++(0.6,0.8) -- ++(-1,0) -- cycle;
  \node at (1.75,0.4) {\footnotesize 5};
  \draw[fill=white](2,5.8) -- ++(1,0) -- ++(0.6,0.8) -- ++(-1,0) -- cycle;
  \node at (2.75,6.2) {\footnotesize 6};
  \draw[fill=white](2,2.9) -- ++(1,0) -- ++(0.6,0.8) -- ++(-1,0) -- cycle;
  \node at (2.75,3.3) {\footnotesize 7};
  \draw[fill=white](2,0) -- ++(1,0) -- ++(0.6,0.8) -- ++(-1,0) -- cycle;
  \node at (2.75,0.4) {\footnotesize 8};
  \draw[fill=white](3,5.8) -- ++(1,0) -- ++(0.6,0.8) -- ++(-1,0) -- cycle;
  \node at (3.75,6.2) {\footnotesize 9};
  \draw[fill=white](3,2.9) -- ++(1,0) -- ++(0.6,0.8) -- ++(-1,0) -- cycle;
  \node at (3.75,3.3) {\footnotesize 10};
  \draw[fill=white](3,0) -- ++(1,0) -- ++(0.6,0.8) -- ++(-1,0) -- cycle;
  \node at (3.75,0.4) {\footnotesize 11};
  \draw[fill=white](0.6,6.6) -- ++(1,0) -- ++(0.6,0.8) -- ++(-1,0) -- cycle;
  \node at (1.35,7) {\footnotesize 12};
  \draw[fill=white](0.6,3.7) -- ++(1,0) -- ++(0.6,0.8) -- ++(-1,0) -- cycle;
  \node at (1.35,4.1) {\footnotesize 13};
  \draw[fill=white](0.6,0.8) -- ++(1,0) -- ++(0.6,0.8) -- ++(-1,0) -- cycle;
  \node at (1.35,1.2) {\footnotesize 14};
  \draw[fill=white](1.6,6.6) -- ++(1,0) -- ++(0.6,0.8) -- ++(-1,0) -- cycle;
  \node at (2.35,7) {\footnotesize 15};
  \draw[fill=white](1.6,3.7) -- ++(1,0) -- ++(0.6,0.8) -- ++(-1,0) -- cycle;
  \node at (2.35,4.1) {\footnotesize 16};
  \draw[fill=white](1.6,0.8) -- ++(1,0) -- ++(0.6,0.8) -- ++(-1,0) -- cycle;
  \node at (2.35,1.2) {\footnotesize 17};
  \draw[fill=white](2.6,6.6) -- ++(1,0) -- ++(0.6,0.8) -- ++(-1,0) -- cycle;
  \node at (3.35,7) {\footnotesize 18};
  \draw[fill=white](2.6,3.7) -- ++(1,0) -- ++(0.6,0.8) -- ++(-1,0) -- cycle;
  \node at (3.35,4.1) {\footnotesize 19};
  \draw[fill=white](2.6,0.8) -- ++(1,0) -- ++(0.6,0.8) -- ++(-1,0) -- cycle;
  \node at (3.35,1.2) {\footnotesize 20};
  \draw[fill=white](3.6,6.6) -- ++(1,0) -- ++(0.6,0.8) -- ++(-1,0) -- cycle;
  \node at (4.35,7) {\footnotesize 21};
  \draw[fill=white](3.6,3.7) -- ++(1,0) -- ++(0.6,0.8) -- ++(-1,0) -- cycle;
  \node at (4.35,4.1) {\footnotesize 22};
  \draw[fill=white](3.6,0.8) -- ++(1,0) -- ++(0.6,0.8) -- ++(-1,0) -- cycle;
  \node at (4.35,1.2) {\footnotesize 23};
  \draw[fill=white](1.2,7.4) -- ++(1,0) -- ++(0.6,0.8) -- ++(-1,0) -- cycle;
  \node at (1.95,7.8) {\footnotesize 24};
  \draw[fill=white](1.2,4.5) -- ++(1,0) -- ++(0.6,0.8) -- ++(-1,0) -- cycle;
  \node at (1.95,4.9) {\footnotesize 25};
  \draw[fill=white](1.2,1.6) -- ++(1,0) -- ++(0.6,0.8) -- ++(-1,0) -- cycle;
  \node at (1.95,2) {\footnotesize 26};
  \draw[fill=white](2.2,7.4) -- ++(1,0) -- ++(0.6,0.8) -- ++(-1,0) -- cycle;
  \node at (2.95,7.8) {\footnotesize 27};
  \draw[fill=white](2.2,4.5) -- ++(1,0) -- ++(0.6,0.8) -- ++(-1,0) -- cycle;
  \node at (2.95,4.9) {\footnotesize 28};
  \draw[fill=white](2.2,1.6) -- ++(1,0) -- ++(0.6,0.8) -- ++(-1,0) -- cycle;
  \node at (2.95,2) {\footnotesize 29};
  \draw[fill=white](3.2,7.4) -- ++(1,0) -- ++(0.6,0.8) -- ++(-1,0) -- cycle;
  \node at (3.95,7.8) {\footnotesize 30};
  \draw[fill=white](3.2,4.5) -- ++(1,0) -- ++(0.6,0.8) -- ++(-1,0) -- cycle;
  \node at (3.95,4.9) {\footnotesize 31};
  \draw[fill=white](3.2,1.6) -- ++(1,0) -- ++(0.6,0.8) -- ++(-1,0) -- cycle;
  \node at (3.95,2) {\footnotesize 32};
  \draw[fill=white](4.2,7.4) -- ++(1,0) -- ++(0.6,0.8) -- ++(-1,0) -- cycle;
  \node at (4.95,7.8) {\footnotesize 33};
  \draw[fill=white](4.2,4.5) -- ++(1,0) -- ++(0.6,0.8) -- ++(-1,0) -- cycle;
  \node at (4.95,4.9) {\footnotesize 34};
  \draw[fill=white](4.2,1.6) -- ++(1,0) -- ++(0.6,0.8) -- ++(-1,0) -- cycle;
  \node at (4.95,2) {\footnotesize 35};
\draw[draw=black,thick](0,0) -- ++(4,0) -- ++(0,5.8) -- ++(-4,0) -- cycle;
\draw[draw=black,thick](0,5.8) -- ++(4,0) -- ++(1.8,2.4) -- ++(-4,0) -- cycle;
\draw[draw=black,thick](4,0) -- ++(1.8,2.4) -- ++(0,5.8) -- ++(-1.8,-2.4) -- cycle;
 \node at (2,-0.75) {$\Y(:,:,:,1)$};
 \end{scope}
 \begin{scope}[xshift=6.3cm]
 \draw[dashed,thick] (1.8,2.4) -- ++(0,5.8);
 \draw[dashed,thick] (5.8,2.4) -- ++(0,5.8);
  \draw[fill=white](0,5.8) -- ++(1,0) -- ++(0.6,0.8) -- ++(-1,0) -- cycle;
  \node at (0.75,6.2) {\footnotesize 36};
  \draw[fill=white](0,2.9) -- ++(1,0) -- ++(0.6,0.8) -- ++(-1,0) -- cycle;
  \node at (0.75,3.3) {\footnotesize 37};
  \draw[fill=white](0,0) -- ++(1,0) -- ++(0.6,0.8) -- ++(-1,0) -- cycle;
  \node at (0.75,0.4) {\footnotesize 38};
  \draw[fill=white](1,5.8) -- ++(1,0) -- ++(0.6,0.8) -- ++(-1,0) -- cycle;
  \node at (1.75,6.2) {\footnotesize 39};
  \draw[fill=white](1,2.9) -- ++(1,0) -- ++(0.6,0.8) -- ++(-1,0) -- cycle;
  \node at (1.75,3.3) {\footnotesize 40};
  \draw[fill=white](1,0) -- ++(1,0) -- ++(0.6,0.8) -- ++(-1,0) -- cycle;
  \node at (1.75,0.4) {\footnotesize 41};
  \draw[fill=white](2,5.8) -- ++(1,0) -- ++(0.6,0.8) -- ++(-1,0) -- cycle;
  \node at (2.75,6.2) {\footnotesize 42};
  \draw[fill=white](2,2.9) -- ++(1,0) -- ++(0.6,0.8) -- ++(-1,0) -- cycle;
  \node at (2.75,3.3) {\footnotesize 43};
  \draw[fill=white](2,0) -- ++(1,0) -- ++(0.6,0.8) -- ++(-1,0) -- cycle;
  \node at (2.75,0.4) {\footnotesize 44};
  \draw[fill=white](3,5.8) -- ++(1,0) -- ++(0.6,0.8) -- ++(-1,0) -- cycle;
  \node at (3.75,6.2) {\footnotesize 45};
  \draw[fill=white](3,2.9) -- ++(1,0) -- ++(0.6,0.8) -- ++(-1,0) -- cycle;
  \node at (3.75,3.3) {\footnotesize 46};
  \draw[fill=white](3,0) -- ++(1,0) -- ++(0.6,0.8) -- ++(-1,0) -- cycle;
  \node at (3.75,0.4) {\footnotesize 47};
  \draw[fill=white](0.6,6.6) -- ++(1,0) -- ++(0.6,0.8) -- ++(-1,0) -- cycle;
  \node at (1.35,7) {\footnotesize 48};
  \draw[fill=white](0.6,3.7) -- ++(1,0) -- ++(0.6,0.8) -- ++(-1,0) -- cycle;
  \node at (1.35,4.1) {\footnotesize 49};
  \draw[fill=white](0.6,0.8) -- ++(1,0) -- ++(0.6,0.8) -- ++(-1,0) -- cycle;
  \node at (1.35,1.2) {\footnotesize 50};
  \draw[fill=white](1.6,6.6) -- ++(1,0) -- ++(0.6,0.8) -- ++(-1,0) -- cycle;
  \node at (2.35,7) {\footnotesize 51};
  \draw[fill=white](1.6,3.7) -- ++(1,0) -- ++(0.6,0.8) -- ++(-1,0) -- cycle;
  \node at (2.35,4.1) {\footnotesize 52};
  \draw[fill=white](1.6,0.8) -- ++(1,0) -- ++(0.6,0.8) -- ++(-1,0) -- cycle;
  \node at (2.35,1.2) {\footnotesize 53};
  \draw[fill=white](2.6,6.6) -- ++(1,0) -- ++(0.6,0.8) -- ++(-1,0) -- cycle;
  \node at (3.35,7) {\footnotesize 54};
  \draw[fill=white](2.6,3.7) -- ++(1,0) -- ++(0.6,0.8) -- ++(-1,0) -- cycle;
  \node at (3.35,4.1) {\footnotesize 55};
  \draw[fill=white](2.6,0.8) -- ++(1,0) -- ++(0.6,0.8) -- ++(-1,0) -- cycle;
  \node at (3.35,1.2) {\footnotesize 56};
  \draw[fill=white](3.6,6.6) -- ++(1,0) -- ++(0.6,0.8) -- ++(-1,0) -- cycle;
  \node at (4.35,7) {\footnotesize 57};
  \draw[fill=white](3.6,3.7) -- ++(1,0) -- ++(0.6,0.8) -- ++(-1,0) -- cycle;
  \node at (4.35,4.1) {\footnotesize 58};
  \draw[fill=white](3.6,0.8) -- ++(1,0) -- ++(0.6,0.8) -- ++(-1,0) -- cycle;
  \node at (4.35,1.2) {\footnotesize 59};
  \draw[fill=white](1.2,7.4) -- ++(1,0) -- ++(0.6,0.8) -- ++(-1,0) -- cycle;
  \node at (1.95,7.8) {\footnotesize 60};
  \draw[fill=white](1.2,4.5) -- ++(1,0) -- ++(0.6,0.8) -- ++(-1,0) -- cycle;
  \node at (1.95,4.9) {\footnotesize 61};
  \draw[fill=white](1.2,1.6) -- ++(1,0) -- ++(0.6,0.8) -- ++(-1,0) -- cycle;
  \node at (1.95,2) {\footnotesize 62};
  \draw[fill=white](2.2,7.4) -- ++(1,0) -- ++(0.6,0.8) -- ++(-1,0) -- cycle;
  \node at (2.95,7.8) {\footnotesize 63};
  \draw[fill=white](2.2,4.5) -- ++(1,0) -- ++(0.6,0.8) -- ++(-1,0) -- cycle;
  \node at (2.95,4.9) {\footnotesize 64};
  \draw[fill=white](2.2,1.6) -- ++(1,0) -- ++(0.6,0.8) -- ++(-1,0) -- cycle;
  \node at (2.95,2) {\footnotesize 65};
  \draw[fill=white](3.2,7.4) -- ++(1,0) -- ++(0.6,0.8) -- ++(-1,0) -- cycle;
  \node at (3.95,7.8) {\footnotesize 66};
  \draw[fill=white](3.2,4.5) -- ++(1,0) -- ++(0.6,0.8) -- ++(-1,0) -- cycle;
  \node at (3.95,4.9) {\footnotesize 67};
  \draw[fill=white](3.2,1.6) -- ++(1,0) -- ++(0.6,0.8) -- ++(-1,0) -- cycle;
  \node at (3.95,2) {\footnotesize 68};
  \draw[fill=white](4.2,7.4) -- ++(1,0) -- ++(0.6,0.8) -- ++(-1,0) -- cycle;
  \node at (4.95,7.8) {\footnotesize 69};
  \draw[fill=white](4.2,4.5) -- ++(1,0) -- ++(0.6,0.8) -- ++(-1,0) -- cycle;
  \node at (4.95,4.9) {\footnotesize 70};
  \draw[fill=white](4.2,1.6) -- ++(1,0) -- ++(0.6,0.8) -- ++(-1,0) -- cycle;
  \node at (4.95,2) {\footnotesize 71};
\draw[draw=black,thick](0,0) -- ++(4,0) -- ++(0,5.8) -- ++(-4,0) -- cycle;
\draw[draw=black,thick](0,5.8) -- ++(4,0) -- ++(1.8,2.4) -- ++(-4,0) -- cycle;
\draw[draw=black,thick](4,0) -- ++(1.8,2.4) -- ++(0,5.8) -- ++(-1.8,-2.4) -- cycle;
 \node at (2,-0.75) {$\Y(:,:,:,2)$};
 \end{scope}
\node[black!50!white,very thick,->] (down) at (-2.25, 4.3) {\footnotesize 0};
\draw[black!50!white,very thick,->] (-2.25,5.8) -- (down);
\node[black!50!white,very thick,->] (right) at (-0.75, 5.8) {\footnotesize 1};
\draw[black!50!white,very thick,->] (-2.25,5.8) -- (right);
\node[black!50!white,very thick,->] (back) at (-1.05, 7) {\footnotesize 2};
\draw[black!50!white,very thick,->] (-2.25,5.8) -- (back);
\end{tikzpicture}
\caption{A $\exsize$ tensor, where labels correspond to linear indices. The arrows show the direction for each mode,
  except the fourth mode which is denoted explicitly in the labels at the bottom.}
\label{fig:tensor_seq_dist}
\end{figure}

Consider the \emph{natural descending} format
that generalizes the column-major matrix layout.
That is, tensor element $\FullIndex$
can be mapped to
\begin{equation}
  \label{eq:idxtolin}
  j'
  = \IdxToLin
  \equiv \sum_{n \in \RangeSet} j_n \cdot \LeftSize,
\end{equation}
which we refer to as the \emph{linear index}.
We can also define the inverse operation, 
$$\FullIndex = \LinToIdx.$$
\Cref{fig:tensor_seq_dist} shows a $\exsize$ tensor
with each entry labeled by its linear index.

\subsection{Unfolded Tensor Data Layout}
\label{sec:unfolded_layout}

{
\setlength\arraycolsep{1.4mm}

If the tensor is stored in the natural descending format, then the data layout
of each unfolding can be thought of as a set of contiguous submatrices, where
each submatrix is stored in row-major ordering.
In other words, there is no reason to do any data movement in memory to
work with the unfolded matrix because BLAS calls can operate on submatrices stored in row- or column-major order, as observed in \cite{AuBaKo16,LBPSV15,HBJT18,PTC13a}.
The number and dimensions of
these submatrices depend on the mode of the unfolding.  For the $n$th-mode
unfolding, the number of submatrices is $\RightSize$, and each submatrix is
of size $\SingleSize \times \LeftSize$.
Let's look at these for the $\exsize$ tensor in \cref{fig:tensor_seq_dist}.
Recall that the values indicate the relative position in the natural descending format.
In the case $n=0$, each submatrix has one column, which corresponds to a global
column-major matrix, so it has 24 submatrices of size $3 \times 1$:
\begin{displaymath}
\Mz{Y}{0} = \left[ \begin{array}{*{23}{r|}r}
0  & 3  & 6  & 9  & 12  & 15  & 18  & 21  & 24  & 27  & 30  & 33  & 36  & 39  & 42  & 45  & 48  & 51  & 54  & 57  & 60  & 63  & 66  & 69  \\ 
1  & 4  & 7  & 10  & 13  & 16  & 19  & 22  & 25  & 28  & 31  & 34  & 37  & 40  & 43  & 46  & 49  & 52  & 55  & 58  & 61  & 64  & 67  & 70  \\ 
2  & 5  & 8  & 11  & 14  & 17  & 20  & 23  & 26  & 29  & 32  & 35  & 38  & 41  & 44  & 47  & 50  & 53  & 56  & 59  & 62  & 65  & 68  & 71  \\ 
\end{array}\right].
\end{displaymath}
Note that this yields a special mode-0 case in the implementation of operations because it is more efficient to treat the data as one column-major block than many blocks of vectors.
The mode-1 unfolding has 6 row-major submatrices of size $4 \times 3$:
\begin{displaymath}
  \Mz{Y}{1} =
  \left[ \begin{array}{*{5}{rrr|}rrr}
0  & 1  & 2  & 12  & 13  & 14  & 24  & 25  & 26  & 36  & 37  & 38  & 48  & 49  & 50  & 60  & 61  & 62  \\ 
3  & 4  & 5  & 15  & 16  & 17  & 27  & 28  & 29  & 39  & 40  & 41  & 51  & 52  & 53  & 63  & 64  & 65  \\ 
6  & 7  & 8  & 18  & 19  & 20  & 30  & 31  & 32  & 42  & 43  & 44  & 54  & 55  & 56  & 66  & 67  & 68  \\ 
9  & 10  & 11  & 21  & 22  & 23  & 33  & 34  & 35  & 45  & 46  & 47  & 57  & 58  & 59  & 69  & 70  & 71  \\ 
\end{array}\right].
\end{displaymath}
The mode-2 unfolding has 2 row-major submatrices of size $3 \times 12$:
\begin{displaymath}
\Mz{Y}{2} = \left[ \begin{array}{rrrrrrrrrrrr|rrrrrrrrrrrr}
0  & 1  & 2  & 3  & 4  & 5  & 6  & 7  & 8  & 9  & 10  & 11  & 36  & 37  & 38  & 39  & 40  & 41  & 42  & 43  & 44  & 45  & 46  & 47  \\ 
12  & 13  & 14  & 15  & 16  & 17  & 18  & 19  & 20  & 21  & 22  & 23  & 48  & 49  & 50  & 51  & 52  & 53  & 54  & 55  & 56  & 57  & 58  & 59  \\ 
24  & 25  & 26  & 27  & 28  & 29  & 30  & 31  & 32  & 33  & 34  & 35  & 60  & 61  & 62  & 63  & 64  & 65  & 66  & 67  & 68  & 69  & 70  & 71  \\ 
\end{array}\right].
\end{displaymath}
In the case $n=N-1$, there is only one submatrix, which corresponds to a global
row-major matrix, so we have one submatrix of size $2 \times 36$:
\begin{displaymath}
\Mz{Y}{3} = \left[ \begin{array}{rrrrrrrrrrrrrrrrrrrrrrr}
0  & 1  & 2  & 3  & 4  & 5  & 6  & 7  & 8  & 9  & 10  & 11  & \cdots & 26  & 27  & 28  & 29  & 30  & 31  & 32  & 33  & 34  & 35  \\ 
36  & 37  & 38  & 39  & 40  & 41  & 42  & 43  & 44  & 45  & 46  & 47  & \cdots & 62  & 63  & 64  & 65  & 66  & 67  & 68  & 69  & 70  & 71  \\ 
\end{array}\right].
\end{displaymath}
}
These layouts become important when we discuss the local operations in \cref{sec:functions}.

\subsection{Global Tensor Data Distribution}
\label{sec:tensor_distribution}

Now we consider how the tensor is distributed on the global processor grid.
We let an overbar indicate local quantities with respect to a specific processor.
If $J_n$ is the global size of mode $n$, then 
$\bar J_n$ is the corresponding \emph{local size} for the subtensor owned by a
processor $\FullIndex*[p]$. %
Processor $\FullIndex*[p]$ owns a block of size $\FullSize*$
where $\bar J_n$ is given by
\begin{equation}\label{eq:local-size}
  \bar J_n
  = \Lsz
  \equiv
  \begin{cases}
    \Dn + 1& \text{if } \bar p_n < \Mn \\
    \Dn & \text{if } \bar p_n \geq \Mn
  \end{cases}
  . 
\end{equation}
\Cref{tab:local-sizes} shows the size of each local subtensor
for a tensor of size $\exsize$ distributed on a processor grid of size $\expsize$.
\Cref{tab:local-sizes} also indicates the \emph{processor linear index},
which is defined analogously to \cref{eq:idxtolin}:
\begin{displaymath}
  \bar p'
  = \IdxToLin{\bar p}{P}
  = \sum_{n \in \RangeSet} \bar p_n \cdot \LeftSize[P]
  \quad \in \RangeSet{\LinearSize[P]}.
\end{displaymath}

\begin{table}[t]
  \caption{Local tensor sizes for a $\exsize$ tensor distributed on a $\expsize$ processor grid, demonstrating how uneven sizes are handled.}
  \label{tab:local-sizes}
  \centering\small
  \begin{tabular}{ccc}
    Proc.~ID & Proc.~Linear ID & Local Size \\
    $(p_0,p_1,p_2,p_3)$ & $p'$ & $\bar J_0 \times \bar J_1 \times \bar J_2 \times \bar J_3$ \\ \hline
    $(0,0,0,0)$ & 0 & $2 \times 2 \times 2 \times 2$ \\
    $(1,0,0,0)$ & 1 & $1 \times 2 \times 2 \times 2$ \\
    $(0,1,0,0)$ & 2 & $2 \times 2 \times 2 \times 2$ \\
    $(1,1,0,0)$ & 3 & $1 \times 2 \times 2 \times 2$ \\
    $(0,0,1,0)$ & 4 & $2 \times 2 \times 1 \times 2$ \\
    $(1,0,1,0)$ & 5 & $1 \times 2 \times 1 \times 2$ \\
    $(0,1,1,0)$ & 6 & $2 \times 2 \times 1 \times 2$ \\
    $(1,1,1,0)$ & 7 & $1 \times 2 \times 1 \times 2$ \\
  \end{tabular}
\end{table}

\begin{figure}
\centering
\subfloat[Processor Linear ID]{\label{fig:procids}
  %
  %
  %
  %
  %
  %
  %
\begin{tikzpicture}[%
  scale=0.5,
  ]
 \begin{scope}[xshift=0cm]
 \draw[dashed,thick] (1.8,2.4) -- ++(0,5.8);
 \draw[dashed,thick] (5.8,2.4) -- ++(0,5.8);
  \draw[fill=color1!50](0,5.8) -- ++(1,0) -- ++(0.6,0.8) -- ++(-1,0) -- cycle;
  \node at (0.75,6.2) {\footnotesize 0};
  \draw[fill=color1!50](0,2.9) -- ++(1,0) -- ++(0.6,0.8) -- ++(-1,0) -- cycle;
  \node at (0.75,3.3) {\footnotesize 0};
  \draw[fill=color2!50](0,0) -- ++(1,0) -- ++(0.6,0.8) -- ++(-1,0) -- cycle;
  \node at (0.75,0.4) {\footnotesize 1};
  \draw[fill=color1!50](1,5.8) -- ++(1,0) -- ++(0.6,0.8) -- ++(-1,0) -- cycle;
  \node at (1.75,6.2) {\footnotesize 0};
  \draw[fill=color1!50](1,2.9) -- ++(1,0) -- ++(0.6,0.8) -- ++(-1,0) -- cycle;
  \node at (1.75,3.3) {\footnotesize 0};
  \draw[fill=color2!50](1,0) -- ++(1,0) -- ++(0.6,0.8) -- ++(-1,0) -- cycle;
  \node at (1.75,0.4) {\footnotesize 1};
  \draw[fill=color3!50](2,5.8) -- ++(1,0) -- ++(0.6,0.8) -- ++(-1,0) -- cycle;
  \node at (2.75,6.2) {\footnotesize 2};
  \draw[fill=color3!50](2,2.9) -- ++(1,0) -- ++(0.6,0.8) -- ++(-1,0) -- cycle;
  \node at (2.75,3.3) {\footnotesize 2};
  \draw[fill=color4!50](2,0) -- ++(1,0) -- ++(0.6,0.8) -- ++(-1,0) -- cycle;
  \node at (2.75,0.4) {\footnotesize 3};
  \draw[fill=color3!50](3,5.8) -- ++(1,0) -- ++(0.6,0.8) -- ++(-1,0) -- cycle;
  \node at (3.75,6.2) {\footnotesize 2};
  \draw[fill=color3!50](3,2.9) -- ++(1,0) -- ++(0.6,0.8) -- ++(-1,0) -- cycle;
  \node at (3.75,3.3) {\footnotesize 2};
  \draw[fill=color4!50](3,0) -- ++(1,0) -- ++(0.6,0.8) -- ++(-1,0) -- cycle;
  \node at (3.75,0.4) {\footnotesize 3};
  \draw[fill=color1!50](0.6,6.6) -- ++(1,0) -- ++(0.6,0.8) -- ++(-1,0) -- cycle;
  \node at (1.35,7) {\footnotesize 0};
  \draw[fill=color1!50](0.6,3.7) -- ++(1,0) -- ++(0.6,0.8) -- ++(-1,0) -- cycle;
  \node at (1.35,4.1) {\footnotesize 0};
  \draw[fill=color2!50](0.6,0.8) -- ++(1,0) -- ++(0.6,0.8) -- ++(-1,0) -- cycle;
  \node at (1.35,1.2) {\footnotesize 1};
  \draw[fill=color1!50](1.6,6.6) -- ++(1,0) -- ++(0.6,0.8) -- ++(-1,0) -- cycle;
  \node at (2.35,7) {\footnotesize 0};
  \draw[fill=color1!50](1.6,3.7) -- ++(1,0) -- ++(0.6,0.8) -- ++(-1,0) -- cycle;
  \node at (2.35,4.1) {\footnotesize 0};
  \draw[fill=color2!50](1.6,0.8) -- ++(1,0) -- ++(0.6,0.8) -- ++(-1,0) -- cycle;
  \node at (2.35,1.2) {\footnotesize 1};
  \draw[fill=color3!50](2.6,6.6) -- ++(1,0) -- ++(0.6,0.8) -- ++(-1,0) -- cycle;
  \node at (3.35,7) {\footnotesize 2};
  \draw[fill=color3!50](2.6,3.7) -- ++(1,0) -- ++(0.6,0.8) -- ++(-1,0) -- cycle;
  \node at (3.35,4.1) {\footnotesize 2};
  \draw[fill=color4!50](2.6,0.8) -- ++(1,0) -- ++(0.6,0.8) -- ++(-1,0) -- cycle;
  \node at (3.35,1.2) {\footnotesize 3};
  \draw[fill=color3!50](3.6,6.6) -- ++(1,0) -- ++(0.6,0.8) -- ++(-1,0) -- cycle;
  \node at (4.35,7) {\footnotesize 2};
  \draw[fill=color3!50](3.6,3.7) -- ++(1,0) -- ++(0.6,0.8) -- ++(-1,0) -- cycle;
  \node at (4.35,4.1) {\footnotesize 2};
  \draw[fill=color4!50](3.6,0.8) -- ++(1,0) -- ++(0.6,0.8) -- ++(-1,0) -- cycle;
  \node at (4.35,1.2) {\footnotesize 3};
  \draw[fill=color5!50](1.2,7.4) -- ++(1,0) -- ++(0.6,0.8) -- ++(-1,0) -- cycle;
  \node at (1.95,7.8) {\footnotesize 4};
  \draw[fill=color5!50](1.2,4.5) -- ++(1,0) -- ++(0.6,0.8) -- ++(-1,0) -- cycle;
  \node at (1.95,4.9) {\footnotesize 4};
  \draw[fill=color6!50](1.2,1.6) -- ++(1,0) -- ++(0.6,0.8) -- ++(-1,0) -- cycle;
  \node at (1.95,2) {\footnotesize 5};
  \draw[fill=color5!50](2.2,7.4) -- ++(1,0) -- ++(0.6,0.8) -- ++(-1,0) -- cycle;
  \node at (2.95,7.8) {\footnotesize 4};
  \draw[fill=color5!50](2.2,4.5) -- ++(1,0) -- ++(0.6,0.8) -- ++(-1,0) -- cycle;
  \node at (2.95,4.9) {\footnotesize 4};
  \draw[fill=color6!50](2.2,1.6) -- ++(1,0) -- ++(0.6,0.8) -- ++(-1,0) -- cycle;
  \node at (2.95,2) {\footnotesize 5};
  \draw[fill=color7!50](3.2,7.4) -- ++(1,0) -- ++(0.6,0.8) -- ++(-1,0) -- cycle;
  \node at (3.95,7.8) {\footnotesize 6};
  \draw[fill=color7!50](3.2,4.5) -- ++(1,0) -- ++(0.6,0.8) -- ++(-1,0) -- cycle;
  \node at (3.95,4.9) {\footnotesize 6};
  \draw[fill=color8!50](3.2,1.6) -- ++(1,0) -- ++(0.6,0.8) -- ++(-1,0) -- cycle;
  \node at (3.95,2) {\footnotesize 7};
  \draw[fill=color7!50](4.2,7.4) -- ++(1,0) -- ++(0.6,0.8) -- ++(-1,0) -- cycle;
  \node at (4.95,7.8) {\footnotesize 6};
  \draw[fill=color7!50](4.2,4.5) -- ++(1,0) -- ++(0.6,0.8) -- ++(-1,0) -- cycle;
  \node at (4.95,4.9) {\footnotesize 6};
  \draw[fill=color8!50](4.2,1.6) -- ++(1,0) -- ++(0.6,0.8) -- ++(-1,0) -- cycle;
  \node at (4.95,2) {\footnotesize 7};
\draw[draw=black,thick](0,0) -- ++(4,0) -- ++(0,5.8) -- ++(-4,0) -- cycle;
\draw[draw=black,thick](0,5.8) -- ++(4,0) -- ++(1.8,2.4) -- ++(-4,0) -- cycle;
\draw[draw=black,thick](4,0) -- ++(1.8,2.4) -- ++(0,5.8) -- ++(-1.8,-2.4) -- cycle;
 \node at (2,-0.75) {$\Y(:,:,:,1)$};
 \end{scope}
 \begin{scope}[xshift=6.3cm]
 \draw[dashed,thick] (1.8,2.4) -- ++(0,5.8);
 \draw[dashed,thick] (5.8,2.4) -- ++(0,5.8);
  \draw[fill=color1!50](0,5.8) -- ++(1,0) -- ++(0.6,0.8) -- ++(-1,0) -- cycle;
  \node at (0.75,6.2) {\footnotesize 0};
  \draw[fill=color1!50](0,2.9) -- ++(1,0) -- ++(0.6,0.8) -- ++(-1,0) -- cycle;
  \node at (0.75,3.3) {\footnotesize 0};
  \draw[fill=color2!50](0,0) -- ++(1,0) -- ++(0.6,0.8) -- ++(-1,0) -- cycle;
  \node at (0.75,0.4) {\footnotesize 1};
  \draw[fill=color1!50](1,5.8) -- ++(1,0) -- ++(0.6,0.8) -- ++(-1,0) -- cycle;
  \node at (1.75,6.2) {\footnotesize 0};
  \draw[fill=color1!50](1,2.9) -- ++(1,0) -- ++(0.6,0.8) -- ++(-1,0) -- cycle;
  \node at (1.75,3.3) {\footnotesize 0};
  \draw[fill=color2!50](1,0) -- ++(1,0) -- ++(0.6,0.8) -- ++(-1,0) -- cycle;
  \node at (1.75,0.4) {\footnotesize 1};
  \draw[fill=color3!50](2,5.8) -- ++(1,0) -- ++(0.6,0.8) -- ++(-1,0) -- cycle;
  \node at (2.75,6.2) {\footnotesize 2};
  \draw[fill=color3!50](2,2.9) -- ++(1,0) -- ++(0.6,0.8) -- ++(-1,0) -- cycle;
  \node at (2.75,3.3) {\footnotesize 2};
  \draw[fill=color4!50](2,0) -- ++(1,0) -- ++(0.6,0.8) -- ++(-1,0) -- cycle;
  \node at (2.75,0.4) {\footnotesize 3};
  \draw[fill=color3!50](3,5.8) -- ++(1,0) -- ++(0.6,0.8) -- ++(-1,0) -- cycle;
  \node at (3.75,6.2) {\footnotesize 2};
  \draw[fill=color3!50](3,2.9) -- ++(1,0) -- ++(0.6,0.8) -- ++(-1,0) -- cycle;
  \node at (3.75,3.3) {\footnotesize 2};
  \draw[fill=color4!50](3,0) -- ++(1,0) -- ++(0.6,0.8) -- ++(-1,0) -- cycle;
  \node at (3.75,0.4) {\footnotesize 3};
  \draw[fill=color1!50](0.6,6.6) -- ++(1,0) -- ++(0.6,0.8) -- ++(-1,0) -- cycle;
  \node at (1.35,7) {\footnotesize 0};
  \draw[fill=color1!50](0.6,3.7) -- ++(1,0) -- ++(0.6,0.8) -- ++(-1,0) -- cycle;
  \node at (1.35,4.1) {\footnotesize 0};
  \draw[fill=color2!50](0.6,0.8) -- ++(1,0) -- ++(0.6,0.8) -- ++(-1,0) -- cycle;
  \node at (1.35,1.2) {\footnotesize 1};
  \draw[fill=color1!50](1.6,6.6) -- ++(1,0) -- ++(0.6,0.8) -- ++(-1,0) -- cycle;
  \node at (2.35,7) {\footnotesize 0};
  \draw[fill=color1!50](1.6,3.7) -- ++(1,0) -- ++(0.6,0.8) -- ++(-1,0) -- cycle;
  \node at (2.35,4.1) {\footnotesize 0};
  \draw[fill=color2!50](1.6,0.8) -- ++(1,0) -- ++(0.6,0.8) -- ++(-1,0) -- cycle;
  \node at (2.35,1.2) {\footnotesize 1};
  \draw[fill=color3!50](2.6,6.6) -- ++(1,0) -- ++(0.6,0.8) -- ++(-1,0) -- cycle;
  \node at (3.35,7) {\footnotesize 2};
  \draw[fill=color3!50](2.6,3.7) -- ++(1,0) -- ++(0.6,0.8) -- ++(-1,0) -- cycle;
  \node at (3.35,4.1) {\footnotesize 2};
  \draw[fill=color4!50](2.6,0.8) -- ++(1,0) -- ++(0.6,0.8) -- ++(-1,0) -- cycle;
  \node at (3.35,1.2) {\footnotesize 3};
  \draw[fill=color3!50](3.6,6.6) -- ++(1,0) -- ++(0.6,0.8) -- ++(-1,0) -- cycle;
  \node at (4.35,7) {\footnotesize 2};
  \draw[fill=color3!50](3.6,3.7) -- ++(1,0) -- ++(0.6,0.8) -- ++(-1,0) -- cycle;
  \node at (4.35,4.1) {\footnotesize 2};
  \draw[fill=color4!50](3.6,0.8) -- ++(1,0) -- ++(0.6,0.8) -- ++(-1,0) -- cycle;
  \node at (4.35,1.2) {\footnotesize 3};
  \draw[fill=color5!50](1.2,7.4) -- ++(1,0) -- ++(0.6,0.8) -- ++(-1,0) -- cycle;
  \node at (1.95,7.8) {\footnotesize 4};
  \draw[fill=color5!50](1.2,4.5) -- ++(1,0) -- ++(0.6,0.8) -- ++(-1,0) -- cycle;
  \node at (1.95,4.9) {\footnotesize 4};
  \draw[fill=color6!50](1.2,1.6) -- ++(1,0) -- ++(0.6,0.8) -- ++(-1,0) -- cycle;
  \node at (1.95,2) {\footnotesize 5};
  \draw[fill=color5!50](2.2,7.4) -- ++(1,0) -- ++(0.6,0.8) -- ++(-1,0) -- cycle;
  \node at (2.95,7.8) {\footnotesize 4};
  \draw[fill=color5!50](2.2,4.5) -- ++(1,0) -- ++(0.6,0.8) -- ++(-1,0) -- cycle;
  \node at (2.95,4.9) {\footnotesize 4};
  \draw[fill=color6!50](2.2,1.6) -- ++(1,0) -- ++(0.6,0.8) -- ++(-1,0) -- cycle;
  \node at (2.95,2) {\footnotesize 5};
  \draw[fill=color7!50](3.2,7.4) -- ++(1,0) -- ++(0.6,0.8) -- ++(-1,0) -- cycle;
  \node at (3.95,7.8) {\footnotesize 6};
  \draw[fill=color7!50](3.2,4.5) -- ++(1,0) -- ++(0.6,0.8) -- ++(-1,0) -- cycle;
  \node at (3.95,4.9) {\footnotesize 6};
  \draw[fill=color8!50](3.2,1.6) -- ++(1,0) -- ++(0.6,0.8) -- ++(-1,0) -- cycle;
  \node at (3.95,2) {\footnotesize 7};
  \draw[fill=color7!50](4.2,7.4) -- ++(1,0) -- ++(0.6,0.8) -- ++(-1,0) -- cycle;
  \node at (4.95,7.8) {\footnotesize 6};
  \draw[fill=color7!50](4.2,4.5) -- ++(1,0) -- ++(0.6,0.8) -- ++(-1,0) -- cycle;
  \node at (4.95,4.9) {\footnotesize 6};
  \draw[fill=color8!50](4.2,1.6) -- ++(1,0) -- ++(0.6,0.8) -- ++(-1,0) -- cycle;
  \node at (4.95,2) {\footnotesize 7};
\draw[draw=black,thick](0,0) -- ++(4,0) -- ++(0,5.8) -- ++(-4,0) -- cycle;
\draw[draw=black,thick](0,5.8) -- ++(4,0) -- ++(1.8,2.4) -- ++(-4,0) -- cycle;
\draw[draw=black,thick](4,0) -- ++(1.8,2.4) -- ++(0,5.8) -- ++(-1.8,-2.4) -- cycle;
 \node at (2,-0.75) {$\Y(:,:,:,2)$};
 \end{scope}
\end{tikzpicture}
}
\hfil
\subfloat[Local linear indices]{\label{fig:localids}
  %
  %
  %
  %
  %
  %
  %
\begin{tikzpicture}[%
  scale=0.5,
  ]
 \begin{scope}[xshift=0cm]
 \draw[dashed,thick] (1.8,2.4) -- ++(0,5.8);
 \draw[dashed,thick] (5.8,2.4) -- ++(0,5.8);
  \draw[fill=color1!50](0,5.8) -- ++(1,0) -- ++(0.6,0.8) -- ++(-1,0) -- cycle;
  \node at (0.75,6.2) {\footnotesize 0};
  \draw[fill=color1!50](0,2.9) -- ++(1,0) -- ++(0.6,0.8) -- ++(-1,0) -- cycle;
  \node at (0.75,3.3) {\footnotesize 1};
  \draw[fill=color2!50](0,0) -- ++(1,0) -- ++(0.6,0.8) -- ++(-1,0) -- cycle;
  \node at (0.75,0.4) {\footnotesize 0};
  \draw[fill=color1!50](1,5.8) -- ++(1,0) -- ++(0.6,0.8) -- ++(-1,0) -- cycle;
  \node at (1.75,6.2) {\footnotesize 2};
  \draw[fill=color1!50](1,2.9) -- ++(1,0) -- ++(0.6,0.8) -- ++(-1,0) -- cycle;
  \node at (1.75,3.3) {\footnotesize 3};
  \draw[fill=color2!50](1,0) -- ++(1,0) -- ++(0.6,0.8) -- ++(-1,0) -- cycle;
  \node at (1.75,0.4) {\footnotesize 1};
  \draw[fill=color3!50](2,5.8) -- ++(1,0) -- ++(0.6,0.8) -- ++(-1,0) -- cycle;
  \node at (2.75,6.2) {\footnotesize 0};
  \draw[fill=color3!50](2,2.9) -- ++(1,0) -- ++(0.6,0.8) -- ++(-1,0) -- cycle;
  \node at (2.75,3.3) {\footnotesize 1};
  \draw[fill=color4!50](2,0) -- ++(1,0) -- ++(0.6,0.8) -- ++(-1,0) -- cycle;
  \node at (2.75,0.4) {\footnotesize 0};
  \draw[fill=color3!50](3,5.8) -- ++(1,0) -- ++(0.6,0.8) -- ++(-1,0) -- cycle;
  \node at (3.75,6.2) {\footnotesize 2};
  \draw[fill=color3!50](3,2.9) -- ++(1,0) -- ++(0.6,0.8) -- ++(-1,0) -- cycle;
  \node at (3.75,3.3) {\footnotesize 3};
  \draw[fill=color4!50](3,0) -- ++(1,0) -- ++(0.6,0.8) -- ++(-1,0) -- cycle;
  \node at (3.75,0.4) {\footnotesize 1};
  \draw[fill=color1!50](0.6,6.6) -- ++(1,0) -- ++(0.6,0.8) -- ++(-1,0) -- cycle;
  \node at (1.35,7) {\footnotesize 4};
  \draw[fill=color1!50](0.6,3.7) -- ++(1,0) -- ++(0.6,0.8) -- ++(-1,0) -- cycle;
  \node at (1.35,4.1) {\footnotesize 5};
  \draw[fill=color2!50](0.6,0.8) -- ++(1,0) -- ++(0.6,0.8) -- ++(-1,0) -- cycle;
  \node at (1.35,1.2) {\footnotesize 2};
  \draw[fill=color1!50](1.6,6.6) -- ++(1,0) -- ++(0.6,0.8) -- ++(-1,0) -- cycle;
  \node at (2.35,7) {\footnotesize 6};
  \draw[fill=color1!50](1.6,3.7) -- ++(1,0) -- ++(0.6,0.8) -- ++(-1,0) -- cycle;
  \node at (2.35,4.1) {\footnotesize 7};
  \draw[fill=color2!50](1.6,0.8) -- ++(1,0) -- ++(0.6,0.8) -- ++(-1,0) -- cycle;
  \node at (2.35,1.2) {\footnotesize 3};
  \draw[fill=color3!50](2.6,6.6) -- ++(1,0) -- ++(0.6,0.8) -- ++(-1,0) -- cycle;
  \node at (3.35,7) {\footnotesize 4};
  \draw[fill=color3!50](2.6,3.7) -- ++(1,0) -- ++(0.6,0.8) -- ++(-1,0) -- cycle;
  \node at (3.35,4.1) {\footnotesize 5};
  \draw[fill=color4!50](2.6,0.8) -- ++(1,0) -- ++(0.6,0.8) -- ++(-1,0) -- cycle;
  \node at (3.35,1.2) {\footnotesize 2};
  \draw[fill=color3!50](3.6,6.6) -- ++(1,0) -- ++(0.6,0.8) -- ++(-1,0) -- cycle;
  \node at (4.35,7) {\footnotesize 6};
  \draw[fill=color3!50](3.6,3.7) -- ++(1,0) -- ++(0.6,0.8) -- ++(-1,0) -- cycle;
  \node at (4.35,4.1) {\footnotesize 7};
  \draw[fill=color4!50](3.6,0.8) -- ++(1,0) -- ++(0.6,0.8) -- ++(-1,0) -- cycle;
  \node at (4.35,1.2) {\footnotesize 3};
  \draw[fill=color5!50](1.2,7.4) -- ++(1,0) -- ++(0.6,0.8) -- ++(-1,0) -- cycle;
  \node at (1.95,7.8) {\footnotesize 0};
  \draw[fill=color5!50](1.2,4.5) -- ++(1,0) -- ++(0.6,0.8) -- ++(-1,0) -- cycle;
  \node at (1.95,4.9) {\footnotesize 1};
  \draw[fill=color6!50](1.2,1.6) -- ++(1,0) -- ++(0.6,0.8) -- ++(-1,0) -- cycle;
  \node at (1.95,2) {\footnotesize 0};
  \draw[fill=color5!50](2.2,7.4) -- ++(1,0) -- ++(0.6,0.8) -- ++(-1,0) -- cycle;
  \node at (2.95,7.8) {\footnotesize 2};
  \draw[fill=color5!50](2.2,4.5) -- ++(1,0) -- ++(0.6,0.8) -- ++(-1,0) -- cycle;
  \node at (2.95,4.9) {\footnotesize 3};
  \draw[fill=color6!50](2.2,1.6) -- ++(1,0) -- ++(0.6,0.8) -- ++(-1,0) -- cycle;
  \node at (2.95,2) {\footnotesize 1};
  \draw[fill=color7!50](3.2,7.4) -- ++(1,0) -- ++(0.6,0.8) -- ++(-1,0) -- cycle;
  \node at (3.95,7.8) {\footnotesize 0};
  \draw[fill=color7!50](3.2,4.5) -- ++(1,0) -- ++(0.6,0.8) -- ++(-1,0) -- cycle;
  \node at (3.95,4.9) {\footnotesize 1};
  \draw[fill=color8!50](3.2,1.6) -- ++(1,0) -- ++(0.6,0.8) -- ++(-1,0) -- cycle;
  \node at (3.95,2) {\footnotesize 0};
  \draw[fill=color7!50](4.2,7.4) -- ++(1,0) -- ++(0.6,0.8) -- ++(-1,0) -- cycle;
  \node at (4.95,7.8) {\footnotesize 2};
  \draw[fill=color7!50](4.2,4.5) -- ++(1,0) -- ++(0.6,0.8) -- ++(-1,0) -- cycle;
  \node at (4.95,4.9) {\footnotesize 3};
  \draw[fill=color8!50](4.2,1.6) -- ++(1,0) -- ++(0.6,0.8) -- ++(-1,0) -- cycle;
  \node at (4.95,2) {\footnotesize 1};
\draw[draw=black,thick](0,0) -- ++(4,0) -- ++(0,5.8) -- ++(-4,0) -- cycle;
\draw[draw=black,thick](0,5.8) -- ++(4,0) -- ++(1.8,2.4) -- ++(-4,0) -- cycle;
\draw[draw=black,thick](4,0) -- ++(1.8,2.4) -- ++(0,5.8) -- ++(-1.8,-2.4) -- cycle;
 \node at (2,-0.75) {$\Y(:,:,:,1)$};
 \end{scope}
 \begin{scope}[xshift=6.3cm]
 \draw[dashed,thick] (1.8,2.4) -- ++(0,5.8);
 \draw[dashed,thick] (5.8,2.4) -- ++(0,5.8);
  \draw[fill=color1!50](0,5.8) -- ++(1,0) -- ++(0.6,0.8) -- ++(-1,0) -- cycle;
  \node at (0.75,6.2) {\footnotesize 8};
  \draw[fill=color1!50](0,2.9) -- ++(1,0) -- ++(0.6,0.8) -- ++(-1,0) -- cycle;
  \node at (0.75,3.3) {\footnotesize 9};
  \draw[fill=color2!50](0,0) -- ++(1,0) -- ++(0.6,0.8) -- ++(-1,0) -- cycle;
  \node at (0.75,0.4) {\footnotesize 4};
  \draw[fill=color1!50](1,5.8) -- ++(1,0) -- ++(0.6,0.8) -- ++(-1,0) -- cycle;
  \node at (1.75,6.2) {\footnotesize 10};
  \draw[fill=color1!50](1,2.9) -- ++(1,0) -- ++(0.6,0.8) -- ++(-1,0) -- cycle;
  \node at (1.75,3.3) {\footnotesize 11};
  \draw[fill=color2!50](1,0) -- ++(1,0) -- ++(0.6,0.8) -- ++(-1,0) -- cycle;
  \node at (1.75,0.4) {\footnotesize 5};
  \draw[fill=color3!50](2,5.8) -- ++(1,0) -- ++(0.6,0.8) -- ++(-1,0) -- cycle;
  \node at (2.75,6.2) {\footnotesize 8};
  \draw[fill=color3!50](2,2.9) -- ++(1,0) -- ++(0.6,0.8) -- ++(-1,0) -- cycle;
  \node at (2.75,3.3) {\footnotesize 9};
  \draw[fill=color4!50](2,0) -- ++(1,0) -- ++(0.6,0.8) -- ++(-1,0) -- cycle;
  \node at (2.75,0.4) {\footnotesize 4};
  \draw[fill=color3!50](3,5.8) -- ++(1,0) -- ++(0.6,0.8) -- ++(-1,0) -- cycle;
  \node at (3.75,6.2) {\footnotesize 10};
  \draw[fill=color3!50](3,2.9) -- ++(1,0) -- ++(0.6,0.8) -- ++(-1,0) -- cycle;
  \node at (3.75,3.3) {\footnotesize 11};
  \draw[fill=color4!50](3,0) -- ++(1,0) -- ++(0.6,0.8) -- ++(-1,0) -- cycle;
  \node at (3.75,0.4) {\footnotesize 5};
  \draw[fill=color1!50](0.6,6.6) -- ++(1,0) -- ++(0.6,0.8) -- ++(-1,0) -- cycle;
  \node at (1.35,7) {\footnotesize 12};
  \draw[fill=color1!50](0.6,3.7) -- ++(1,0) -- ++(0.6,0.8) -- ++(-1,0) -- cycle;
  \node at (1.35,4.1) {\footnotesize 13};
  \draw[fill=color2!50](0.6,0.8) -- ++(1,0) -- ++(0.6,0.8) -- ++(-1,0) -- cycle;
  \node at (1.35,1.2) {\footnotesize 6};
  \draw[fill=color1!50](1.6,6.6) -- ++(1,0) -- ++(0.6,0.8) -- ++(-1,0) -- cycle;
  \node at (2.35,7) {\footnotesize 14};
  \draw[fill=color1!50](1.6,3.7) -- ++(1,0) -- ++(0.6,0.8) -- ++(-1,0) -- cycle;
  \node at (2.35,4.1) {\footnotesize 15};
  \draw[fill=color2!50](1.6,0.8) -- ++(1,0) -- ++(0.6,0.8) -- ++(-1,0) -- cycle;
  \node at (2.35,1.2) {\footnotesize 7};
  \draw[fill=color3!50](2.6,6.6) -- ++(1,0) -- ++(0.6,0.8) -- ++(-1,0) -- cycle;
  \node at (3.35,7) {\footnotesize 12};
  \draw[fill=color3!50](2.6,3.7) -- ++(1,0) -- ++(0.6,0.8) -- ++(-1,0) -- cycle;
  \node at (3.35,4.1) {\footnotesize 13};
  \draw[fill=color4!50](2.6,0.8) -- ++(1,0) -- ++(0.6,0.8) -- ++(-1,0) -- cycle;
  \node at (3.35,1.2) {\footnotesize 6};
  \draw[fill=color3!50](3.6,6.6) -- ++(1,0) -- ++(0.6,0.8) -- ++(-1,0) -- cycle;
  \node at (4.35,7) {\footnotesize 14};
  \draw[fill=color3!50](3.6,3.7) -- ++(1,0) -- ++(0.6,0.8) -- ++(-1,0) -- cycle;
  \node at (4.35,4.1) {\footnotesize 15};
  \draw[fill=color4!50](3.6,0.8) -- ++(1,0) -- ++(0.6,0.8) -- ++(-1,0) -- cycle;
  \node at (4.35,1.2) {\footnotesize 7};
  \draw[fill=color5!50](1.2,7.4) -- ++(1,0) -- ++(0.6,0.8) -- ++(-1,0) -- cycle;
  \node at (1.95,7.8) {\footnotesize 4};
  \draw[fill=color5!50](1.2,4.5) -- ++(1,0) -- ++(0.6,0.8) -- ++(-1,0) -- cycle;
  \node at (1.95,4.9) {\footnotesize 5};
  \draw[fill=color6!50](1.2,1.6) -- ++(1,0) -- ++(0.6,0.8) -- ++(-1,0) -- cycle;
  \node at (1.95,2) {\footnotesize 2};
  \draw[fill=color5!50](2.2,7.4) -- ++(1,0) -- ++(0.6,0.8) -- ++(-1,0) -- cycle;
  \node at (2.95,7.8) {\footnotesize 6};
  \draw[fill=color5!50](2.2,4.5) -- ++(1,0) -- ++(0.6,0.8) -- ++(-1,0) -- cycle;
  \node at (2.95,4.9) {\footnotesize 7};
  \draw[fill=color6!50](2.2,1.6) -- ++(1,0) -- ++(0.6,0.8) -- ++(-1,0) -- cycle;
  \node at (2.95,2) {\footnotesize 3};
  \draw[fill=color7!50](3.2,7.4) -- ++(1,0) -- ++(0.6,0.8) -- ++(-1,0) -- cycle;
  \node at (3.95,7.8) {\footnotesize 4};
  \draw[fill=color7!50](3.2,4.5) -- ++(1,0) -- ++(0.6,0.8) -- ++(-1,0) -- cycle;
  \node at (3.95,4.9) {\footnotesize 5};
  \draw[fill=color8!50](3.2,1.6) -- ++(1,0) -- ++(0.6,0.8) -- ++(-1,0) -- cycle;
  \node at (3.95,2) {\footnotesize 2};
  \draw[fill=color7!50](4.2,7.4) -- ++(1,0) -- ++(0.6,0.8) -- ++(-1,0) -- cycle;
  \node at (4.95,7.8) {\footnotesize 6};
  \draw[fill=color7!50](4.2,4.5) -- ++(1,0) -- ++(0.6,0.8) -- ++(-1,0) -- cycle;
  \node at (4.95,4.9) {\footnotesize 7};
  \draw[fill=color8!50](4.2,1.6) -- ++(1,0) -- ++(0.6,0.8) -- ++(-1,0) -- cycle;
  \node at (4.95,2) {\footnotesize 3};
\draw[draw=black,thick](0,0) -- ++(4,0) -- ++(0,5.8) -- ++(-4,0) -- cycle;
\draw[draw=black,thick](0,5.8) -- ++(4,0) -- ++(1.8,2.4) -- ++(-4,0) -- cycle;
\draw[draw=black,thick](4,0) -- ++(1.8,2.4) -- ++(0,5.8) -- ++(-1.8,-2.4) -- cycle;
 \node at (2,-0.75) {$\Y(:,:,:,2)$};
 \end{scope}
\end{tikzpicture}

%
%
%
%
}
\caption{The $\exsize$ tensor from \cref{fig:tensor_seq_dist} distributed on a $2 \times 2 \times 2 \times 1$
processor grid. Each of the eight processors is represented by a different color.}
\label{fig:tensor_mpi_dist}
\end{figure}
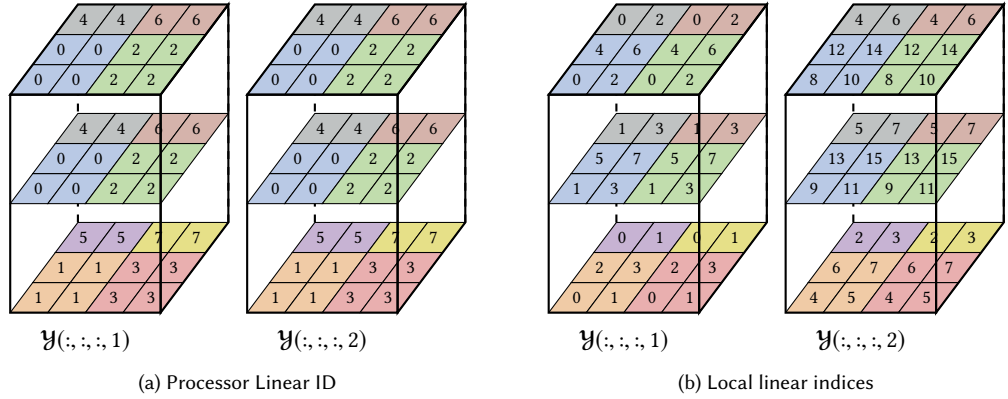

The tensor element with global index $\FullIndex$ is mapped to processor $\FullIndex*[p]$ where $\bar p_n$ is given by
\begin{multline}\label{eq:global-to-processor-index}
  \bar p_n =
  \GblToPrc \equiv
  \begin{cases}
    \left \lfloor \frac{j_n}{D_n+1} \right \rfloor & \text{if } j_n < M_n(D_n+1), \\
    \left \lfloor \frac{j_n - \, M_n(D_n+1)}{D_n} \right \rfloor + M_n &\text{otherwise},
  \end{cases}
  \\
  \text{where} \quad
  D_n = \Dn \qtext{and} M_n = \Mn.
\end{multline}
\Cref{fig:procids} shows the mapping of elements to processors for the tensor in \cref{fig:tensor_seq_dist}
using a processor grid of size $\expsize$.

We define the set of mode-$n$ global indices mapped to processor $\bar p_n$ to be
\begin{multline}
  \label{eq:processor-map}
  \bar{\mathcal{J}}_n =
  \PrcMap \equiv \set{ \bar p_n D_n + \min\{\bar p_n,M_n\}, \dots, (\bar p_n\!+\!1)D_n + \min\{\bar p_n\!+\!1,M_n\} - 1}.
  \\
  \text{where} \quad
  D_n = \Dn \qtext{and} M_n = \Mn.
\end{multline}
The per-mode assignments can be combined to get all the indices assigned
to a specific processor.
One interesting note is that elements that are contiguous in the
global natural descending linear index are not necessarily mapped to the same
processor.  For instance, in the example in
\cref{fig:tensor_mpi_dist}, consider that processor $\bar p'=2$ owns
the following global linear indices:
\begin{displaymath}
  \bar{\mathcal{J}} = \set{ 6,7,9,10,18,19,21,22},
\end{displaymath}
which is not contiguous in the global linear ordering.

The local index is denoted $\FullIndex*$ where $j_n$ is given by
\begin{equation}\label{eq:global-to-local-index}
  \bar j_n =
  \GblToLcl \equiv
  j_n -  \left( \bar p_n   \Dn + \min \left\{ \bar p_n , \Mn \right\} \right).
\end{equation}
Locally, the subtensors are stored contiguously according to the natural descending
index and in the same order as their global linear index.
\Cref{fig:localids} shows the local linear indices for the example in \Cref{fig:procids}.
This equation can be reversed to find the global index $\FullIndex$ where $j_n$ is given by
\begin{equation}\label{eq:local-to-global-index}
  j_n =
  \LclToGbl \equiv
  \bar j_n +  \left( \bar p_n   \Dn + \min \left\{ \bar p_n , \Mn \right\} \right).
\end{equation}

\subsection{Global Unfolding Data Distribution}
\label{sec:glob-unfold-data}

In the parallel case, the unfolded tensor is distributed among the processors in a block fashion.
For instance, the mode-1 unfolding is here color-coded to the processor
(using the same colors as in \cref{fig:tensor_mpi_dist}):
\begin{displaymath}\setlength\arraycolsep{1.4mm}%
\Mz{Y}{1} = \left[ \begin{array}{*{23}{r|}r}
\cellcolor{color1!50} 0  & \cellcolor{color1!50} 1  & \cellcolor{color2!50} 2  & \cellcolor{color1!50} 12  & \cellcolor{color1!50} 13  & \cellcolor{color2!50} 14  & \cellcolor{color5!50} 24  & \cellcolor{color5!50} 25  & \cellcolor{color6!50} 26  & \cellcolor{color1!50} 36  & \cellcolor{color1!50} 37  & \cellcolor{color2!50} 38  & \cellcolor{color1!50} 48  & \cellcolor{color1!50} 49  & \cellcolor{color2!50} 50  & \cellcolor{color5!50} 60  & \cellcolor{color5!50} 61  & \cellcolor{color6!50} 62  \\ 
\cellcolor{color1!50} 3  & \cellcolor{color1!50} 4  & \cellcolor{color2!50} 5  & \cellcolor{color1!50} 15  & \cellcolor{color1!50} 16  & \cellcolor{color2!50} 17  & \cellcolor{color5!50} 27  & \cellcolor{color5!50} 28  & \cellcolor{color6!50} 29  & \cellcolor{color1!50} 39  & \cellcolor{color1!50} 40  & \cellcolor{color2!50} 41  & \cellcolor{color1!50} 51  & \cellcolor{color1!50} 52  & \cellcolor{color2!50} 53  & \cellcolor{color5!50} 63  & \cellcolor{color5!50} 64  & \cellcolor{color6!50} 65  \\ 
\cellcolor{color3!50} 6  & \cellcolor{color3!50} 7  & \cellcolor{color4!50} 8  & \cellcolor{color3!50} 18  & \cellcolor{color3!50} 19  & \cellcolor{color4!50} 20  & \cellcolor{color7!50} 30  & \cellcolor{color7!50} 31  & \cellcolor{color8!50} 32  & \cellcolor{color3!50} 42  & \cellcolor{color3!50} 43  & \cellcolor{color4!50} 44  & \cellcolor{color3!50} 54  & \cellcolor{color3!50} 55  & \cellcolor{color4!50} 56  & \cellcolor{color7!50} 66  & \cellcolor{color7!50} 67  & \cellcolor{color8!50} 68  \\ 
\cellcolor{color3!50} 9  & \cellcolor{color3!50} 10  & \cellcolor{color4!50} 11  & \cellcolor{color3!50} 21  & \cellcolor{color3!50} 22  & \cellcolor{color4!50} 23  & \cellcolor{color7!50} 33  & \cellcolor{color7!50} 34  & \cellcolor{color8!50} 35  & \cellcolor{color3!50} 45  & \cellcolor{color3!50} 46  & \cellcolor{color4!50} 47  & \cellcolor{color3!50} 57  & \cellcolor{color3!50} 58  & \cellcolor{color4!50} 59  & \cellcolor{color7!50} 69  & \cellcolor{color7!50} 70  & \cellcolor{color8!50} 71  \\ 
\end{array}\right].
\end{displaymath}
Note that a single MPI process may own multiple contiguous pieces of the unfolded tensor.
In particular, the data distribution of the unfoldings are not a
standard distribution such as blocked or block-cyclic.
However, the distribution is still a blocked matrix distribution, equivalent to unfolding the local tensor.
Here we show the \emph{local} linear index of each entry:
\begin{displaymath}\setlength\arraycolsep{1.4mm}%
\Mz{Y}{1} = \left[ \begin{array}{*{5}{rrr|}rrr}
\cellcolor{color1!50} 0  & \cellcolor{color1!50} 1  & \cellcolor{color2!50} 0  & \cellcolor{color1!50} 4  & \cellcolor{color1!50} 5  & \cellcolor{color2!50} 2  & \cellcolor{color5!50} 0  & \cellcolor{color5!50} 1  & \cellcolor{color6!50} 0  & \cellcolor{color1!50} 8  & \cellcolor{color1!50} 9  & \cellcolor{color2!50} 4  & \cellcolor{color1!50} 12  & \cellcolor{color1!50} 13  & \cellcolor{color2!50} 6  & \cellcolor{color5!50} 4  & \cellcolor{color5!50} 5  & \cellcolor{color6!50} 2  \\ 
\cellcolor{color1!50} 2  & \cellcolor{color1!50} 3  & \cellcolor{color2!50} 1  & \cellcolor{color1!50} 6  & \cellcolor{color1!50} 7  & \cellcolor{color2!50} 3  & \cellcolor{color5!50} 2  & \cellcolor{color5!50} 3  & \cellcolor{color6!50} 1  & \cellcolor{color1!50} 10  & \cellcolor{color1!50} 11  & \cellcolor{color2!50} 5  & \cellcolor{color1!50} 14  & \cellcolor{color1!50} 15  & \cellcolor{color2!50} 7  & \cellcolor{color5!50} 6  & \cellcolor{color5!50} 7  & \cellcolor{color6!50} 3  \\ 
\cellcolor{color3!50} 0  & \cellcolor{color3!50} 1  & \cellcolor{color4!50} 0  & \cellcolor{color3!50} 4  & \cellcolor{color3!50} 5  & \cellcolor{color4!50} 2  & \cellcolor{color7!50} 0  & \cellcolor{color7!50} 1  & \cellcolor{color8!50} 0  & \cellcolor{color3!50} 8  & \cellcolor{color3!50} 9  & \cellcolor{color4!50} 4  & \cellcolor{color3!50} 12  & \cellcolor{color3!50} 13  & \cellcolor{color4!50} 6  & \cellcolor{color7!50} 4  & \cellcolor{color7!50} 5  & \cellcolor{color8!50} 2  \\ 
\cellcolor{color3!50} 2  & \cellcolor{color3!50} 3  & \cellcolor{color4!50} 1  & \cellcolor{color3!50} 6  & \cellcolor{color3!50} 7  & \cellcolor{color4!50} 3  & \cellcolor{color7!50} 2  & \cellcolor{color7!50} 3  & \cellcolor{color8!50} 1  & \cellcolor{color3!50} 10  & \cellcolor{color3!50} 11  & \cellcolor{color4!50} 5  & \cellcolor{color3!50} 14  & \cellcolor{color3!50} 15  & \cellcolor{color4!50} 7  & \cellcolor{color7!50} 6  & \cellcolor{color7!50} 7  & \cellcolor{color8!50} 3  \\ 
\end{array}\right].
\end{displaymath}
Notice that they are contiguous for each processor.
This means that we can work with the locally unfolded tensor for the local operations in distributed computations.

\subsection{Processor Fibers}
\label{sec:processor-fibers}

Just as for the data tensor, there is a corresponding processor grid ``unfolding'' that corresponds to the tensor unfoldings. Instead
of an $N$-way processor grid, we can think of the processors as rearranged into a 2-way grid. So, if we revisit the example
in the previous subsection, the mode-1 processor unfolding for the $\expsize$ processor grid is:
\begin{displaymath}
  \begin{array}{cccc}
 \cellcolor{color1!50} (0,0,0,0) &  \cellcolor{color2!50} (1,0,0,0) &  \cellcolor{color5!50} (0,0,1,0) &  \cellcolor{color6!50} (1,0,1,0) \\
 \cellcolor{color3!50} (0,1,0,0) &  \cellcolor{color4!50} (1,1,0,0) &  \cellcolor{color7!50} (0,1,1,0) &  \cellcolor{color8!50} (1,1,1,0) \\
  \end{array}.
\end{displaymath}
We see the processors in the same column match in every index except the mode-1 index, since this is the mode-1 unfolding.
We see later that certain operations require collective communications within each processor column in this unfolding.
We refer to these column groups of processors as \emph{mode-$n$ processor fibers}.
We refer to row groups of processors as \emph{mode-$n$ processor slices}.

%
%
%
%
%

\section{Local Kernels}
\label{sec:functions}

In \cref{alg:sthosvd}, there are three key kernels:
Gram of the mode-$n$ unfolded tensors in \cref{line:hosvd:gram},
the $I_n \times I_n$ eigenproblem in \cref{line:hosvd:evecs}, and
tensor-times-matrix to shrink in mode $n$ in \cref{line:hosvd:ttm}.
Here we explain the serial implementation, which is also used for the local
computations in the parallelized version.
We explain all the functions with respect to a generic tensor $\Y$ of size $\FullSize$.

\subsection{Local Gram}

We want to compute $\M{S} = \UnfoldGram$ so $\M{S}$ will be of size $J_n \times J_n$.
The arithmetic cost is $\bigO(\LinearSize J_n)$.
Although the computation is mathematically one matrix
operation, the data layout of the unfolding of the input tensor prevents a
single call to the \texttt{syrk} BLAS subroutine, which requires strided row- or column-major ordering.
Thus, the algorithm works on the native row-major submatrices as described in \cref{sec:unfolded_layout},
except in the case of $n=0$ where the algorithm works on the native column-major matrices.
For $n>0$, $\Mz{Y}{n}$ has $\RightSize$ submatrices of size
$J_n \times \LeftSize$. We denote the
$j$th block-column submatrix as $\ColumnBlock[\Mz{Y}{n}]{j} \equiv \Mz{Y}{n}(\,:\,,j\!\cdot\!
\LeftSize[J]:j\!\cdot\!\LeftSize[J] + \LeftSize[J]-1)$,
which is stored in row-major form.
The algorithm is shown in \cref{alg:gram}.

\begin{algorithm}
\caption{Local Gram (compute gram matrix of unfolding in mode $n$)}
\label{alg:gram}
\begin{algorithmic}
\Procedure{$\M{S}=$ gram}{$\Y,n$}
  \If{$n=0$}
  \State $\M{S}=\UnfoldGram$
  \Comment Call to \texttt{syrk}, $\Mz{Y}{n}$ in column-major format
  \Else
  \State $\M{S} \gets \M{0}$
  \For{$j \in \RangeSet{\RightSize}$}
  \State $\M{S} \gets \M{S} + \ColumnBlock{j} \ColumnBlock{j}^\Tra$
  \Comment Call to \texttt{syrk}, $\ColumnBlock[\Mz{Y}{n}]{j}$ denotes $j$th block column in row-major format
  \EndFor
  \EndIf
\EndProcedure
\end{algorithmic}
\end{algorithm}

\subsection{Local Eigenvalue Decomposition}
\label{sec:local-eigenv-decomp}

We need to compute the eigenvalue decomposition of a $J_n \times J_n$ matrix $\M{S}$.
The computational cost is $\bigO(J_n^3)$.
We use the the LAPACK direct eigenvalue computation routine \texttt{syev}
to compute all the eigenvalues and all the eigenvectors.

A couple of notes are in order.
The cost of the full eigenvector decomposition is  approximately $\frac{10}{3}
J_n^3$.
One alternative approach would be to compute the full set of eigenvalues at a
cost of $\frac{4}{3} J_n^3$, and then compute only the leading
eigenvectors with $O(J_n^2 R_n)$ flops.
Iterative methods, such as subspace iteration, would also work well in
this case.
However, because this phase of computation has never been a bottleneck for our
applications, we have not implemented these cheaper approaches.
If $J_n$ is relatively large compared to the
product of the other dimensions, $\UnfoldSize$, then the eigenvalue
decomposition may become a bottleneck, so we leave this as a topic for future work.

\subsection{Local TTM}

We want to compute the product $\T{Z} = \Y \times_n \M{V}$ where $\M{V}$ is a
matrix of size $K \times J_n$. Recall that the TTM operation
is defined in \cref{sec:tensor-times-matrix}.
The arithmetic cost of TTM is $\bigO(\LinearSize[J]K)$.

As in the case of Gram, although TTM is mathematically a matrix multiplication ($\Mz{Z}{n} = \M{V}\Mz{Y}{n}$), the data layouts of 
$\Mz{Y}{n}$ and $\Mz{Z}{n}$ prevent a single call to the 
\texttt{gemm} subroutine in BLAS because BLAS requires
strided row- or column-major access to the matrices.
We use the same notation as for Gram, so that $\ColumnBlock[\Mz{Y}{n}]{j}$ is the $j$th block column of $\Mz{Y}{n}$ stored in row-major form.
Likewise, $\Mz{Z}{n}$  is organized into row-major block column submatrices of size $K \times \LeftSize$,
and the $j$th submatrix is denoted as $\ColumnBlock[\Mz{Z}{n}]{j}$.
In the case of $n=0$, both $\Mz{Y}{n}$ and $\Mz{Z}{n}$ are 
natively in column-major mode.
The algorithm is shown in \cref{alg:ttm}.

\begin{algorithm}
\caption{Local TTM (tensor-times-matrix in mode $n$)}
\label{alg:ttm}
\begin{algorithmic}
  \Procedure{$\T{Z}=$ ttm}{$\Y,n,\M{V}$}
  \Comment{$\Y$ is tensor of size $J_0 \times \dots \times J_{N-1}$ and $\M{V}$ is matrix of size $K \times J_n$}
  \If{$n=0$}
  \State $\Mz{Z}{n}=\M{V}\Mz{Y}{n}$
  \Comment Call to \texttt{gemm}, $\Mz{Z}{n}$ and $\Mz{Y}{n}$ in column-major format
  \Else
  \For{$j \in \RangeSet{\RightSize[J]}$}
  \State $\ColumnBlock[\Mz{Z}{n}]{j} \gets \M{V} \ColumnBlock{j}$
  \Comment Call to \texttt{gemm}, $\ColumnBlock[\Mz{Z}{n}]{j}$ and $\ColumnBlock[\Mz{Y}{n}]{j}$ denote $j$th block column in row-major
format
  \EndFor
  \EndIf
\EndProcedure
\end{algorithmic}
\end{algorithm}

%
%
%
%
%

\section{Distributed Kernels}
\label{sec:par_algs}

To parallelize STHOSVD (\cref{alg:sthosvd}), we use parallel algorithms for the two key kernels: Gram and TTM.
All other computations are performed redundantly on each processor, including the eigendecomposition.

Throughout this section, we consider a generic
tensor $\Y$ of size $\FullSize$, distributed on a processor grid of
size $\FullSize[P]$, as described in \cref{sec:tensor_distribution}.
We use an overbar to denote the \emph{local} portions/versions/sizes
of distributed variables. For instance, $\Y*$ denotes the local
portion of $\Y$, and $\FullSize*$ is its size. Note that local
quantities, including their sizes, may vary from processor to processor.

\subsection{Assumptions on Collective Communication}
\label{sec:collectives}

To analyze our algorithms, we use the MPI model of
distributed-memory parallel computation.
We assume a fully
connected network of $P$ processors and therefore do not model network
contention.  For simplicity of discussion in our analysis, we assume that $P$ is
a power of two and that optimal collective communication algorithms are used.
The cost to \emph{send/receive} a message of
size $W$ words between any two processors is $\alpha+W\beta$, where $\alpha$ is
the latency cost and $\beta$ is the per-word transfer cost.
The cost of the collective communications used in this work are given in \cref{tab:communication_model}, with $\gamma$ corresponding to the time per floating point operation (flop).
For simplicity of presentation, we will ignore the flop cost of the reductions in later analysis, as they are typically dominated by the bandwidth costs.
For more discussion of the model and descriptions of efficient collectives, see
\cite{CH+07,TRG05}.  
\begin{table}[ht]
  \centering
  \begin{tabular}{|l|l|} \hline
    Send/Receive & $\alpha + \beta W$ \\
    Reduce/All-Reduce & $2 \alpha \log P + (2\beta + \gamma) \frac{P-1}{P} W$ \\
    Reduce-Scatter & $\alpha \log P + (\beta+\gamma) \frac{P-1}{P} W$ \\
    All-to-All & $\alpha(P{-}1) + \beta \frac{P-1}{P} W$ \\
    \hline
  \end{tabular}
  \caption{Communication costs in MPI model, where $W$ is the local input data
  size and $P$ is the number of processors.}
  \label{tab:communication_model}
\end{table}

\subsection{Parallel Gram}
\label{sec:par_gram}

The goal here is to compute $\M{S} = \UnfoldGram$ where $\Y$ is
distributed as described in \cref{sec:tensor_distribution}.  Each processor owns $\Y*$ (local
portion of $\Y$) of size $\FullSize*$ where $\bar J_n = \Lsz$ and the
mode-$n$ indices correspond to the global indices in
$\bar{\mathcal{J}}_n = \PrcMap$.
In the end, each processor will own the entirety of $\M{S}$, which is only of size $J_n \times J_n$.

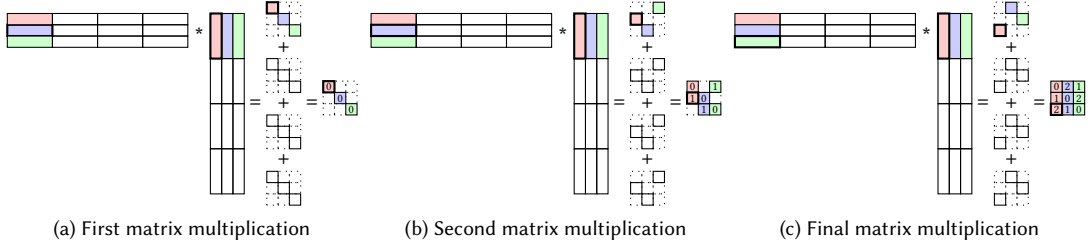
\begin{figure}
\centering
\subfloat[First matrix multiplication]{
\begin{tikzpicture}[scale=0.15,namenode/.style={scale=.75}]

\filldraw[fill=blue!20,thick] (0,15) rectangle(4,16);
\filldraw[fill=green!20] (0,14) rectangle(4,15);
\filldraw[fill=red!20] (0,16) rectangle(4,17);

\draw (4,16) rectangle(8,17);
\draw (4,15) rectangle(8,16);
\draw (4,14) rectangle(8,15);

\draw (8,16) rectangle(12,17);
\draw (8,15) rectangle(12,16);
\draw (8,14) rectangle(12,15);

\draw (12,16) rectangle(16,17);
\draw (12,15) rectangle(16,16);
\draw (12,14) rectangle(16,15);

\path (17,15) node {*};

\filldraw[fill=blue!20] (19,13) rectangle(20,17);
\filldraw[fill=green!20] (20,13) rectangle(21,17);
\filldraw[fill=red!20, thick] (18,13) rectangle(19,17);

\draw (18,9) rectangle(19,13);
\draw (19,9) rectangle(20,13);
\draw (20,9) rectangle(21,13);

\draw (18,5) rectangle(19,9);
\draw (19,5) rectangle(20,9);
\draw (20,5) rectangle(21,9);

\draw (18,1) rectangle(19,5);
\draw (19,1) rectangle(20,5);
\draw (20,1) rectangle(21,5);

\path (22,9) node {=};

\draw[dotted] (23,16) rectangle (24,17);
\draw[dotted] (23,15) rectangle (24,16);

\draw[dotted] (24,17) rectangle (25,18);
\draw[dotted] (24,15) rectangle(25,16);

\draw[dotted] (25,17) rectangle (26,18);
\draw[dotted] (25,16) rectangle (26,17);

\filldraw[fill=red!20, thick] (23,17) rectangle (24,18);
\filldraw[fill=blue!20] (24,16) rectangle (25,17);
\filldraw[fill=green!20] (25,15) rectangle (26,16);

\path(24.5,14) node {+};

\draw[dotted] (23,11) rectangle (24,12);
\draw[dotted] (23,10) rectangle (24,11);

\draw[dotted] (24,12) rectangle (25,13);
\draw[dotted] (24,10) rectangle(25,11);

\draw[dotted] (25,12) rectangle (26,13);
\draw[dotted] (25,11) rectangle (26,12);

\draw (23,12) rectangle (24,13);
\draw (24,11) rectangle (25,12);
\draw (25,10) rectangle (26,11);

\path (24.5,9) node {+};

\draw[dotted] (23,6) rectangle (24,7);
\draw[dotted] (23,5) rectangle (24,6);

\draw[dotted] (24,7) rectangle (25,8);
\draw[dotted] (24,5) rectangle(25,6);

\draw[dotted] (25,7) rectangle (26,8);
\draw[dotted] (25,6) rectangle (26,7);

\draw (23,7) rectangle (24,8);
\draw (24,6) rectangle (25,7);
\draw (25,5) rectangle (26,6);

\path (24.5,4) node {+};

\draw[dotted] (23,1) rectangle (24,2);
\draw[dotted] (23,0) rectangle (24,1);

\draw[dotted] (24,2) rectangle (25,3);
\draw[dotted] (24,0) rectangle(25,1);

\draw[dotted] (25,2) rectangle (26,3);
\draw[dotted] (25,1) rectangle (26,2);

\draw (23,2) rectangle (24,3);
\draw (24,1) rectangle (25,2);
\draw (25,0) rectangle (26,1);

\path (27,9) node {=};

\draw[dotted] (28,9) rectangle (29,10);
\draw[dotted] (28,8) rectangle (29,9);

\draw[dotted] (29,10) rectangle (30,11);
\draw[dotted] (29,8) rectangle(30,9);

\draw[dotted] (30,10) rectangle (31,11);
\draw[dotted] (30,9) rectangle (31,10);

\filldraw[fill=red!20, thick] (28,10) rectangle (29,11) 
         node[pos=.5, scale=.5] {0}; 
\filldraw[fill=blue!20] (29,9) rectangle (30,10)
         node[pos=.5, scale=.5] {0}; 
\filldraw[fill=green!20] (30,8) rectangle (31,9)
         node[pos=.5, scale=.5] {0}; 

\end{tikzpicture}

%
%
%
%
}
\subfloat[Second matrix multiplication]{
\begin{tikzpicture}[scale=0.15,namenode/.style={scale=.75}]

\filldraw[fill=green!20] (0,14) rectangle(4,15);
\filldraw[fill=red!20] (0,16) rectangle(4,17);
\filldraw[fill=blue!20, thick] (0,15) rectangle(4,16);

\draw (4,16) rectangle(8,17);
\draw (4,15) rectangle(8,16);
\draw (4,14) rectangle(8,15);

\draw (8,16) rectangle(12,17);
\draw (8,15) rectangle(12,16);
\draw (8,14) rectangle(12,15);

\draw (12,16) rectangle(16,17);
\draw (12,15) rectangle(16,16);
\draw (12,14) rectangle(16,15);

\path (17,15) node {*};

\filldraw[fill=blue!20] (19,13) rectangle(20,17);
\filldraw[fill=green!20] (20,13) rectangle(21,17);
\filldraw[fill=red!20, thick] (18,13) rectangle(19,17);

\draw (18,9) rectangle(19,13);
\draw (19,9) rectangle(20,13);
\draw (20,9) rectangle(21,13);

\draw (18,5) rectangle(19,9);
\draw (19,5) rectangle(20,9);
\draw (20,5) rectangle(21,9);

\draw (18,1) rectangle(19,5);
\draw (19,1) rectangle(20,5);
\draw (20,1) rectangle(21,5);

\path (22,9) node {=};

\draw[dotted] (23,17) rectangle (24,18);
\draw[dotted] (23,15) rectangle (24,16);

\draw[dotted] (24,17) rectangle (25,18);
\draw[dotted] (24,16) rectangle(25,17);

\draw[dotted] (25,15) rectangle (26,16);
\draw[dotted] (25,16) rectangle (26,17);

\filldraw[fill=red!20, thick] (23,16) rectangle (24,17);
\filldraw[fill=blue!20] (24,15) rectangle (25,16);
\filldraw[fill=green!20] (25,17) rectangle (26,18);

\path(24.5,14) node {+};

\draw (23,11) rectangle (24,12);
\draw[dotted] (23,10) rectangle (24,11);

\draw[dotted] (24,12) rectangle (25,13);
\draw (24,10) rectangle(25,11);

\draw (25,12) rectangle (26,13);
\draw[dotted] (25,11) rectangle (26,12);

\draw[dotted] (23,12) rectangle (24,13);
\draw[dotted] (24,11) rectangle (25,12);
\draw[dotted] (25,10) rectangle (26,11);

\path (24.5,9) node {+};

\draw(23,6) rectangle (24,7);
\draw[dotted] (23,5) rectangle (24,6);

\draw[dotted] (24,7) rectangle (25,8);
\draw (24,5) rectangle(25,6);

\draw (25,7) rectangle (26,8);
\draw[dotted] (25,6) rectangle (26,7);

\draw[dotted] (23,7) rectangle (24,8);
\draw[dotted] (24,6) rectangle (25,7);
\draw[dotted] (25,5) rectangle (26,6);

\path (24.5,4) node {+};

\draw (23,1) rectangle (24,2);
\draw[dotted] (23,0) rectangle (24,1);

\draw[dotted] (24,2) rectangle (25,3);
\draw (24,0) rectangle(25,1);

\draw (25,2) rectangle (26,3);
\draw[dotted] (25,1) rectangle (26,2);

\draw[dotted] (23,2) rectangle (24,3);
\draw[dotted] (24,1) rectangle (25,2);
\draw[dotted] (25,0) rectangle (26,1);

\path (27,9) node {=};

\draw[dotted] (28,8) rectangle (29,9);

\draw[dotted] (29,10) rectangle (30,11);
\filldraw[fill=blue!20] (29,8) rectangle(30,9)
         node[pos=.5, scale=.5] {1}; 

\filldraw[fill=green!20] (30,10) rectangle (31,11)
         node[pos=.5, scale=.5] {1};  
         
\draw[dotted] (30,9) rectangle (31,10);

\filldraw[fill=red!20] (28,10) rectangle (29,11) 
         node[pos=.5, scale=.5] {0}; 
\filldraw[fill=blue!20] (29,9) rectangle (30,10)
         node[pos=.5, scale=.5] {0}; 
\filldraw[fill=green!20] (30,8) rectangle (31,9)
         node[pos=.5, scale=.5] {0}; 
         
\filldraw[fill=red!20, thick] (28,9) rectangle (29,10)
         node[pos=.5, scale=.5] {1}; 
         
\end{tikzpicture}

%
%
%
%
}
\subfloat[Final matrix multiplication]{
\begin{tikzpicture}[scale=0.15,namenode/.style={scale=.75}]

\filldraw[fill=red!20] (0,16) rectangle(4,17);
\filldraw[fill=blue!20] (0,15) rectangle(4,16);
\filldraw[fill=green!20, thick] (0,14) rectangle(4,15);

\draw (4,16) rectangle(8,17);
\draw (4,15) rectangle(8,16);
\draw (4,14) rectangle(8,15);

\draw (8,16) rectangle(12,17);
\draw (8,15) rectangle(12,16);
\draw (8,14) rectangle(12,15);

\draw (12,16) rectangle(16,17);
\draw (12,15) rectangle(16,16);
\draw (12,14) rectangle(16,15);

\path (17,15) node {*};

\filldraw[fill=blue!20] (19,13) rectangle(20,17);
\filldraw[fill=green!20] (20,13) rectangle(21,17);
\filldraw[fill=red!20, thick] (18,13) rectangle(19,17);

\draw (18,9) rectangle(19,13);
\draw (19,9) rectangle(20,13);
\draw (20,9) rectangle(21,13);

\draw (18,5) rectangle(19,9);
\draw (19,5) rectangle(20,9);
\draw (20,5) rectangle(21,9);

\draw (18,1) rectangle(19,5);
\draw (19,1) rectangle(20,5);
\draw (20,1) rectangle(21,5);

\path (22,9) node {=};

\draw[dotted] (23,17) rectangle (24,18);
\draw[dotted] (23,16) rectangle (24,17);

\draw[dotted] (24,15) rectangle (25,16);
\draw[dotted] (24,16) rectangle(25,17);

\draw[dotted] (25,15) rectangle (26,16);
\draw[dotted] (25,17) rectangle (26,18);

\filldraw[fill=red!20, thick] (23,15) rectangle (24,16);
\filldraw[fill=blue!20] (24,17) rectangle (25,18);
\filldraw[fill=green!20] (25,16) rectangle (26,17);

\path(24.5,14) node {+};

\draw[dotted] (23,11) rectangle (24,12);
\draw (23,10) rectangle (24,11);

\draw (24,12) rectangle (25,13);
\draw[dotted] (24,10) rectangle(25,11);

\draw[dotted] (25,12) rectangle (26,13);
\draw (25,11) rectangle (26,12);

\draw[dotted] (23,12) rectangle (24,13);
\draw[dotted] (24,11) rectangle (25,12);
\draw[dotted] (25,10) rectangle (26,11);

\path (24.5,9) node {+};

\draw[dotted] (23,6) rectangle (24,7);
\draw (23,5) rectangle (24,6);

\draw (24,7) rectangle (25,8);
\draw[dotted] (24,5) rectangle(25,6);

\draw[dotted] (25,7) rectangle (26,8);
\draw (25,6) rectangle (26,7);

\draw[dotted] (23,7) rectangle (24,8);
\draw[dotted] (24,6) rectangle (25,7);
\draw[dotted] (25,5) rectangle (26,6);

\path (24.5,4) node {+};

\draw[dotted] (23,1) rectangle (24,2);
\draw (23,0) rectangle (24,1);

\draw (24,2) rectangle (25,3);
\draw[dotted] (24,0) rectangle(25,1);

\draw[dotted] (25,2) rectangle (26,3);
\draw (25,1) rectangle (26,2);

\draw[dotted] (23,2) rectangle (24,3);
\draw[dotted] (24,1) rectangle (25,2);
\draw[dotted] (25,0) rectangle (26,1);

\path (27,9) node {=};

\filldraw[fill=blue!20] (29,10) rectangle (30,11)
         node[pos=.5, scale=.5] {2}; 
\filldraw[fill=blue!20] (29,8) rectangle(30,9)
         node[pos=.5, scale=.5] {1}; 

\filldraw[fill=green!20] (30,10) rectangle (31,11)
         node[pos=.5, scale=.5] {1};  
         
\filldraw[fill=green!20] (30,9) rectangle (31,10)
         node[pos=.5, scale=.5] {2}; 

\filldraw[fill=red!20] (28,10) rectangle (29,11) 
         node[pos=.5, scale=.5] {0}; 
\filldraw[fill=blue!20] (29,9) rectangle (30,10)
         node[pos=.5, scale=.5] {0}; 
\filldraw[fill=green!20] (30,8) rectangle (31,9)
         node[pos=.5, scale=.5] {0}; 
         
\filldraw[fill=red!20] (28,9) rectangle (29,10)
         node[pos=.5, scale=.5] {1}; 
         
\filldraw[fill=red!20, thick] (28,8) rectangle (29,9)
         node[pos=.5, scale=.5] {2}; 
         
\end{tikzpicture}

%
%
%
%
}
\caption{Round-robin variant of parallel Gram matrix computation with $\SingleSize[P]=3$ \cite{AuBaKo16}}
\label{fig:gram_old}
\end{figure}

Austin et al.~\cite{AuBaKo16} previously proposed a parallel Gram
as illustrated in \cref{fig:gram_old}. 
In this algorithm, each processor fiber, which owns a column block of $\Mz{Y}{n}$, computes a contribution $\M{V}$ to the result $\M{S}$; $\M{V}$ is distributed across the processor fiber.
In order to compute $\M{V}$ within the fiber, the processors rotate their tensor data around in a round-robin fashion, and at each step, each processor computes a $\SingleSize*\times\SingleSize*$ block of $\M{V}$.
In order to compute $\M{S}$, the processors perform an All-Reduce across processor slices, so that $\M{S}$ is redundantly stored on every processor fiber but distributed across the processors within the fiber.
We refer to the older version as
the \emph{round-robin} variant.

We propose a new version of parallel Gram that is nearly always faster, which we refer
to as the \emph{redistribution} variant.
\Cref{alg:par_gram} presents the new parallel algorithm for computing the Gram
matrix corresponding to a particular mode.
The algorithm assumes that the input tensor $\Y$ is block distributed;
at the end of the algorithm, the output matrix $\M S=\UnfoldGram$ is
redundantly stored on every processor.
The parallel Gram algorithm presented here differs from the one described in
\cite{AuBaKo16}, depicted in \cref{fig:gram_old}.
When the number of processors in the specified mode $P_n$ is 1, the
algorithms are identical.
However, when $P_n>1$, the previous algorithm uses $P_n-1$
communication steps within the processor fiber and then communicates across the
processor slice.
As we describe in more detail below, \Cref{alg:par_gram} works by first
redistributing the tensor data with an All-to-All collective within the
processor fiber and then performing a reduction across all processors.
We compare the communication costs of the two algorithms at the end of the
section.

\begin{figure}
\centering
\begin{tikzpicture}[scale=0.25,namenode/.style={scale=.75}]

\filldraw[fill=red!20] (0,11) rectangle(3,12);
\filldraw[fill=blue!20] (0,10) rectangle(3,11);
\filldraw[fill=green!20] (0,9) rectangle(3,10);

\draw (3,9) rectangle(6,10);
\draw (3,10) rectangle(6,11);
\draw (3,11) rectangle(6,12);

\draw (6,9) rectangle(9,10);
\draw (6,10) rectangle(9,11);
\draw (6,11) rectangle(9,12);

\draw (9,9) rectangle(12,10);
\draw (9,10) rectangle(12,11);
\draw (9,11) rectangle(12,12);

\path (13,10) node {*};

\filldraw[fill=red!20] (14,9) rectangle(15,12);
\filldraw[fill=blue!20] (15,9) rectangle(16,12);
\filldraw[fill=green!20] (16,9) rectangle(17,12); 

\draw (14,6) rectangle(15,9);
\draw (15,6) rectangle(16,9);
\draw (16,6) rectangle(17,9);

\draw (14,3) rectangle(15,6);
\draw (15,3) rectangle(16,6);
\draw (16,3) rectangle(17,6);

\draw (14,0) rectangle(15,3);
\draw (15,0) rectangle(16,3);
\draw (16,0) rectangle(17,3);

\draw[->] (18,6) -- (19,6);

\filldraw[fill=red!20] (20,9) rectangle(21,12);
\filldraw[fill=blue!20] (21,9) rectangle(22,12);
\filldraw[fill=green!20] (22,9) rectangle(23,12);

\draw (23,9) rectangle(24,12);
\draw (24,9) rectangle(25,12);
\draw (25,9) rectangle(26,12);

\draw (26,9) rectangle(27,12);
\draw (27,9) rectangle(28,12);
\draw (28,9) rectangle(29,12);

\draw (29,9) rectangle(30,12);
\draw (30,9) rectangle(31,12);
\draw (31,9) rectangle(32,12);

\path (33,10) node {*};

\filldraw[fill=red!20] (34,11) rectangle(37,12);
\filldraw[fill=blue!20] (34,10) rectangle(37,11);
\filldraw[fill=green!20] (34,9) rectangle(37,10); 

\draw (34,8) rectangle(37,9);
\draw (34,7) rectangle(37,8);
\draw (34,6) rectangle(37,7);

\draw (34,5) rectangle(37,6);
\draw (34,4) rectangle(37,5);
\draw (34,3) rectangle(37,4);

\draw (34,2) rectangle(37,3);
\draw (34,1) rectangle(37,2);
\draw (34,0) rectangle(37,1);

\path (38,6) node {=};

\filldraw[fill=white] (44.5,5.5) rectangle(47.5,8.5);
\filldraw[fill=white] (44,5) rectangle(47,8);
\filldraw[fill=white] (43.5,4.5) rectangle(46.5,7.5);

\filldraw[fill=white] (43,4) rectangle(46,7);
\filldraw[fill=white] (42.5,3.5) rectangle(45.5,6.5);
\filldraw[fill=white] (42,3) rectangle(45,6);

\filldraw[fill=white] (41.5,2.5) rectangle(44.5,5.5);
\filldraw[fill=white] (41,2) rectangle(44,5);
\filldraw[fill=white] (40.5,1.5) rectangle(43.5,4.5);

\filldraw[fill=green!20] (40,1) rectangle(43,4);
\filldraw[fill=blue!20] (39.5,0.5) rectangle(42.5,3.5);
\filldraw[fill=red!20] (39,0) rectangle(42,3);

\draw[<->] (39,4) -- node[above, sloped] {sum along} (44.5,9.5);

\path (49,6) node {=};

\filldraw[fill=gray!20] (50,4.5) rectangle(53,7.5);

\end{tikzpicture}

%
%
%
%
\caption{New redistribution variant of parallel Gram matrix computation}
\label{fig:gram_new}
\end{figure}
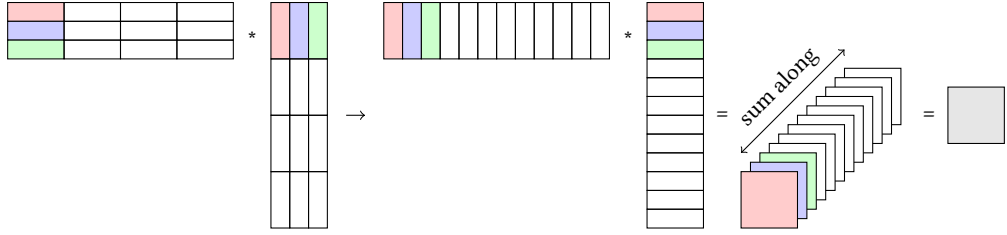

As in the case of \cref{alg:par_ttm},
\cref{line:par_gram:myprocid,line:par_gram:myprocfiber} of \cref{alg:par_gram} define the processor's
index and the set of processors within the processor's $n$th-mode processor fiber.
The goal of \cref{line:par_gram:alltoall} is to obtain a 1D parallel
distribution of the tensor. 
With this distribution, all processors can perform
Gram computations with their local data, and the only remaining communication is
to sum up the results over all processors.
\Cref{fig:gram_new} illustrates the redistribution.
Because each processor fiber stores a set of columns of the matricized tensor
(distributed row-wise), the redistribution occurs within each fiber
independently and converts the row-wise distribution to a column-wise
distribution.

Again, a key issue in the implementation is the need to pack and unpack buffers for the All-to-All collective, depending on $n$.
The input buffer must be arranged so that the data to be received by each processor is stored contiguously, and the result buffer is ordered so that the data each processor sent is contiguous.
In the case $n=0$, the local matricized tensor is column-major, so no packing is necessary.
After the All-to-All, the result buffer consists of $\SingleSize[P]$ contiguous column-major blocks, where each column has length $\SingleSize*$.
The unpacking consists of collecting the $\SingleSize[P]$ chunks of each mode-$n$ fiber (of length $\SingleSize$) into contiguous columns.
In the case $0<n<N-1$, the local matricized tensor is stored as an array of row-major submatrices.
The packing consists of converting every row-major submatrix to column-major ordering, which makes every local mode-$n$ fiber contiguous.
The unpacking is the same as for $n=0$: for each mode-$n$ fiber, the $\SingleSize[P]$ chunks are made contiguous.
We note that the local ordering of the columns is not consistent with a matricized tensor format, but the Gram computation is invariant under column permutations, so this does not affect the result.
In the case $n=N-1$, the local matricized tensor is row-major.
Instead of converting the entire matrix to column-major, each contiguous row is broken up into $\SingleSize[P]$ chunks, and the row ordering is maintained.
After the All-to-All, no unpacking is necessary because the result buffer is a (vertical) concatenation of row-major matrices and is thus also row-major.

\begin{algorithm}
\caption{Parallel Gram (redistribution variant)}
\label{alg:par_gram}
\begin{algorithmic}[1]
\Procedure{$\M{S}=$ par\_gram}{$\Y*,n,\FullIndex[P]$}
  \State $\texttt{myProcID} \gets \FullIndex*[p]$
    \label{line:par_gram:myprocid}
  \State $\texttt{myProcFiber} \gets
  (\bar p_{0}, \dots, \bar p_{n-1}, \, : \,, \bar p_{n+1}, \dots, \bar p_{N-1})$
  \label{line:par_gram:myprocfiber}
  \Comment{Processor group of size $P_n$}
  \State $\texttt{allProcs} \gets $ all $\LinearSize[P]$ processor ids
    \label{line:par_gram:allprocs}
  \State $\M[\bar]{Z} =
  \textproc{AllToAll}(\Mz{\local{Y}}{n},\texttt{myProcFiber})$
  \label{line:par_gram:alltoall}
  \Comment{Change from within-fiber block row to block column distribution}
  \State $\M[\bar]{W} = \M[\bar]{Z}\M[\bar]{Z}^\Tra$
  \label{line:par_gram:seq_gram}
  \Comment{Local computation}
  \State $\M{S} = \textproc{AllReduce}(\M[\bar]{W},\texttt{allProcs})$
  \label{line:par_gram:allred}
  \Comment{All processors own a copy of the same matrix}
\EndProcedure
\end{algorithmic}
\end{algorithm}

After obtaining a 1D distribution of the matricized tensor, each processor
performs its local computation in \cref{line:par_gram:seq_gram}, computing the
Gram matrix associated with its subset of the columns of the matricized tensor.
In order to compute the Gram matrix of the entire matricized tensor, all
processors participate in an All-Reduce (\cref{line:par_gram:allred}), which
results in (symmetric) $\M{S}$ being redundantly stored on all processors.

The cost of \cref{alg:par_gram} is
\begin{align*}
C_{\textsc{gram}} &=
  \underbrace{(P_n-1)\lt( \alpha + \beta
  \frac{\LinearSize*}{P_n} \rt) }
  _{\text{\cref{line:par_gram:alltoall}}} +
  \underbrace{ \gamma J_n \LinearSize*}
  _{\text{\cref{line:par_gram:seq_gram} }}
   +  \underbrace{2\alpha \log \LinearSize[P] + \beta J_n^2}
  _{\text{\cref{line:par_gram:allred}}} \\
  &= \alpha \cdot O \lt(P_n\rt) + \beta \cdot O
  \lt(\LinearSize* + J_n^2 \rt) + \gamma
  \cdot O \lt( J_n\LinearSize* \rt).
\end{align*}
For comparison, the cost of the previous
parallel Gram algorithm \cite{AuBaKo16} is $$\alpha \cdot O \lt(P_n\rt)
+ \beta \cdot O \lt(P_n\LinearSize* +
\frac{J_n^2}{P_n} \rt) + \gamma \cdot O \lt(
J_n\LinearSize* \rt), $$ where the major
difference is in the bandwidth cost.
In particular, \cref{alg:par_gram} is more efficient when $P_n
\LinearSize* \geq J_n^2$.
In this case, both bandwidth cost terms of the new algorithm are smaller than
the first term of the old algorithm.
(If the converse is true, then both bandwidth cost terms of the old algorithm
are smaller than the second term of the new algorithm.) We expect the inequality
to hold in nearly all cases (except for extreme strong-scaling cases) because
$\LinearSize*$ itself is the size of the local input tensor,
while $J_n^2$ is the size of the $n$th-mode Gram matrix.

The temporary memory requirement of \cref{alg:par_gram} (besides the input and
output data) is $2\LinearSize* + J_n^2$
words.
Two temporary arrays of the size of the local tensor are required for the send
and receive buffers in the All-to-All (local data has to be reordered to match
the requirements of the input buffer), and $J_n^2$ space is required
for $\M W$ (All-Reduce requires separate input and output buffers).
For comparison, the temporary memory requirement of the previous Gram algorithm
is $\LinearSize* + \bar J_n^2$.

\subsection{Parallel TTM}
\label{sec:par_ttm}

We consider the problem of computing the TTM $\T{Z} = \Y \times_n \M{V}$
where the input $\Y$ of size $\FullSize$ is block distributed,
the matrix $\M{V}$ of size $K_n \times J_n$ with $K_n < J_n$ is stored redundantly on every processor,
and the output tensor $\T{Z}$ of size
$J_1 \times \cdots \times J_{n-1} \times K_n \times J_{n+1} \times \cdots \times J_N$
will be block distributed.
Specifically, the data is distributed so that processor $\FullIndex*[p]$ owns the following:
\begin{itemize}
\item $\Y*$ (local portion of $\Y$) is of size $\FullSize*$ where
  $\bar J_n = \Lsz$ and the mode-$n$ indices correspond to the global
  indices in $\bar{\mathcal{J}}_n = \PrcMap$.
\item $\T[\bar]{Z}$ (local portion of $\T{Z}$) is of size
  $\bar J_1 \times \cdots \times \bar J_{n-1} \times \bar K_n \times
  \bar J_{n+1} \times \cdots \times \bar J_N$ where
  $\bar K_n = \Lsz{K_n}$; it is distributed the same as $\Y*$ except that the mode-$n$ indices correspond to the
  global indices in $\bar{\mathcal{K}}_n = \PrcMap{K_n}$.
\item $\M[\bar]{V}$ is of size $K_n \times \bar J_n$ and is the submatrix of $\M{V}$ corresponding to the columns in the set $\bar{\mathcal{J}}_n$.
  (Although every processor owns all of $\M{V}$, the distributed TTM only needs $\M[\bar]{V}$.)
\end{itemize}
\Cref{alg:par_ttm} presents the parallel algorithm for distributed TTM.
This is the same algorithm as previously presented by Austin et al.~\cite{AuBaKo16}, but here we provide additional implementation details
and analysis.

The computation reduces to a large matrix-matrix product, i.e.,
$\Mz{Z}{n} = \M{V}\Mz{Y}{n}$. If $\Mz{Y}{n}$ is partitioned into
column blocks, the computation can be computed separately in each
block. Therefore, since each mode-$n$ processor fiber owns a separate
column block of $\Mz{Y}{n}$, they are independent.
For this reason, 
\cref{line:par_ttm:myprocfiber} defines the
local processor fiber, and all communication for that processor
occurs within the $P_n$ nodes of that fiber.
There are $\UnfoldSize[P]$ independent column fibers.

\begin{algorithm}[ht]
\caption{Parallel TTM \cite{AuBaKo16}}
\label{alg:par_ttm}
\begin{algorithmic}[1]
\Procedure{$\T[\bar]{Z}=$ par\_ttm}{$\Y*,n,\M[\bar]{V},\FullIndex[P]$}
\Comment{$\Y*$ is local portion of the tensor and $\M[\bar]{V}$ is column block of $\M{V}$}
  \State $\texttt{myProcID} \gets \FullIndex*[p]$
    \label{line:par_ttm:myprocid} 
  \State $\texttt{myProcFiber} \gets
  (\bar p_{0}, \dots, \bar p_{n-1},\, : \,, \bar p_{n+1},
  \dots, \bar p_{N-1})$
  \label{line:par_ttm:myprocfiber}
  \Comment{Processor group of size $P_n$}
  \If{$K_n \leq \lfloor J_n / P_n \rfloor $}
  \label{line:par_ttm:decision} \Comment{Reduce-scatter variant}
    \State $\T[\bar]{W} = \textproc{ttm}(\Y*,n, \M[\bar]{V})$
      \label{line:par_ttm:mono_ttm}
    \State $\T[\bar]{Z} =
    \textproc{ReduceScatter}(\T[\bar]{W},\texttt{myProcFiber})$
      \label{line:par_ttm:reducescatter}
  \Else \Comment{Multiple-reduction variant}
  \State Partition $\M[\bar]{V}$ into row blocks $\ColumnBlock[\M[\bar]{V}]{\ell}$
  of size $\hat K_{\ell} = \Lsz{K_n}{\ell} \times \hat J_n$ for $\ell \in \RangeSet{P_n}$
    \For{$\ell \in \RangeSet{P_n}$}
      \State $\T[\bar]{W} =
      \textproc{ttm}(\Y*,n,\ColumnBlock[\M[\bar]{V}]{\ell})$
        \label{line:par_ttm:blocked_ttm}
      \State $\T[\bar]{Z} =
      \textproc{Reduce}(\T[\bar]{W},\texttt{myProcFiber},\ell)$ \Comment{Root for Reduce is processor
      $(\bar p_{0}, \dots, \bar p_{n-1}, \ell , \bar p_{n+1}, \dots, \bar p_{N-1})$}
        \label{line:par_ttm:reduce}
    \EndFor
  \EndIf
\EndProcedure
\end{algorithmic}
\end{algorithm}

The algorithm chooses between two methods based on the size of $K_n$.
The decision is based on the size of the intermediate quantities that are computed.
The choice of method ensures that the temporary memory never exceeds the memory of
$\Y*$.

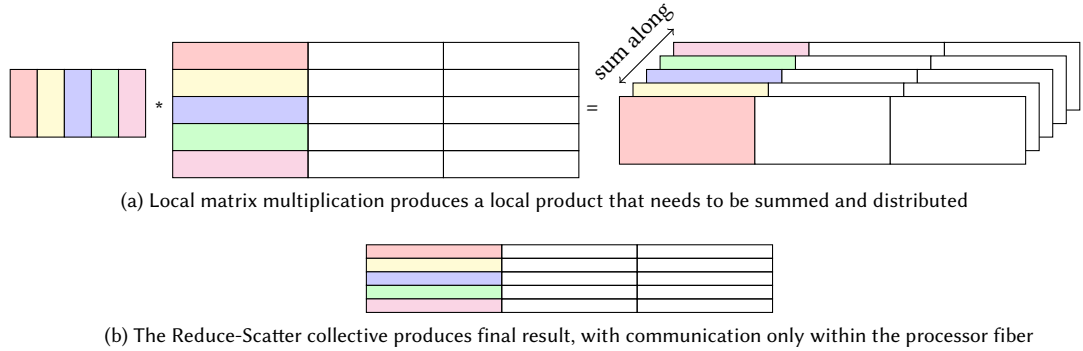
\begin{figure}[th]
\centering
\subfloat[Local matrix multiplication produces a local product that needs to be summed and distributed]{
\begin{tikzpicture}[scale=0.18,namenode/.style={scale=.75}]

\filldraw[fill=red!20] (0,3) rectangle (2,8);
\filldraw[fill=yellow!20] (2,3) rectangle (4,8);
\filldraw[fill=blue!20] (4,3) rectangle (6,8);
\filldraw[fill=green!20] (6,3) rectangle (8,8);
\filldraw[fill=magenta!20] (8,3) rectangle (10,8);

\filldraw[fill=red!20] (12,8) rectangle(22,10);
\filldraw[fill=yellow!20] (12,6) rectangle(22,8);
\filldraw[fill=blue!20] (12,4) rectangle(22,6);
\filldraw[fill=green!20] (12,2) rectangle(22,4);
\filldraw[fill=magenta!20] (12,0) rectangle(22,2);

\draw (22,0) rectangle(32,2);
\draw (22,2) rectangle(32,4);
\draw (22,4) rectangle(32,6);
\draw (22,6) rectangle(32,8);
\draw (22,8) rectangle(32,10);

\draw (32,0) rectangle(42,2);
\draw (32,2) rectangle(42,4);
\draw (32,4) rectangle(42,6);
\draw (32,6) rectangle(42,8);
\draw (32,8) rectangle(42,10);

\filldraw[fill=magenta!20] (49,5) rectangle(59,10);
\filldraw[fill=green!20] (48,4) rectangle(58,9);
\filldraw[fill=blue!20] (47,3) rectangle(57,8);
\filldraw[fill=yellow!20] (46,2) rectangle(56,7);
\filldraw[fill=red!20] (45,1) rectangle(55,6);

\filldraw[fill=white] (59,5) rectangle(69,10);
\filldraw[fill=white] (58,4) rectangle(68,9);
\filldraw[fill=white] (57,3) rectangle(67,8);
\filldraw[fill=white] (56,2) rectangle(66,7);
\filldraw[fill=white] (55,1) rectangle(65,6);

\filldraw[fill=white] (69,5) rectangle(79,10);
\filldraw[fill=white] (68,4) rectangle(78,9);
\filldraw[fill=white] (67,3) rectangle(77,8);
\filldraw[fill=white] (66,2) rectangle(76,7);
\filldraw[fill=white] (65,1) rectangle(75,6);

\path (11,5) node(x) {*};
\path (43,5) node(y) {=};
\draw[<->] (45,7) -- node[above, sloped] {sum along} (49,11);

\end{tikzpicture}

%
%
%
%
}

\subfloat[The Reduce-Scatter collective produces final result, with communication only within the processor fiber]{
  \hspace{2in}
\begin{tikzpicture}[scale=0.18,namenode/.style={scale=.75}]

\filldraw[fill=red!20] (44,7) rectangle(54,8);
\filldraw[fill=yellow!20] (44,6) rectangle(54,7);
\filldraw[fill=blue!20] (44,5) rectangle(54,6);
\filldraw[fill=green!20] (44,4) rectangle(54,5);
\filldraw[fill=magenta!20] (44,3) rectangle(54,4);

\draw (54,3) rectangle(64,4);
\draw (54,4) rectangle(64,5);
\draw (54,5) rectangle(64,6);
\draw (54,6) rectangle(64,7);
\draw (54,7) rectangle(64,8);

\draw (64,3) rectangle(74,4);
\draw (64,4) rectangle(74,5);
\draw (64,5) rectangle(74,6);
\draw (64,6) rectangle(74,7);
\draw (64,7) rectangle(74,8);

\end{tikzpicture}

%
%
%
%
  \hspace{2in}
}
\caption{Reduce-Scatter variant of TTM, used if temporary storage does not exceed original data. The data belonging to five processors (in the same processor fiber) is color-coded for tracking.}
\label{fig:ttm_reduce_scatter}
\end{figure}

In the \emph{Reduce-Scatter variant}, each processor computes the local matrix-matrix product,
$\M[\bar]{W} = \M[\bar]{V} \M[\bar]{Y}_{(n)}$, and then sums and distributes the result using a Reduce-Scatter.
The temporary object $\M[\bar]{W}$ is of size $K_n \times \UnfoldSize*$, 
which may be too large to store on the processor. We assume that the processor has enough extra memory
to store something the same size as $\M[\bar]{Y}_{(n)}$ (i.e., $\Y*$), which is of size $\bar J_n \times \UnfoldSize[\bar J]$,
and so we use this variant if $K_n \leq \lfloor J_n / P_n \rfloor \approx \bar J_n$.
This process is demonstrated in \cref{fig:ttm_reduce_scatter} where five processors are shaded (in the first processor fiber) to understand that data ownership of each processor. 

A key issue in the implementation is
that the entries' ordering in $\T[\bar]{W}$ is not correct for a 
Reduce-Scatter and so must be reorganized when the data is packed for the Reduce-Scatter
to obtain the proper layout of $\T[\bar]{Z}$ at
the end of the call.
The input buffer for Reduce-Scatter must be arranged so that contributions to each processor's result $\T[\bar]{Z}$ are contiguous.
The ordering of $\T[\bar]{W}$ consists of $\RightSize*$ row-major submatrices of dimension $K \times \LeftSize*$, and it must be reordered into $\SingleSize[P]$ contiguous subblocks, each consisting of $\RightSize*$ row-major submatrices of dimension $K/\SingleSize[P] \times \LeftSize*$ (assuming $\SingleSize[P]$ divides $K$ evenly).
If $n=N-1$ (the last dimension), no reordering is necessary, and no unpacking is necessary after the Reduce-Scatter for any dimension.
 
\begin{figure}[th]
\centering
\begin{tikzpicture}[scale=0.13,namenode/.style={scale=.75}]

\filldraw[fill=red!5, dotted] (0,3) rectangle (2,8);
\filldraw[fill=yellow!5, dotted] (2,3) rectangle (4,8);
\filldraw[fill=blue!5, dotted] (4,3) rectangle (6,8);
\filldraw[fill=green!5, dotted] (6,3) rectangle (8,8);
\filldraw[fill=magenta!5, dotted] (8,3) rectangle (10,8);

\filldraw[fill=red!20] (0,4) rectangle (2,5);
\filldraw[fill=yellow!20] (2,4) rectangle (4,5);
\filldraw[fill=blue!20] (4,4) rectangle (6,5);
\filldraw[fill=green!20] (6,4) rectangle (8,5);
\filldraw[fill=magenta!20] (8,4) rectangle (10,5);
\draw[thick] (0,4) rectangle(10,5);

\filldraw[fill=red!20] (12,8) rectangle(22,10);
\filldraw[fill=yellow!20] (12,6) rectangle(22,8);
\filldraw[fill=blue!20] (12,4) rectangle(22,6);
\filldraw[fill=green!20] (12,2) rectangle(22,4);
\filldraw[fill=magenta!20] (12,0) rectangle(22,2);

\draw (22,0) rectangle(32,2);
\draw (22,2) rectangle(32,4);
\draw (22,4) rectangle(32,6);
\draw (22,6) rectangle(32,8);
\draw (22,8) rectangle(32,10);

\draw (32,0) rectangle(42,2);
\draw (32,2) rectangle(42,4);
\draw (32,4) rectangle(42,6);
\draw (32,6) rectangle(42,8);
\draw (32,8) rectangle(42,10);

\filldraw[fill=magenta!20] (44,1) rectangle(54,2);
\filldraw[fill=green!20] (44,3) rectangle(54,4);
\filldraw[fill=blue!20] (44,5) rectangle(54,6);
\filldraw[fill=yellow!20] (44,7) rectangle(54,8);
\filldraw[fill=red!20] (44,9) rectangle(54,10);

\filldraw[fill=white] (54,1) rectangle(64,2);
\filldraw[fill=white] (54,3) rectangle(64,4);
\filldraw[fill=white] (54,5) rectangle(64,6);
\filldraw[fill=white] (54,7) rectangle(64,8);
\filldraw[fill=white] (54,9) rectangle(64,10);

\filldraw[fill=white] (64,1) rectangle(74,2);
\filldraw[fill=white] (64,3) rectangle(74,4);
\filldraw[fill=white] (64,5) rectangle(74,6);
\filldraw[fill=white] (64,7) rectangle(74,8);
\filldraw[fill=white] (64,9) rectangle(74,10);

\filldraw[fill=red!5, dotted] (76,7) rectangle(86,8);
\filldraw[fill=yellow!5, dotted] (76,6) rectangle(86,7);
\filldraw[fill=blue!5, dotted] (76,5) rectangle(86,6);
\filldraw[fill=green!20] (76,4) rectangle(86,5);
\filldraw[fill=magenta!5, dotted] (76,3) rectangle(86,4);

\draw[dotted] (86,7) rectangle(96,8);
\draw[dotted] (86,6) rectangle(96,7);
\draw[dotted] (86,5) rectangle(96,6);
\draw[dotted] (86,4) rectangle(96,5);
\draw[dotted] (86,3) rectangle(96,4);

\draw[dotted] (96,7) rectangle(106,8);
\draw[dotted] (96,6) rectangle(106,7);
\draw[dotted] (96,5) rectangle(106,6);
\draw[dotted] (96,4) rectangle(106,5);
\draw[dotted] (96,3) rectangle(106,4);

\draw[thick] (76,4) rectangle(106,5);

\path (11,5) node {*};
\path (43,5) node {=};
\path (59,2.5) node {+};
\path (59,4.5) node {+};
\path (59,6.5) node {+};
\path (59,8.5) node {+};
\path (75,5) node {=};

\end{tikzpicture}

%
%
%
%
\caption{Multiple-reduction variant of TTM, used to obtain a smaller memory footprint. The data belonging to five processors (in the same processor fiber) is color-coded for tracking.}
\label{fig:ttm_reduce}
\end{figure}
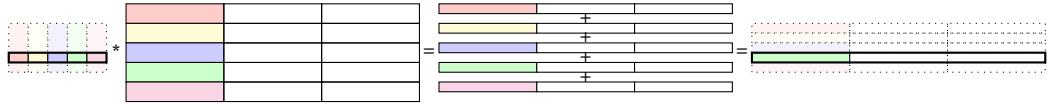

In the \emph{multiple-reduction variant}, the algorithm uses a
blocked approach that involves $P_n$ local TTMs and
$P_n$ collective communications.
Here, each iteration computes the contribution to one processor's output (within the
processor fiber --- all processor fibers are working concurrently)
and uses a Reduce collective to compute the
sum across all processors in the fiber and store the result on the $\ell$th
processor.
The matrix $\M[\bar]{V}$ is divided into block rows so that $\M[\bar]{V}[\ell]$ owns
the rows in the set $\PrcMap{\ell, K_n, P_n}$.
In this case, the dimensions of the temporary tensor $\T[\bar]{W}$ in the $\ell$th iteration
is bounded above by
$\bar J_{0} \times \cdots \times \lceil K_n / P_n \rceil
\times \cdots \times \bar J_{N-1}$, which is essentially the
same size as $\T[\bar]{Z}$
(recall that $\hat K_{\ell}$ is within 1 of $\bar K_n$).
This process is demonstrated in \cref{fig:ttm_reduce}, where we highlight
the contribution to the 4th processor in the 1st column fiber.

The two variants perform the same number of flops and communicate almost the same
amount of data (to within a factor of 2).
The number of flops is $\bigO(\LinearSize[J] K_n / \LinearSize[P])$, which is the cost of the local TTM(s).

The data communication costs differ by only a factor of 2.
In the Reduce-Scatter variant, the amount of data communicated is the cost of one Reduce-Scatter collective over the mode-$n$ fiber: %
$\prod_{m\neq n} \lceil J_m / P_m \rceil  \lceil K_n / P_n \rceil (P_n-1) \approx
\UnfoldSize[J] K_n (P_{n}-1) / \LinearSize[P]$.
In the multiple-reduction variant, there are $P_n$ Reduce collectives of a data size that is smaller by a factor of $P_n$, which yields the same asymptotic cost, but incurs an extra factor of 2 due to the cost of the Reduce collective.

The number of messages is fewer for the Reduce-Scatter variant, which has 
only one collective, at a cost of $\log P_n$ messages.
The multiple-reduction variant involves $P_n$
collectives, at a total cost of $2P_n \log P_n$ messages.

The temporary memory ($\T[\bar]{W}$) of the multiple-reduction variant is much lower:
$\lceil K_n/P_n \rceil \UnfoldSize* $ words
for the multiple-reduction variant
versus
$K_n\UnfoldSize*$ words 
for the  Reduce-Scatter variant.
As an aside, we note the block size of the blocked algorithm can be chosen arbitrarily,
navigating a tradeoff between latency cost and memory footprint; we used a version
that corresponded to the result size 
for simplicity, but it assumes the available remaining memory
is at least the size of the required memory for the problem.

\section{Parallel ST-HOSVD Cost Analysis}
\label{sec:distributed-st-hosvd}

\begin{table}
\begin{tabular}{|c|cccc|}
\hline
 & \textbf{Flops} & \textbf{Bandwidth} & \textbf{Latency} & \textbf{Memory} \\
\hline
\begin{tabular}{c} TTM \\ (Reduce-Scatter) \end{tabular} &
$\frac{\LinearSize[J]}{\LinearSize[P]}K$ &
$\frac{\UnfoldSize[J]}{\UnfoldSize[P]}K$
& $\log P_n$ &
$\frac{\UnfoldSize[J]}{\UnfoldSize[P]}K$ \\
\begin{tabular}{c} TTM \\ (Reduce) \end{tabular} &
$\frac{\LinearSize[J]}{\LinearSize[P]}K$ &
$\frac{\UnfoldSize[J]}{\UnfoldSize[P]}K$ &
$P_n \log P_n$ &
$\frac{\UnfoldSize[J]}{\LinearSize[P]}K$  \\
\hline
\begin{tabular}{c} Gram \\ (Original) \end{tabular} &
$\frac{\LinearSize[J]}{\LinearSize[P]}J_n$ &
$\frac{\LinearSize[J]}{\LinearSize[P]}P_n +
\frac{J_n^2}{P_n}$ & $P_n$ &
$\frac{\LinearSize[J]}{\LinearSize[P]} +
\lt(\frac{J_n}{P_n}\rt)^2$ \\
\begin{tabular}{c} Gram \\ (New) \end{tabular} &
$\frac{\LinearSize[J]}{\LinearSize[P]}J_n$ &
$\frac{\LinearSize[J]}{\LinearSize[P]} + J_n^2$ & $P_n$ &
$\frac{\LinearSize[J]}{\LinearSize[P]} + J_n^2$ \\
\hline
\end{tabular}
\caption{Asymptotic costs of algorithms for TTM and Gram with respect to mode
$n$ for a tensor with global dimensions $\FullSize[J]$ and processor grid with
dimensions $\FullSize[P]$.
We omit leading constant factors and lower order terms.  
The columns correspond to per-processor costs: number of floating point
operations, number of words communicated, number of messages communicated, and
amount of temporary local memory required, respectively.
Recall $\LinearSize[J] = \prod_{n \in \RangeSet{N}} J_n$ and $\UnfoldSize[J] = \LinearSize[J]/J_n$,
with analogous definitions for $\LinearSize[P]$ and $\UnfoldSize[P]$.
\label{tbl:costs}}
\end{table}

In this section we analyze the computation, communication, and temporary memory requirements of the ST-HOSVD algorithm.
We note that the computation and bandwidth costs are sensitive to mode order, while the latency cost and memory requirements are not.
This analysis assumes that the mode order used by the algorithm is increasing by mode index; the costs for other mode orders can be derived by relabeling modes.
If the core ranks are specified a priori, then an optimal (in terms of flops or communication) mode ordering can be determined similar to the case of reconstruction, as described in \Cref{sec:reconstruction}. 
We also note that the communication costs are sensitive to the processor grid, but the computation cost and memory requirements are not.
In this analysis, we use the new Gram algorithm and allow for the choice of TTM algorithm based on the relative sizes of dimensions as described in \cref{sec:par_ttm}.
To simplify the analysis, we provide an upper bound on the communication costs by assuming the  Reduce version, which sends more messages, is used for each mode.

The $n$th mode Gram performs $\LeftSize[R]I_n^2\RightSize[I]/\LinearSize[P]$ flops (exploiting the symmetry of the output), the $n$th mode eigenvalue computation requires $(10/3)I_n^3$ flops, and the $n$th mode TTM performs $2\LeftSize[R]R_nI_n\RightSize[I]/\LinearSize[P]$ flops.
Thus, the leading order terms in the flop costs are
$$\gamma \cdot  \lt( \frac{1}{\LinearSize[P]} \sum_{n=0}^{N-1} \lt(I_n^2 + 2 R_nI_n\rt) \LeftSize[R] \RightSize[I] + \frac{10}{3} \sum_{n=0}^{N-1}I_n^3 \rt).$$

The communication cost of the $n$th mode (new) Gram computation is $\beta \cdot (\LeftSize[R]I_n\RightSize[I]/\LinearSize[P] \cdot ((P_n {-} 1)/P_n) + I_n^2) + \alpha \cdot (P_n {-} 1 + 2\log \LinearSize[P])$.
The communication cost of the $n$th mode (Reduce) TTM is $\beta \cdot (\LeftSize[R]R_n\RightSize[I]/\UnfoldSize[P] \cdot ((P_n {-} 1)/P_n)) + \alpha \cdot (2P_n \log P_n)$.
Thus, the leading order terms in the bandwidth costs, assuming we use the new Gram algorithm, are
$$\beta \cdot  \lt( \frac{1}{\LinearSize[P]} \sum_{n=0}^{N-1} \lt(I_n \frac{P_n {-} 1}{P_n} + 2R_n(P_n {-} 1)\rt) \LeftSize[R] \RightSize[I] + \sum_{n=0}^{N-1} I_n^2 \rt).$$
The leading order terms in the latency costs, conservatively assuming we use the Reduce TTM algorithm at each step, are
$$\alpha \cdot  \lt( 2N \log \LinearSize[P] + \sum_{n=0}^{N-1} 2P_n \log P_n \rt).$$

The temporary memory required for the $n$th Gram computation is twice the size of the current local tensor data, which is $\LeftSize[R] I_n \RightSize[I] / \LinearSize[P]$.
Because this memory can be re-used and $R_n\leq I_n$ for each $n$, this cost is dominated by the first mode, requiring $2\LinearSize[I]/\LinearSize[P]$ words.
The Gram computation also requires space for storing the output Gram matrix, which is of size $I_n^2/2$ and can be re-used across modes.
The eigenvalue computation requires as much as $I_n^2$ extra memory if all eigenvectors are computed.
The temporary memory required for TTM is guaranteed to be smaller than the input tensor, which is always bigger than the output tensor in the case of ST-HOSVD, so the temporary memory required for the $n$th TTM never exceeds the size required of the $n$th Gram computation.
Thus, the leading order terms of the total temporary memory required on each processor is 
$$ 2 \max\left\{ \frac{\LinearSize[I]}{\LinearSize[P]}, \max_n I_n^2\right\}.$$

%
%
%
%
%

\section{Optimized reconstruction}
\label{sec:reconstruction}

After the data has been compressed using ST-HOSVD, the user may wish to move the
compressed data to another machine and reconstruct an approximation of the
original data there.  Since the full reconstructed data set would take up just
as much space as the original, we provide the user an option to reconstruct
sub-tensors of the original tensor.  
Thanks to the structure of the Tucker model, linear operations can be applied cheaply to each mode.
For example, the user may want to
downsample one of the spatial dimensions (mode 2) and select a single variable (mode 3) at a
single timestep (mode 4) for the purposes of visualization.  This can be accomplished
through the following series of TTMs.

\begin{equation*}
\T{Z} = \G \times_0 \FacMat{0} \times_1 \FacMat{1} \times_2
{\FacMat[C]{2}}^T
\FacMat{2} \times_3 {\FacMat[C]{3}}^T \FacMat{3} \times_4 {\FacMat[C]{4}}^T
\FacMat{4}
\end{equation*}
where
\begin{equation*}
\FacMat[C]{2} = \left[\begin{array}{cccc}
.5 & 0 & \cdots & 0 \\
.5 & 0 & \cdots & 0 \\
0 & .5 & \cdots & 0 \\
0 & .5 & \cdots & 0 \\
\vdots & \vdots & \ddots & \vdots
\end{array}\right], \qquad
\FacMat[C]{3} = \left[\begin{array}{c}
0 \\ \vdots \\ 1 \\ 0 \\ \vdots
\end{array}
\right], \qquad
\FacMat[C]{4} = \left[\begin{array}{c}
0 \\ \vdots \\ 1 \\ 0 \\ \vdots
\end{array}\right].
\end{equation*}

The five TTMs can be done in any order to obtain the result, but the \mTTM\ ordering can have a large effect on computational cost, memory footprint, and communication cost (in the parallel case).
We demonstrate the effects of reconstruction mode ordering on run time and memory in \cref{sec:exp-rec}.
In the code, the user can specify the mode ordering or allow the software to automatically select the ordering that minimizes either computational cost or memory footprint.
In the case of parallel reconstruction with linear operations applied to the factor matrices, each processor redundantly stores each $\FacMat[C]{n}$ and computes ${\FacMat[C]{n}}^T\FacMat[U]{n}$ locally.

To determine the optimal mode ordering, the TuckerMPI code exhaustively searches over all $N!$ permutations to obtain the optimal ordering, but the optimal ordering can be determined in $O(N\log N)$ time by sorting with a specific comparator \cite{Chakravarthy17}.
For example, consider the $N=2$ case, which corresponds to a product of 3 matrices, with input core matrix dimensions $R_0\times R_1$ and output subtensor dimensions $K_0\times K_1$.
Computing the mode-0 product followed by the mode-1 product requires $K_0R_0R_1+K_0R_1K_1$ scalar multiplications, while computing the products in the opposite order requires $R_0R_1K_1+R_0K_0K_1$ scalar multiplications, so to minimize flops we order the products based on the comparison of these two costs.

More generally, let the input core tensor $\G$ have dimensions $\FullSize[R]$ and the output subtensor $\T{Z}$ have dimensions $\FullSize[K]$.
The key insight is that optimal ordering of all $N$ modes is the one such that every pair of modes is ordered correctly according to the $N=2$ case.
That is, to minimize flops, mode $i$ should precede mode $j$ in the ordering if $K_iR_iR_j+K_iR_jK_j < R_iR_jK_j+R_iK_iK_j$.
This can be shown, as argued by Chakravarthy \cite{Chakravarthy17}, by recognizing that this comparator yields a total ordering on modes and arguing by contradiction.
Suppose the optimal ordering is not sorted by this comparator, then there exists two consecutive modes $n$ and $n+1$ that are out of order.
Swapping the two consecutive modes will reduce the overall cost because the costs of the first $n-1$ TTMs and the last $N-n-1$ TTMs are equivalent, but the cost of the $n$ and $n+1$ TTMs are reduced by the property of the comparator, and we have a contradiction.

Similar arguments can be made for communication (bandwidth) cost and memory footprint.
Given the bandwidth cost of \Cref{alg:par_ttm}, to minimize words moved, mode $i$ should precede mode $j$ if $R_jK_i(P_i-1)+K_iK_j(P_j-1) < R_iK_j(P_j-1)+K_iK_j(P_i-1)$.
To minimize temporary memory, mode $i$ should precede mode $j$ if $R_i/K_i < R_j/K_j$.

%
%
%
%
%

\section{Experimental Results}
\label{sec:experiments}

\subsection{Experimental Platform}
\label{sec:exper-platf}
We run all experiments on Skybridge, a Sandia supercomputer consisting of
1,848 dual-socket 8-core Intel Sandy Bridge (2.6 GHz) compute nodes.  Each node
has 64 GB of memory, a peak flop rate of 332.8 GFLOPS (i.e., 20.8 GFLOPS per core), and the nodes are connected by an Infiniband interconnect.
We use Intel compilers and the MKL for BLAS and LAPACK subroutines. We
execute 16 MPI processes per node with 1 thread per process unless
otherwise stated.
Data files are stored on a Lustre file system, and Skybridge's I/O is shared with other clusters.
All reported timings in this section are averages over multiple runs.
All reported memory requirements are computed by the application, using a wrapper around memory allocation calls, and does not include memory used by the operating system or by the MPI implementation.

\subsection{Data Description}
\label{sec:data_description}

We consider two large-scale simulation datasets which were produced by S3D
\cite{CC+09}, a massively parallel direct numerical simulation of
compressible reacting flows, developed at Sandia National
Laboratories.  The datasets are as follows:
\begin{itemize}
\item {\bfseries SP:} This 5-way data tensor is of size $500 \times 500 \times
500 \times 11 \times 400$ and corresponds to a cubic $500 \times 500 \times 500$
spatial grid for 11 variables over 400 time steps.  Each time step requires 11
GB, so the entire dataset is 4.4 TB.  The SP dataset is from the simulation of a
3D statistically steady planar turbulent premixed flame of
methane-air combustion \cite{KZCS16}.  
The first 50 time steps (a 550 GB dataset) was used in previous work \cite{AuBaKo16}.
\item {\bfseries JICF:} This 5-way data tensor of size $1500 \times 2080 \times
1500 \times 18 \times 10$ comes from a $1500 \times 2080 \times 1500$ spatial
grid with 18 variables over 10 time steps.  Each time step requires 674 GB
storage, so the entire dataset is 6.7 TB. 
The JICF dataset is from a jet in crossflow simulation, which is a canonical configuration for many combustion systems \cite{LW+15}.
\end{itemize}

The general experimental setup is described in \cref{tab:setup}.
For each dataset, we use the same number of nodes for all experiments:  250 for SP
and 350 JICF.
Since each node has 64~GB of RAM and runs 16 threads/processes, storage of the full tensor requires
a little more than $1/4$ of the memory per node (our data is stored in double precision).
For each dataset, we consider three different processor grid configurations (A/B/C), as specified in~\cref{tab:partitions}.
They vary primarily in how the processors are distributed in the first three modes with scenario A having fewer processors in mode 0, scenario B being more evenly divided on modes 0--2, and scenario C having more processors on mode 0.

\begin{table}
  \centering
  \caption{Experimental setup}\label{tab:setup}%
  \subfloat[Data sets to be compressed and the number of parallel nodes to be used in the compression experiments.]{%
    \label{tab:datasets}%
    \begin{tabular}{|c|c|c|c|c|c|c|}
      \hline
      \bf Dataset &  \bf Overall & \bf Total & \multicolumn{2}{c|}{\bf Number of} & \multicolumn{2}{c|}{\bf Storage per} \\ 
      \bf Name & \bf Tensor Size & \bf Storage & \bf  Nodes & \bf Processes & \bf Node & \bf Process \\ \hline
      SP & \spsize & 4.4~TB & 250 & 4000 & 17.6~GB & 1.1~GB \\ \hline
      JICF & \jisize & 6.7~TB & 350 & 5600 & 19.3~GB & 1.2~GB \\ \hline
    \end{tabular}}\\
  \subfloat[Three different processor grid configurations per dataset, to test the efficiency of the compression. The local tensor size may vary, but here we list
  the largest local size in each dimension.]{%
  \label{tab:partitions}%
  \begin{tabular}{|c|c|c|c|}    \hline
    \bf Dataset & \bf Proc.~Config.~Name & \bf Processor Grid & \bf Local Tensor Size \\ \hline
    \multirow{3}{*}{SP} & A & \spgrida & $500 \times 500 \times 13 \times 11 \times 4$ \\ 
    & B & \spgridb & $50 \times 63 \times 100 \times 11 \times 40$ \\ 
    & C & \spgridc & $13 \times 50 \times 500 \times 11 \times 40$ \\ \hline
    \multirow{3}{*}{JICF}  & A & \jigrida & $1500 \times 130 \times 43 \times 18 \times 1$ \\ 
                & B & \jigridb & $150 \times 260 \times 215 \times 18 \times 1$ \\ 
                & C & \jigridc & $43 \times 130 \times 1500 \times 18 \times 1$ \\ \hline
  \end{tabular}}
\end{table}

For testing and debugging, TuckerMPI also provides both sequential and parallel synthetic tensor generators for the users'
convenience.
The user specifies the desired size and rank of a tensor, denoted
by $\FullIndex[I]$ and $\FullIndex[R]$ respectively, and the amount of relative noise to be
added, denoted $\eta$. 
We construct a core tensor $\G$ with dimensions defined by
$\FullIndex[R]$; the numbers are drawn from a standard normal distribution.  The factor matrices
$\FacMat{0} \cdots \FacMat{N-1}$ are generated in a similar fashion, where
$\FacMat$ is $\FacSize$.  By performing the tensor
times matrix mulitplications $\G \times_0 \FacMat{0} \times_1 \FacMat{1}
\cdots \times_{N-1} \FacMat{N-1}$, we obtain a tensor of size $\FullIndex[I]$ with
rank $\FullIndex[R]$, which we refer to as $\T{M}$.

After obtaining the tensor $\T{M}$, we wish to add noise to it so that it
is not exactly rank $\M{R}$.  Let $\X = \T{M} + \eta
\frac{\left\|\T{M}\right\|}{\left\|\T{N}\right\|} \T{N}$, where
$\T{N}$ is a randomly generated tensor of noise whose values are also
obtained from a standard normal distribution.  This gives us
$\frac{\left\|\X-\T{M}\right\|}{\left\|\X\right\|} \approx
\eta$.
Note that we do not explicitly construct and store $\T{N}$ due to its large
size, so we approximate its norm using its expected value: $\left\|\T{N}\right\| \approx \sqrt{\LinearSize[I]}$.

\subsection{Comparison of Gram Algorithms}
\label{sec:comp-gram-algor}

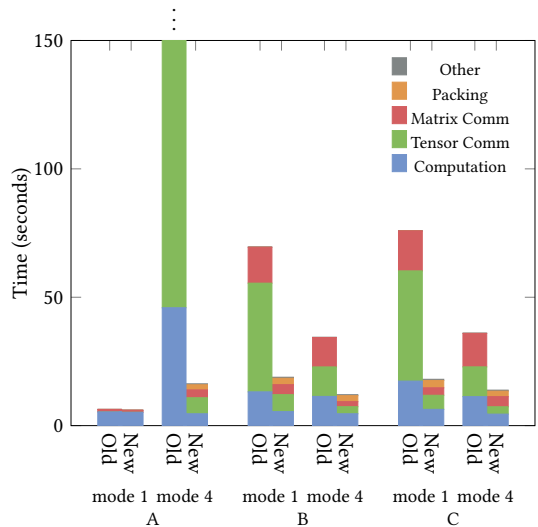
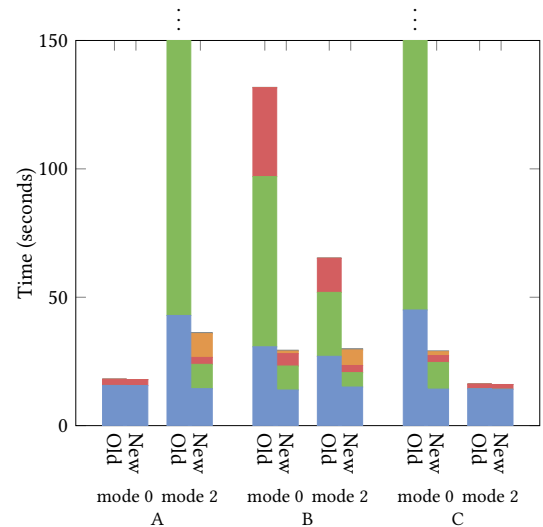
\begin{figure}
  \centering
  \subfloat[Total times and memory usage per process. ]{\label{tab:gram_compare}%
  \begin{tabular}{|c|c|c|*{3}{c|}*{3}{c|}} \hline
    & \bf Gram & \bf Proc. & \multicolumn{3}{c|}{\bf Run Times (sec)} & \multicolumn{3}{c|}{\bf Memory Usage (GB)} \\ 
    \bf Dataset & \bf Mode & \bf Config. & \bf Old & \bf New & \bf Ratio & \bf Old & \bf New & \bf Ratio \\ \hline
    \multirow{6}{*}{SP} & \multirow{3}{*}{1} & A & 6.3 & 6.0 & 1.0 & 1.15 & 1.15 & 1.0 \\ %
    && B & 69.6 & \bf 18.9 & 3.7 & 2.22 & 3.32 & 1.5 \\ 
    && C & 75.9 & \bf 18.1 & 4.2 & 2.29 & 3.34 & 1.5 \\ \cline{2-9}
    &\multirow{3}{*}{4} & A & 788.9 & \bf 16.3 & 48.4 & 2.29 & 3.43 & 1.5 \\ 
    && B & 34.4 & \bf 12.1 & 2.8 & 2.22 & 3.33 & 1.5 \\ 
    && C & 36.0 & \bf 13.8 & 2.6 & 2.29 & 3.43 & 1.5 \\ \hline
    \multirow{6}{*}{JICF} & \multirow{3}{*}{0} & A & 18.1 & 18.0 & 1.0 & 1.26 & 1.24 & 1.0 \\ 
    && B & 131.7 & \bf 29.4 & 4.5 & 2.44 & 3.62 & 1.5 \\ 
    && C & 434.2 & \bf 29.1 & 14.9 & 2.43 & 3.61 & 1.5 \\ \cline{2-9}
    & \multirow{3}{*}{2} & A & 390.5 & \bf 36.2 & 10.8 & 2.43 & 3.62 & 1.5 \\ 
    && B & 65.3 & \bf 30.0 & 2.2 & 2.44 & 3.62 & 1.5 \\ 
    && C & 16.2 & 16.0 & 1.0 & 1.26 & 1.24 & 1.0 \\ \hline
  \end{tabular}}\\
\subfloat[Breakdown for SP dataset]{
\centering
\begin{tikzpicture}[scale=.9]
\renewcommand{\datafile}{data/gram-sp.dat}
\renewcommand{\pga}{1x1x40x1x100}
\renewcommand{\pgb}{10x8x5x1x10}
\renewcommand{\pgc}{40x10x1x1x10}
\renewcommand{\pgalabel}{A}
\renewcommand{\pgblabel}{B}
\renewcommand{\pgclabel}{C}
\renewcommand{\ma}{1}
\renewcommand{\mb}{4}
\legendtrue
\ylabeltrue
\makegramplot
\end{tikzpicture}
\label{fig:gram-sp}
}\hfill
\subfloat[Breakdown JICF dataset]{
\centering
\begin{tikzpicture}[scale=.9]
\renewcommand{\datafile}{data/gram-jicf.dat}
\renewcommand{\pga}{1x16x35x1x10}
\renewcommand{\pgb}{10x8x7x1x10}
\renewcommand{\pgc}{35x16x1x1x10}
\renewcommand{\pgalabel}{A}
\renewcommand{\pgblabel}{B}
\renewcommand{\pgclabel}{C}
\renewcommand{\ma}{0}
\renewcommand{\mb}{2}
\legendfalse
\ylabeltrue
\makegramplot
\end{tikzpicture}
\label{fig:gram-jicf}
}
\caption{Gram total run time and breakdowns for old (round robin variant from \cite{AuBaKo16}) and new (redistribution variant) Gram algorithms on two datasets with different processor configurations as detailed in \cref{tab:partitions}. %
  The ``Gram Mode'' or ``Mode'' refers to the unfolding mode, i.e., $n$.  In the table, differences of more that $2\times$ between the old and new are highlighted in boldface.
  In the breakdowns, ``Packing'' only occurs for the new method and refers to the reordering of the data in memory before communication with the other processors, ``Tensor Comm'' refers to communication of the tensor data which happens every step of the round robin procedure for the old method and only once in the redistribution for the new method, and  ``Matrix Comm'' refers to the all-reduce which happens after each step in the round robin procedure for the old method and just once in the new method.
Note that some bars go past the y-axis limit as indicated by vertical dots.}
\label{fig:gram-perf}
\end{figure}

We compare the two versions of the Gram algorithms described in \cref{sec:par_gram}: the old round robin variant from \cite{AuBaKo16} and the new redistribution variant from \cref{alg:par_gram}.
We use the two datasets and pick two representative modes, with corresponding experimental conditions detailed in \cref{tab:setup}.
The results are shown in \Cref{tab:gram_compare}.
The new algorithm is up to 48 times faster than the old algorithm and never slower. The least speedup occurs when there is only one processor in the Gram mode fiber and no communication is performed in either case. The most speedup is when there are 100 processors in each fiber of the Gram mode.
Ignoring the cost of the All-Reduce, the communication cost ratio between the old and new algorithms for a Gram operation in mode $n$ is $P_n$, which is an upper bound on the possible speedup.
We see speedups of around 20-50\% of that upper bound because of time that both algorithms spend on computation and the increased cost of the All-Reduce for the new algorithm. With respect to processor configurations, there is much less variance in the new algorithm than in the old. The timing for the new algorithm varies by no more than $3\times$, whereas the timing for the old algorithm varies between 12--62$\times$ depending on the processor configuration. 

A breakdown of the run time for each of the experiments is shown in \Cref{fig:gram-perf}. 
We see that the main speedup comes from a reduction in communication of the tensor, but we also see a reduction in computation time.
This is the result of the new algorithm making a single BLAS call rather than the old algorithm's $P_n$ BLAS calls on smaller subproblems.
The new algorithm does include some overhead for packing and unpacking the data, but it is negligible compared to the benefits of reduced communication.

\Cref{tab:gram_compare} also reports the per-process memory requirement of each algorithm.
 Because the new algorithm requires re-packing the data and performing an All-to-All collective, it requires space for 2 extra copies of the local tensor data.
 The old algorithm requires temporary space for only one extra copy of the local tensor data, to perform its round robin exchange of data.
 The ratio of three total copies to two total copies yields the memory footprint ratio of 1.5 as reported in the table.
 When the number of processors in the mode of the Gram computation is one, no communication is necessary and no extra memory is required, so the memory footprint ratio is 1 in those cases.

\subsection{ST-HOSVD}
\label{sec:st-hosvd-1}
We analyze the parallel ST-HOSVD (\cref{alg:sthosvd}) using the new redistribution version of the Gram algorithm.
First, we consider its performance on two real-world datasets, varying the processor grid but not the number of processors.
Second, we do strong and weak scaling studies varying the number of processors.

\subsubsection{Compression of combustion data}
\label{sec:compr-comb-data}

\begin{figure}
\centering
\subfloat[Original]{\label{fig:iso_orig}
\includegraphics[scale=.125]{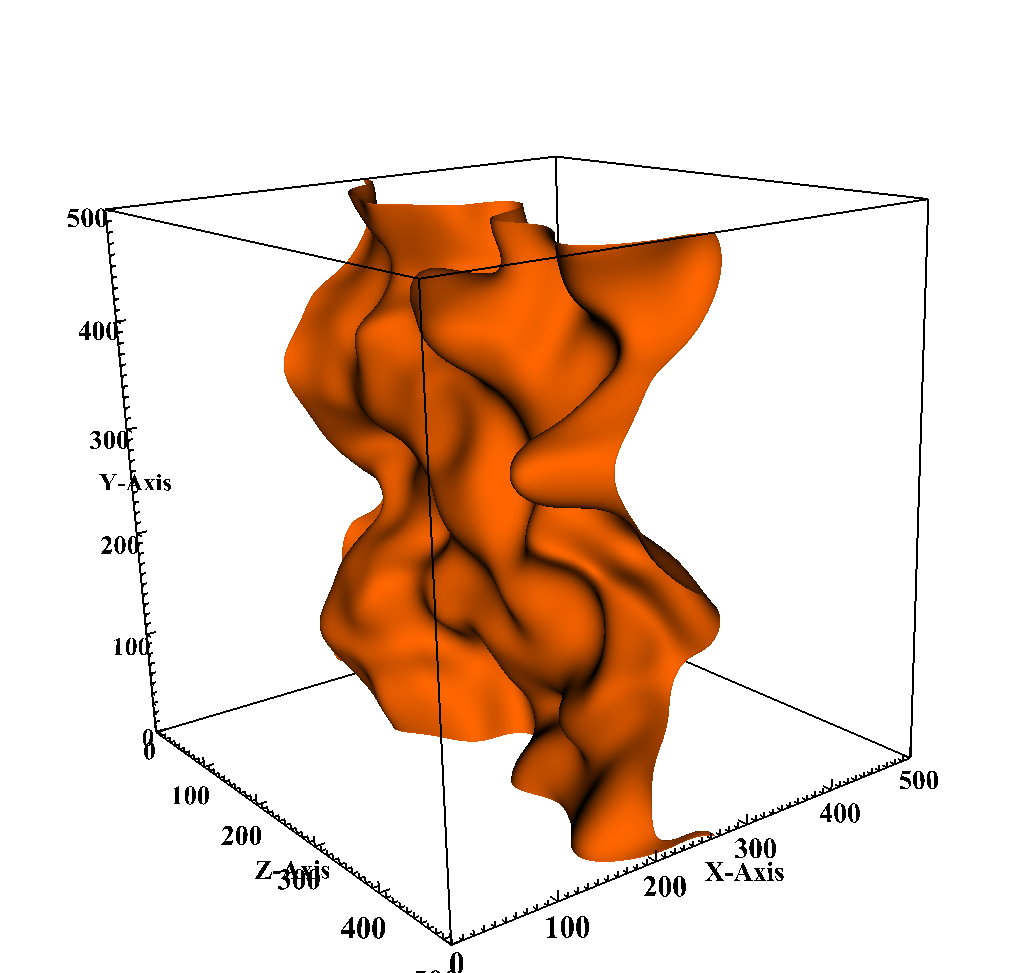}
}
\hfil
\subfloat[Low compression ($\epsilon$=1e-4)]{\label{fig:iso_low}
\includegraphics[scale=.125]{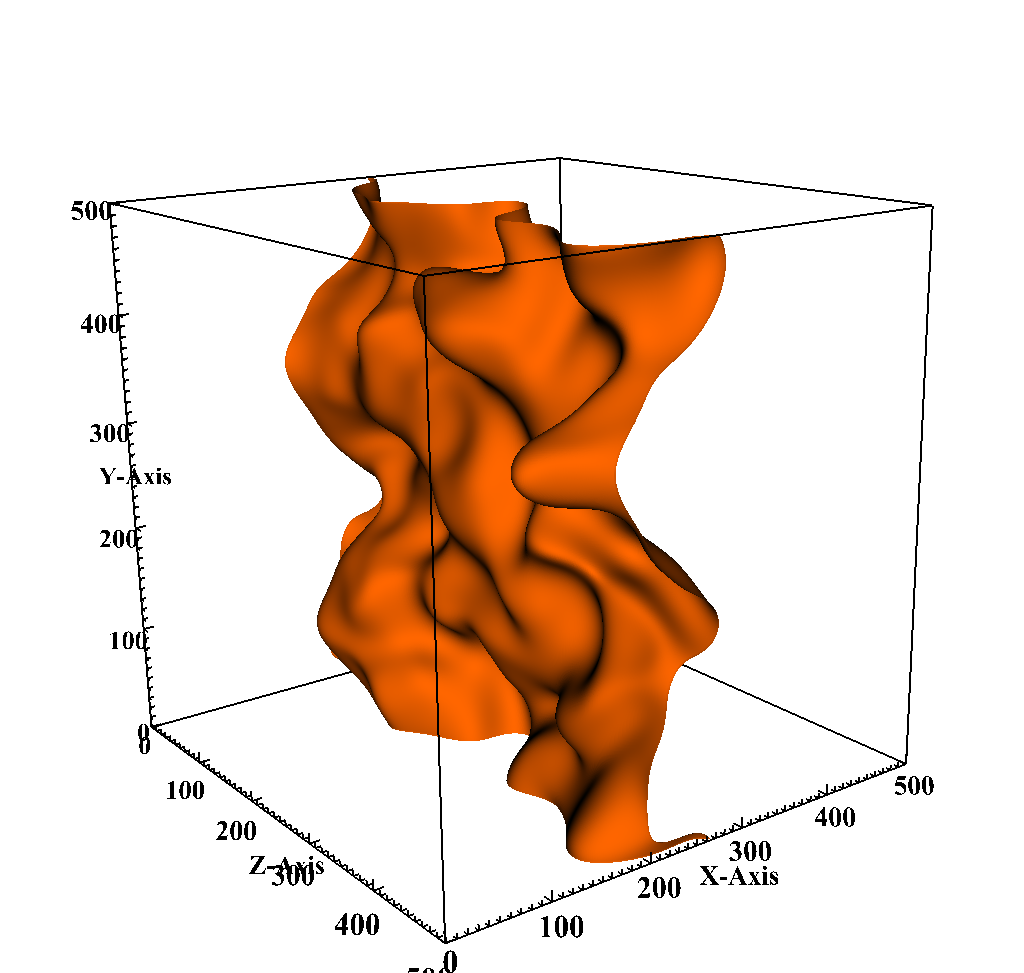}
}
\hfil
\subfloat[High compression ($\epsilon$=1e-2)]{\label{fig:iso_high}
\includegraphics[scale=.125]{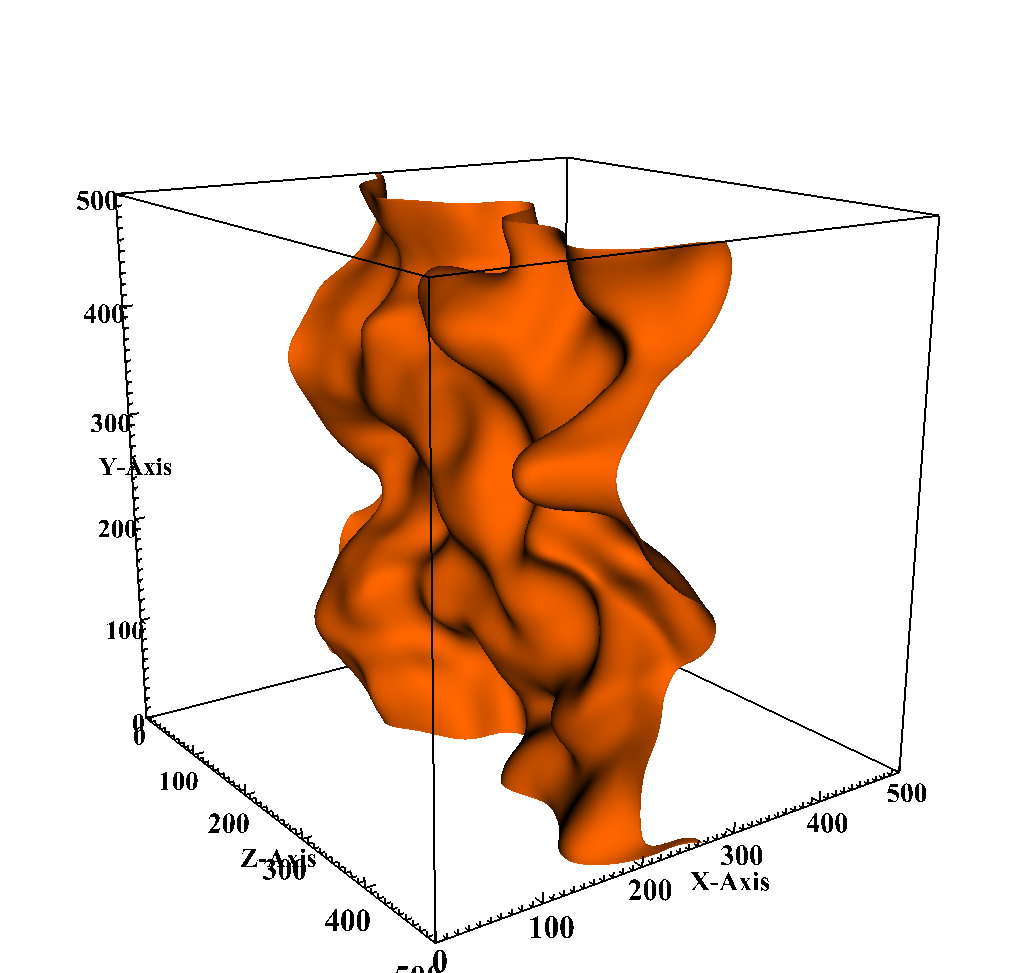}
}
\caption{Temperature isosurfaces at a given timestep in SP data computed from the original and compressed (and reconstructed) data sets.}
\label{fig:isosurface}
\end{figure}

We use TuckerMPI to compress the SP and JICF datasets described in \cref{tab:datasets}
using the number of processors and grids specified in \cref{tab:partitions}.
In terms of the pre-processing described in \cref{sec:pre-post-processing},
there was no preprocessing on the SP data since it was already scaled, and max scaling was applied to the JICF data.  
For each dataset, we consider two relative error tolerances: 1e-2 and 1e-4, which we
refer to  as ``High'' and ``Low'' compression, respectively.
The scenarios are summarized in \cref{tab:scenarios} along with the
compressed core size, total storage for the core and factor matrices,
and overall compression ratio.
For real-world datasets such as these, the compression potential depends on the amount of redundancy that
is inherent in the data. For time-evolving simulations, there may be spatial regions with minimal change, and so these parts
can be highly compressed. In our code, the user specifies the desired relative error tolerance ($\epsilon$),
from which the level of compression is determined on the fly.
(Alternatively, the software allows the user to specify the desired final core size, from which the final relative error is determined.)
For our two datasets, 
the high-compression scenario yields reductions in size of 4--5 orders of magnitude.
We note that our subject-matter experts have deemed the high compression datasets to be faithful representations
that are scientifically useful.
For instance, \Cref{fig:isosurface} shows a visual comparison between the original and compressed versions. Specifically, \cref{fig:isosurface} shows a temperature isosurface at the 201st (middle) time point, used to track the boundaries of a flame during the simulation.
The errors in the reconstructions of the compressed versions are unobservable in this visualization.
As we see from \cref{fig:sthosvd-perf}, the resulting compressed data sets are small enough to be easily shared across high-speed networks and/or analyzed on workstations as described in \cref{sec:exp-rec}.

\begin{figure}[th]
\centering
\subfloat[ST-HOSVD compression results (independent of processor grid)]{ \label{tab:scenarios}%
  \begin{tabular}{|c|c|c|c|c|c|} \hline
                & \bf Scenario & \bf Relative & \bf Compressed & \bf Total & \bf Compression \\
    \bf Dataset & \bf Name     & \bf Error & \bf Core Size & \bf Storage & \bf Ratio \\ \hline
    \multirow{2}{*}{SP} & High & 1e-2 & $30 \times 38 \times 35 \times 6 \times 11$ & 21.5~MB & $2 \times 10^5$ \\
                & Low & 1e-4 & $95 \times 129 \times 125 \times 7 \times 125$ & 10.7~GB & $4 \times 10^2$ \\ \hline
    \multirow{2}{*}{JICF} & High & 1e-2 & $90 \times 61 \times 48 \times 13 \times 6$ & 167~MB & $4 \times 10^4$ \\
    & Low & 1e-4 & $424 \times 387 \times 261 \times 18 \times 10$ & 45.7~GB & $1 \times 10^2$ \\ \hline
  \end{tabular}}\\
\subfloat[ST-HOSVD on SP dataset with different processor grids]{
\centering
\begin{tikzpicture}[scale=.85]
\renewcommand{\datafile}{data/sthosvd-sp.dat}
\renewcommand{\pga}{1x1x40x1x100}
\renewcommand{\pgb}{10x8x5x1x10}
\renewcommand{\pgc}{40x10x1x1x10}
\renewcommand{\pgalabel}{A}
\renewcommand{\pgblabel}{B}
\renewcommand{\pgclabel}{C}
\renewcommand{\tola}{1e-2}
\renewcommand{\tolb}{1e-4}
\renewcommand{\prep}{None}
\makesthosvdplot
\end{tikzpicture}
\label{fig:sthosvd-sp}
}\hfill
\subfloat[ST-HOSVD on JICF dataset with different processor grids]{
\centering
\begin{tikzpicture}[scale=.85]
\renewcommand{\datafile}{data/sthosvd-jicf.dat}
\renewcommand{\pga}{1x16x35x1x10}
\renewcommand{\pgb}{10x8x7x1x10}
\renewcommand{\pgc}{35x16x1x1x10}
\renewcommand{\pgalabel}{A}
\renewcommand{\pgblabel}{B}
\renewcommand{\pgclabel}{C}
\renewcommand{\tola}{1e-2}
\renewcommand{\tolb}{1e-4}
\renewcommand{\prep}{Max}
\makesthosvdplot
\end{tikzpicture}
\label{fig:sthosvd-jicf}
}\\
\subfloat[STHOSVD per-core memory usage, I/O time, and computation time]{  \label{tab:mem_io}%
    \begin{tabular}[ht]{|c|c|c|c|c|c|c|c|c|c|} \hline
    & & \multicolumn{4}{c|}{\bf SP dataset} & \multicolumn{4}{c|}{\bf JICF dataset} \\
    \bf Proc.   & \bf Compression & \bf Memory & \multicolumn{2}{c|}{\bf I/O Time (s)} & \bf Comp. & \bf Memory & \multicolumn{2}{c|}{\bf I/O Time (s)} & \bf Comp. \\
    \bf Config. & \bf Scenario    & \bf Usage (GB) & \bf Input & \bf Output            & \bf Time (s) & \bf Usage (GB) & \bf Input & \bf Output & \bf Time (s) \\ \hline
    \multirow{2}{*}{A} & High $\epsilon$=1e-2& 1.22 & \multirow{2}{*}{370} & 0.2 & 6  & 1.43 & \multirow{2}{*}{2308} & 2 & 57 \\
    & Low  $\epsilon$=1e-4 & 1.42 & & 22 & 13 & 2.24 & & 424 & 106 \\ \hline
    \multirow{2}{*}{B} & High $\epsilon$=1e-2 & 3.33 & \multirow{2}{*}{877} & 0.9 & 24 & 3.62 & \multirow{2}{*}{2187} & 4 & 137 \\
    & Low  $\epsilon$=1e-4 & 3.33 & & 983 & 38 & 3.62 & & 18517 & 187 \\ \hline
    \multirow{2}{*}{C} & High $\epsilon$=1e-2 & 3.34 & \multirow{2}{*}{861} & 0.8 & 37 & 3.61 & \multirow{2}{*}{2077} & 5 & 180 \\
    & Low  $\epsilon$=1e-4 & 3.34 & & 13470 & 79 & 3.61 & & DNC & 244 \\ \hline    
  \end{tabular}  
}
\caption{ST-HOSVD compression, run time breakdown, memory usage, and I/O cost for different choices of $\epsilon$ (in in \cref{alg:sthosvd}) and processor grids.
The different processor grids (A/B/C) are given in \cref{tab:partitions}. In the breakdowns, since these are 5-way datasets, there are five iterations of ST-HOSVD, each of which calls Gram, Evecs, and TTM. These are stacked from mode 0 (bottom) to mode 4 (top). The mode-0 calls are the most expensive because the tensor is reduced in size for subsequent iterations.}
\label{fig:sthosvd-perf}
\end{figure}

\Cref{fig:sthosvd-perf} shows the run time results of ST-HOSVD using the different compression scenarios in \cref{tab:scenarios} and processor grid configurations in \cref{tab:partitions}.
Note that the degree of compression has no dependence whatsoever on the processors grid configuration; the configuration impacts only the run time.
In the figure, the run times are broken down into color-coded segments corresponding to the three main kernels: Gram, Evecs, and TTM.
As these are five-way tensors, there are five calls to each kernel.
The times corresponding to mode zero are at the bottom and the most expensive.
Since the tensor is reduced at each iteration (via the call to TTM), the calls become significantly cheaper so that the breakdown is hardly apparent
past the first 2-3 iterations/modes.
Comparing the ``high'' and ``low'' scenarios, the initial Gram computations are roughly equivalent, but all other computations in the 1e-2 case finish more quickly than the 1e-4 case because the data is compressed more drastically at each step.
Comparing SP and JICF datasets, the Evecs computation is negligible for SP but is noticeable for JICF because its largest mode sizes are 3--4 times larger than for SP (the eigenvector computation is not parallelized).
Comparing across processor grid configurations (A/B/C), loading more processors onto later modes generally improves run time, as the heavy communication steps are performed on data that can be orders of magnitude smaller than the initial tensor.
The fastest of the processor grids are  3--6$\times$ faster than the slowest of the processor grids.
This is not included in the figure, but the average time taken for preprocessing the JICF data (scaling each mode-3 slice by the inverse of the maximum entry) is 8 seconds and varied minimally across processor grids.
This amounts to at most 15\% of the run time of ST-HOSVD after the tensor is loaded in memory.
The SP data in our experiments has already been scaled, so TuckerMPI does no pre-processing.

We note that this experiment does not consider alternative mode orderings, which can have a significant effect on run time.
In the case of ST-HOSVD with a specified tolerance, the core tensor size is not known a priori, so it is not possible to pick an optimal ordering as discussed in \cref{sec:reconstruction}.
However, one can use some heuristics to pick a mode ordering.
For example, starting with a mode whose processor grid dimension is 1 avoids communication of the tensor in the first Gram (which is typically the most expensive operation).
Indeed, in this experiment, the fastest processor grid (A) has one processor in mode 0 for both SP and JICF data sets, but using the natural mode ordering 01234 is not necessarily optimal, even for that processor grid.

\Cref{tab:mem_io} shows the memory footprints and I/O times of the different cases.
The input SP tensor requires about 1.1~GB per core on 4000 cores, and TuckerMPI requires about 3$\times$ that much memory to complete its computations, which is dominated by the memory required by the initial Gram computation. 
When a processor grid with only one processor in mode 0 is used, the memory footprint is only about 10-30\% more than the initial data, depending on how much compression is achieved through the algorithm.
The JICF data on 5600 cores requires about 1.2 GB per core, and again we see a memory footprint of about $3\times$ that amount.
In terms of I/O time,
the TuckerMPI code uses the MPI I/O interface for reading and writing binary files of multidimensional arrays.
From the input results, we see a file reading bandwidth of 3--10 GB/sec., requiring $O(10)$ minutes to read the input tensors from disk.
In comparison, computing the ST-HOSVD takes $O(1)$ minute per \cref{tab:mem_io}.
For instance, the SP-High-A scenario is 100$\times$ faster than reading the data from from disk.
Writing to disk is much more expensive than reading.
As a result, writing the compressed representation takes a disproportionate amount of time.
We believe this is an artifact of the underlying MPI I/O implementation with the Lustre filesystem on Skybridge.
One experiment, labeled `DNC', did not complete due to an error within the MPI I/O implementation.
Improving I/O performance for the low compression scenario is a topic of future work.

\subsubsection{Strong scaling with synthetic data}
\label{sec:strong-scaling-with}

In this section, we demonstrate the scalability of our code on a synthetic 4D tensor with dimensions $256 \times 256 \times 256 \times 256$.  
The experimental setup is as follows.
We fix the core dimensions to $32\times 32\times 32\times 32$ (4096$\times$ compression) so that all mode orderings are equivalent, and we scale from 1 node up to 128 nodes (2048 cores), using 1 MPI process per core.
Since all the dimensions are the same, the mode ordering is irrelevant.
We used the following processor grids: $1{\times}1{\times}2{\times}8$, $1{\times}1{\times}2{\times}16$, $1{\times}1{\times}4{\times}16$, $1{\times}1{\times}8{\times}16$, $1{\times}1{\times}16{\times}16$, $1{\times}2{\times}16{\times}16$, $1{\times}4{\times}16{\times}16$, $1{\times}4{\times}16{\times}32$, chosen heuristically to have fewer processors in the first and second modes to minimize communication in the early Gram computations, which dominate the run time.

The run times are reported in \cref{fig:strong-scaling}.
The input tensor is 32~GB in size, about half the memory available on a single node.
The performance scales well up to 16 nodes (256 cores), achieving $9\times$ speedup over the single node, because the run time is dominated by the local Gram computation in the first node, which is perfectly parallelized.
The degradation of performance after 16 nodes is caused in large part by the local computation in the first Gram computation (a single call to \texttt{syrk}) failing to scale perfectly.
This is due to the local dimensions becoming too small and performance variability across nodes.
We note that performance reported in \cite{AuBaKo16} shows strong scaling of nearly the same algorithm on nearly the same problem to 256 nodes, but it was benchmarked on a different parallel computer.

\subsubsection{Weak scaling with synthetic data}
\label{sec:weak-scaling-with}

In this section, we demonstrate the weak scalability of our code on synthetic 4D tensors whose sizes scale with the number of processors so that the portion per processor remains constant.
We ran our code on $2^k$ nodes (for $1 \leq k \leq 5$), using 16 MPI processes per node, with a $1 \times 1 \times 4k^2 \times 4k^2$ processor grid.  
We take a tensor of order $200k \times 200k \times 200k \times 200k$  and reduce it to a core size of $20k \times 20k \times 20k \times 20k$.
This means that the local data on each node is approximately 0.8~GB.
The processor grid was chosen as the best performing among $2k\times 2k\times 2k\times 2k$, $k\times k\times 4k\times 4k$ and $1\times k\times 4k\times 4k^2$.
The largest speedup of the $1 \times 1 \times 4k^2 \times 4k^2$ grid over the other grids was about 70\%, and the $1\times k\times 4k\times 4k^2$ grid performed nearly as well.

\begin{figure}[th]
  \centering
\subfloat[Strong scaling performance of ST-HOSVD for $256\times256\times256\times256$ tensor compressed to size $32\times32\times32\times32$ (4096$\times$ compression), using $2^k$ nodes for $0 \leq k \leq 7$. The processor grids are $1{\times}1{\times}2{\times}8$, $1{\times}1{\times}2{\times}16$, $1{\times}1{\times}4{\times}16$, $1{\times}1{\times}8{\times}16$, $1{\times}1{\times}16{\times}16$, $1{\times}2{\times}16{\times}16$, $1{\times}4{\times}16{\times}16$, $1{\times}4{\times}16{\times}32$.]{
\centering
\begin{tikzpicture}[scale=.85]
\renewcommand{\datafile}{data/strong-scaling.dat}
\makestrongscalingplot
\end{tikzpicture}
\label{fig:strong-scaling}
}%
\hfill%
\subfloat[Weak scaling performance of ST-HOSVD for $200k\times200k\times200k\times200k$ tensor with reduced size $20k\times20k\times20k\times20k$, using $k^4$ nodes and a $1\times1\times4k^2\times4k^2$ processor grid for $1 \leq k \leq 5$. For reference, the peak GFLOPS per core is 20.8.]{
\centering
\begin{tikzpicture}[scale=.85]
\renewcommand{\datafile}{data/weak-scaling.dat}
\makeweakscalingplot
\end{tikzpicture}
\label{fig:weak-scaling}
}
\caption{Parallel scaling experiments on synthetic data. There are 16 cores per node.}
\label{fig:scaling}
\end{figure}
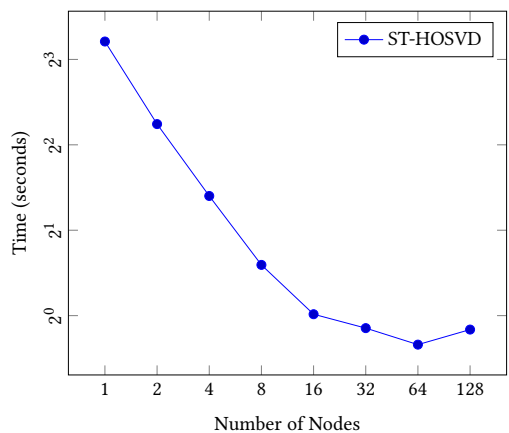
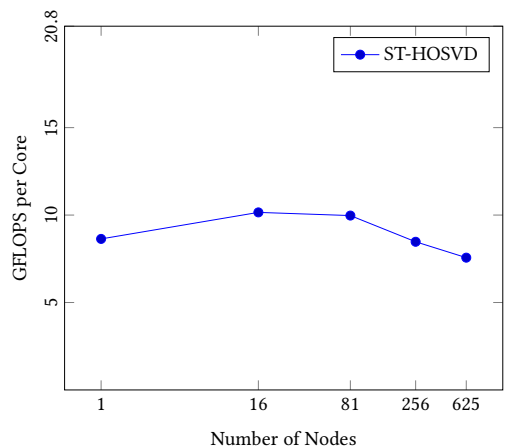

\Cref{fig:weak-scaling} reports the performance in terms of GFLOPS per core.
We observe that weak-scaling performance is generally preserved up to 625 nodes, with the code achieving 40-50\% of peak performance throughout.
In this experiment, the amount of local computation is held fixed, while the communication is increased with the number of processors.
The weak scaling is possible because the run time remains dominated by computation.
On 1 node ($k=1$), the tensor size is 12.8~GB, and on 625 nodes ($k=5$) it is 8~TB. 
Compared to the performance reported in \cite{AuBaKo16}, we observe similar overall performance relative to the architecture.

\subsection{Reconstruction}
\label{sec:exp-rec}

Compressing data makes it cheaper to store and to transmit, but ultimately it needs to be reconstructed to be useful. We expect that most users will do only \emph{partial} reconstructions. These can be used to visualize a portion of the data, extract summary statistics, etc. Moreover, this can oftentimes be done on a workstation rather than a parallel computer, making analysis much simpler.
We can also reconstruct the \emph{entire} dataset, which is primarily useful for comparison to the original dataset for quality control. 
In either case, reconstruction is a \mTTM\ operation as described in \cref{sec:reconstruction}.

\subsubsection{Partial Reconstruction}
\label{sec:part-reconstr}
Our aim in this section is to show that we can do \emph{partial}
reconstructions on a laptop or workstation, so we run all experiments
sequentially using only one MPI process (and thus only one node and one thread).
We note that each node on Skybridge has 64 GB, and that the sequential execution has access to all 64 GB.
We consider the SP dataset, as described in \cref{sec:data_description}.

\paragraph{Visualization of Single Time Step}
One partial reconstruction for visualization is to extract the entire $500 \times 500 \times 500$ grid corresponding to a \emph{single} variable (out of 11) at a \emph{single} timestep (out of 400). 
This results in a tensor of size $500 \times 500 \times 500 \times 1 \times 1$ which requires 1~GB of storage.
This is the first step used to generate the visualizations of the compressed data in \cref{fig:isosurface}, for example.

As discussed in \cref{sec:reconstruction}, the mode ordering is important in the reconstruction. 
We never want the intermediate size to be bigger than the larger of the input and output tensor sizes, but this can happen for the wrong mode ordering. Consider the high compression scenario and the intermediate tensor sizes that result as shown in \cref{tab:reconstruction_sizes}. Using the ordering 01234 results in a 66~GB tensor after the third TTM in the reconstruction. But ordering 43120 never has an intermediate result larger than the final reconstruction.
We stress that any ordering produces the same result (up to floating point error) and the difference is in the size of the intermediate results and the run time.

\begin{table}[th]
  \centering
  \begin{tabular}{|c|c|c|c|c|}
    \hline
        & \multicolumn{2}{c|}{\bf Ordering 01234} & \multicolumn{2}{c|}{\bf Ordering 43120} \\
    \bf TTM & \bf Dimensions & \bf Intermed.~Size (GB) & \bf Dimensions & \bf Intermed.~Size (GB) \\ \hline
    None & $30 \times 38 \times 35 \times 6 \times 11$ & 0.02  & $30 \times 38 \times 35 \times 6 \times 11$ & 0.02 \\ \hline
    1st & $500 \times 38 \times 35 \times 6 \times 11$ & 0.35 & $30 \times 38 \times 35 \times 6 \times 1$ & $<$0.01 \\ \hline 
    2nd & $500 \times 500 \times 35 \times 6 \times 11$ & 4.62 & $30 \times 38 \times 35 \times 1 \times 1$ & $<$0.01 \\ \hline 
    3rd & $500 \times 500 \times 500 \times 6 \times 11$ & 66.0 & $30 \times 500 \times 35 \times 1 \times 1$ & $<$0.01 \\ \hline 
    4th & $500 \times 500 \times 500 \times 1 \times 11$ & 11.0 & $30 \times 500 \times 500 \times 1 \times 1$ & 0.06 \\ \hline 
    5th & $500 \times 500 \times 500 \times 1 \times 1$ & 1.00 & $500 \times 500 \times 500 \times 1 \times 1$ & 1.00 \\ \hline 
  \end{tabular}
  \caption{Intermediate tensor sizes in the partial reconstruction for two different TTM orderings, using the high compression scenario ($\epsilon$=1e-2) for the SP dataset. The TTM ordering can make a dramatic different in memory usage.}
  \label{tab:reconstruction_sizes}
\end{table}

In \cref{tab:partial}, we report the maximum memory usage, the compute time (\mTTM), and the I/O times for different mode orders. 
The sizes of the core tensors on disk are 21 MB and 11 GB per \cref{tab:scenarios}, and the size of the reconstructed output is 1 GB.
The I/O time shows a bandwidth rate of about 1 GB/sec.
In the high compression case, the partial reconstruction is larger than then the input; in contrast, the reverse is true for the low compression scenario.
For both the high and low compression scenarios, the unique minimizer of the flops is mode order 43120, and it is one of the memory minimizers.
Compared with the other orders benchmarked in the experiment, the optimal order runs as much as an order of magnitude faster, and some orders, including the straightforward 01234 order, fail due to out-of-memory errors.

\begin{table}[th]
  \centering
  \begin{tabular}{|c|c|c|c|c|c|}
    \hline
    \bf Compression & \bf Mode  & \bf Max Memory     & \bf Compute    & \multicolumn{2}{c|}{\bf I/O Time (sec)} \\
    \bf Scenario    & \bf Order & \bf Usage (GB) & \bf Time (sec) & \bf Input & \bf Output \\ \hline
    \multirow{5}{*}{\shortstack{High\\$\epsilon$=1e-2}} & 01234 & \multicolumn{2}{c|}{\em out of memory} & \multirow{5}{*}{\shortstack{0.05\\(21.5~MB)}} & \multirow{10}{*}{\shortstack{1.01\\(1~GB)}} \\ \cline{2-4}
                    & 03421 & 1.08 & 1.55 & & \\ \cline{2-4}
                    & 24013 & 7.00 & 8.88 & & \\ \cline{2-4}
                    & 34021 & 7.00 & 1.19 & & \\ \cline{2-4}
                    & 43120 & 1.08 & 1.05 & & \\ \cline{1-5}
    \multirow{5}{*}{\shortstack{Low\\$\epsilon$=1e-4}} & 01234 & \multicolumn{2}{c|}{\em out of memory} & \multirow{5}{*}{\shortstack{12.23\\(10.7~GB)}} & \\ \cline{2-4}
                    & 03421 & \multicolumn{2}{c|}{\em out of memory} & & \\ \cline{2-4}
                    & 24013 & 53.62 & 129.70 & & \\ \cline{2-4}
                    & 34021 & 12.26 & 5.83 & & \\ \cline{2-4}
                    & 43120 & 10.81 & 4.72 & & \\ \hline
  \end{tabular}
  \caption{Partial reconstruction results on the SP dataset with a \emph{single} MPI process, computing a partial reconstruction of size $500 \times 500 \times 500 \times 1 \times 1$ (1~GB).}
  \label{tab:partial}
\end{table}

\paragraph{Computing a Summary Statistic}
As an example summary statistic for the SP data, we can compute the average fraction of carbon dioxide across all space and time in a few seconds.
Mathematically, 
in the notation of \cref{sec:reconstruction}, we set $\FacMat[C]{0}$, $\FacMat[C]{1}$, $\FacMat[C]{2}$, and $\FacMat[C]{4}$ to be all-ones vectors and $\FacMat[C]{3}$ (the mode corresponding to variables) to be all zeros except for a one in the index that corresponds to carbon dioxide.
\Cref{tab:gen-partial} shows the results of this computation on a single node for both compression scenarios.
The average fraction of carbon dioxide across all space and time is 0.0534 (calculated using the original data).%
\footnote{In preprocessing, the data was rescaled to have a maximum value of 0.5, and the mean based on the rescaled data was 0.3014.}
In comparison,
the relative error of the averages computed from the compressed versions are 6.7e-6 (high compression) and 5.7e-8 (low compression).
The memory required is only that of storing the compressed data, and the time is dominated by the cost of reading the core tensor from disk.
For the high compression case, the computation took less than 1 second in total; for the low compression case, it required less than 16 seconds.
Recall that working with the original data set requires a parallel computer and 100s of nodes just to read the data.

\begin{table}[th]
  \centering
  \begin{tabular}{|c|c|c|c|c|c|c|}
    \hline
    \bf Compression & \bf Mode  & \bf Max Memory     & \bf Compute    & \multicolumn{2}{c|}{\bf I/O Time (sec)} & \bf Relative \\
    \bf Scenario    & \bf Order & \bf Usage (GB) & \bf Time (sec) & \bf Input & \bf Output & \bf Error \\ \hline
    High ($\epsilon$=1e-2) & 01234 & 0.022 & 0.16 & 0.69 (21.5~MB) &  \multirow{2}{*}{0.04 (8~B)} & 6.7e-6 \\ 
    Low ($\epsilon$=1e-4) & 01234 & 10.84 & 2.63 & 13.02 (10.7~GB) & & 5.7e-8 \\ 
    \hline
  \end{tabular}
  \caption{Summary statistic results on the SP dataset with a \emph{single} MPI process, computing the average fraction of carbon dioxide across all time steps and entire physical grid. Relative error is in comparison to the value computed based on the original data.}
  \label{tab:gen-partial}
\end{table}

\subsubsection{Full reconstruction of the SP data}
\label{sec:full-reconstr-sp}

As mentioned above, we do not expect users to employ full reconstruction very often, but it is useful as a diagnostic tool to check the quality of the approximation.
Hence, we reconstructed the full SP dataset for both the high and low compression scenarios using a variety of mode orderings and processor configuration A ($1 \times 1 \times 40 \times 1 \times 100$) since it resulted in the fastest compression time.
The results are reported in \cref{tab:reconstruction}.

The I/O timings are averaged over the runs with different mode orderings. 
As the output is about 400 times larger than the larger of the two inputs, the overall time for the experiment is dominated by writing the output to disk, which takes almost an hour.
(This supports the idea of avoiding reading and writing the full data sets to disk.)
As compared with the bandwidth rate of one MPI process (see \cref{tab:partial}), the parallel I/O bandwidth rate is slower for the smaller inputs and about 50\% faster for the larger output.

As discussed in \cref{sec:reconstruction}, the computational and communication costs for the \mTTM, as well as the temporary memory footprint, are all dependent on the mode ordering of the individual TTMs.
Furthermore, the mode ordering that minimizes computation need not be the same as the one that minimizes communication or memory.
For the high compression scenario, mode order 41203 minimizes computation, 42103 minimizes communication, and 34120 minimizes memory.
We also experiment with ordering 12034, which requires about 60\% more temporary memory than the optimal orders and yields an out-of-memory error in the high compression scenario. 
The 34120 order yields minimum max memory usage for both scenarios,
and  42103 is the fastest for both scenarios.

\begin{table}[th]
  \centering
  \begin{tabular}{|c|c|c|c|c|c|}
    \hline
    \bf Compression & \bf Mode  & \bf Max Memory     & \bf Compute    & \multicolumn{2}{c|}{\bf I/O Time (sec)} \\
    \bf Scenario    & \bf Order & \bf Usage (GB) & \bf Time (sec) & \bf Input & \bf Output \\ \hline
    \multirow{4}{*}{\shortstack{High\\$\epsilon$=1e-2}} & 12034 & 2.57 & 320.34 & \multirow{4}{*}{\shortstack{4.02\\(21.5~MB)}} & \multirow{8}{*}{\shortstack{3118.57\\(4.4~TB)}} \\ \cline{2-4}
                    & 34120 & 1.21 & 27.51 & & \\ \cline{2-4}
                    & 41203 & 1.77 & 16.39 & & \\ \cline{2-4}
                    & 42103 & 1.77 & 5.04 & & \\ \cline{1-5}
    \multirow{4}{*}{\shortstack{Low\\$\epsilon$=1e-4}} & 12034 & \multicolumn{2}{c|}{\em out of memory} & \multirow{4}{*}{\shortstack{15.24\\(10.7~GB)}} & \\ \cline{2-4}
                    & 34120 & 1.37 & 82.46 & & \\ \cline{2-4}
                    & 41203 & 1.88 & 53.74 & & \\ \cline{2-4}
                    & 42103 & 1.88 & 19.43 & & \\ \hline
  \end{tabular}
  \caption{Full reconstruction results of the SP dataset using 250 nodes/4000 MPI processes. The processor grid is size $1 \times 1 \times 40 \times 1 \times 100$. The ``Max Memory Usage'' is per process.}
  \label{tab:reconstruction}
\end{table}

%
%
%
%
%

\section{Conclusion}
\label{sec:conclusion}

The Tucker tensor decomposition is useful for compression of many
large-scale datasets because it uncovers latent low-dimensional
structure.
For multi-terabyte datasets that arise in direct numerical simulation
of combustion reactions, we demonstrate that the Tucker decomposition
can obtain five orders of magnitude in compression by reducing a
4.4~TB dataset to 21.5~MB. The compression time is an order of magnitude faster than simply reading the data.
Austin, Bader, and Kolda \cite{AuBaKo16} proposed the first parallel Tucker implementation.
In this paper, we build upon their work, explaining the details of the parallel and serial
algorithms as well as the local and distributed data layouts.
Additionally, we have improved upon the algorithm in \cite{AuBaKo16}
and so can now run on much larger datasets.
We also explain in detail how to reconstruct portions of the data, one
of the most important benefits of the Tucker decomposition, without
requiring any parallel resources.
We show that it is possible to reconstruct portions of the data on a
single processor in only a few seconds and using no more memory than the size
of the input or output (whichever is larger).

Other parallel algorithms and implementations for computing Tucker decompositions of dense tensors using Higher-Order Orthogonal Iteration include \cite{CC+17}, which uses TTM-trees and performs dynamic regridding to optimize computation and communication costs, and \cite{MS18}, which uses TTM-trees and approximate information to update factor matrices in each iteration.
Choi, Lui, and Chakaravarthy \cite{CLC18} implement ST-HOSVD for dense tensors on a cluster equipped with GPUs, and they also use randomized SVD algorithms to improve run time.
In the sparse case, there have also been several algorithmic innovations to improve the parallel performance of Higher-Order Orthogonal Iteration \cite{KU16,SK17,OPLK18}; these optimizations exploit the sparsity of the data tensor and avoid memory overhead of temporary dense data structures.

In future work, we hope to create a library for \emph{in situ}
compression that can compress at each time step in a
simulation. Ideally, the simulation would never write the full dataset
to disk but only compressed versions. Although these compressed
versions cannot be used for restart in the event of a failure, they
are a sort of ``thumbnail'' of the full simulation. This would save
both time (for I/O) and disk space, not to mention making the sharing
of data much easier.

We have not compared our approach to other methods of compression but hope to do so in future studies.
Ballester-Ripoll, Lindstrom, and Pajarola \cite{BaLiPa19} have recently compared Tucker compression (computed in serial)
to the ZFP, SZ, and SQ compression methods and determined that this method ``typically produces renderings that are already close to visually indistinguishable to the original data set.''
Most other compression methods are focused on pointwise errors and achieve only about $O(10)$ compression and moreover divide up the space into blocks and compress the blocks individually.
The Tucker approach we use here bounds the \emph{overall} error and looks for large-scale patterns with the advantage being that it can get much higher compression, e.g., up to $O(10^5)$.
A comparison study would need to determine how to compare across different types of compression metrics (e.g., overall versus pointwise) and have a common large-scale dataset for the comparisons.
We note that we can also potentially combine compression methods, using other methods to compress the core that results from the Tucker compression.
Indeed, the ``readme'' file for the SZ Fast Error-Bounded Scientific Data Compressor lists Tucker compression as an option.\footnote{See https://github.com/disheng222/SZ/blob/master/README (ver. 12c9356), line 133.}

Li~et~al.~\cite{LiGrPoCl15} evaluate the impact of wavelet compression (up to 512$\times$) for turbulent-flow data on visualization and analysis tasks, and such analysis would be interesting to consider also in the case of Tucker compression.

The data we compress in our study corresponds to a regular rectilinear grid. Another
topic for future work is dealing with unstructured and/or
non-rectilinear grids that are not neatly represented as a tensor. We
also assume that the data is dense, but there are applications in data
mining that have sparse tensors so this is another potential topic for study.

%
%
%
%

\appendix
\section*{Acknowledgments}
We would like to thank the anonymous referees for their comments and suggestions on the paper.
We are indebted to Woody Austin for his work on the original prototype software on which this work is based.
We thank Hemanth Kolla for creating \cref{fig:isosurface} and his overall support of this work.
We are also grateful to our colleagues Gavin Baker, Casey Battaglino, Cannada (Drew) Lewis, Jed Duersch, Alex Gorodetsky, and Prashant Rai for discussions about this software. 

This material is based upon work supported in part by the U.S. Department of Energy, Office of Science, Office of Advanced Scientific Computing Research, Applied Mathematics program.
Sandia National Laboratories is a multimission laboratory managed and operated by National Technology and Engineering Solutions of Sandia, LLC., a wholly owned subsidiary of Honeywell International, Inc., for the U.S. Department of Energy's National Nuclear Security Administration under contract DE-NA-0003525.

\bibliographystyle{siamplain}


%

\end{document}